\documentclass[12pt, oneside]{article}
\pdfoutput=1
\usepackage[export]{adjustbox}
\usepackage[paper=a4paper, total={16cm,23cm}]{geometry}
\usepackage[T1]{fontenc}
\usepackage{setspace} 
\usepackage[hang]{footmisc} 
\usepackage{titling} 
\usepackage{tocloft} 
\usepackage{parskip} 
\usepackage{hyperref}
\usepackage{fourier-orns}

\usepackage{amsfonts}
\usepackage{amsmath}
\usepackage{amssymb}
\usepackage{mathrsfs, dsfont, amsthm}
\usepackage{physics}
\usepackage{mathtools}
\numberwithin{equation}{section} 

\usepackage[dvipsnames]{xcolor}
\definecolor{dark-red}{rgb}{0.50,0.12,0.12}
\definecolor{mblue}{rgb}{0.30, 0.45, 0.70}
\definecolor{mred}{rgb}{0.70, 0.20, 0.20}
\definecolor{mgray}{rgb}{0.63, 0.63, 0.63}
\definecolor{myWhite}{RGB}{255,255,243}

\newcommand*\justify{%
  \fontdimen2\font=0.4em
  \fontdimen3\font=0.2em
  \fontdimen4\font=0.1em
  \fontdimen7\font=0.1em
  \hyphenchar\font=`\-
}

\renewcommand{\texttt}[1]{%
  \begingroup
  \ttfamily
\begingroup\lccode`~=`/\lowercase{\endgroup\def~}{/\discretionary{}{}{}}%
\begingroup\lccode`~=`[\lowercase{\endgroup\def~}{[\discretionary{}{}{}}%
\begingroup\lccode`~=`.\lowercase{\endgroup\def~}{.\discretionary{}{}{}}%
\catcode`/=\active\catcode`[=\active\catcode`.=\active
  \justify\scantokens{#1\noexpand}%
  \endgroup
}

\usepackage{hyperref} 
\hypersetup{
colorlinks=true,
allcolors=dark-red,
linktoc=page,
pdftitle={SOFIA: Singularities of Feynman Integrals Automatized},
pdfauthor={},
pdfdisplaydoctitle=true,
pdfstartview=FitH
}

\usepackage{cite}

\usepackage{tikz}
\usepackage{tikzlings}
\usetikzlibrary{calc, patterns, decorations, decorations.markings, shapes, positioning, intersections, quotes, angles}
\usepackage[subrefformat=parens]{subcaption}
\usepackage{pgfplots}
\pgfplotsset{compat=newest}
\newcommand{\mathdefault}[1][]{}
\usetikzlibrary{positioning}
\usetikzlibrary{calc}
\definecolor{mylightred}{RGB}{211,79,73}
\definecolor{mydarkred}{RGB}{199,44,38}
\definecolor{mylightgreen}{RGB}{78,153,67}
\definecolor{mydarkgreen}{RGB}{43,129,33}
\definecolor{mylightpurple}{RGB}{150,107,178}
\definecolor{mydarkpurple}{RGB}{126,78,160}
\definecolor{mylightblue}{RGB}{49,101,205}
\definecolor{mydarkblue}{RGB}{20,92,205}
\tikzset{
  juliadot/.style args={#1,#2}{shape=circle,line width=0.03ex,minimum width=0.4ex,fill=#1,draw=#2}
}

\makeatletter
\newenvironment{sqcases}{
  \matrix@check\sqcases\env@sqcases
}{
  \endarray\right.
}
\def\env@sqcases{
  \let\@ifnextchar\new@ifnextchar
  \left\lbrack
  \def\arraystretch{1.2}
  \array{@{}l@{\quad}l@{}}
}
\makeatother

\def \d   {\mathrm{d}}

\newcommand{\D}{\mathrm{D}}
\newcommand{\ed}{\mathrm{d}}

\makeatletter
\newcommand{\subalign}[1]{
  \vcenter{
    \Let@ \restore@math@cr \default@tag
    \baselineskip\fontdimen10 \scriptfont\tw@
    \advance\baselineskip\fontdimen12 \scriptfont\tw@
    \lineskip\thr@@\fontdimen8 \scriptfont\thr@@
    \lineskiplimit\lineskip
    \ialign{\hfil$\m@th\scriptstyle##$&$\m@th\scriptstyle{}##$\hfil\crcr
      #1\crcr
    }
  }
}
\makeatother

\usepackage{mathtools}
\usepackage{comment}
\newcommand{\beq}{\begin{equation}}
\newcommand{\eeq}{\end{equation}}

\newcommand{\G}{\mathcal{G}}
\renewcommand{\le}{\leqslant}
\renewcommand{\leq}{\leqslant}
\renewcommand{\ge}{\geqslant}
\renewcommand{\geq}{\geqslant}
\def\mylongmapsto#1{%
\begin{tikzpicture}
\draw (0,0.5mm) -- (0,-0.5mm);
\newlength\mylength
\setlength{\mylength}{\widthof{#1}}
\draw[->] (0,0) -- (1.2\mylength,0) node[above,midway] {#1};
\end{tikzpicture}
}

\usepackage{tcolorbox}
\usepackage{colortbl} 
\usepackage{listings}
\usepackage{setspace} 
\definecolor{identifiercolor}{rgb}{.4,.6,.56}
\definecolor{stringcolor}{gray}{0.5}
\definecolor{inactivecolor}{rgb}{0.2,0.2,0.2}

\newcommand{\SOFIA}{\adjustbox{valign=t,scale=0.7,raise=0.05em}{\includegraphics[height=1em]{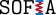}}}
\newcommand{\version}{v1.0.0}

\def\mg#1{}
\def\sm#1{}
\def\mc#1{}

\begin{document}
\raggedbottom
\begin{titlingpage}
    \vspace*{3em}
    \onehalfspacing
    \begin{center}
        \textbf{\Large 
      {\SOFIA} \\ Singularities of Feynman Integrals Automatized}
    \end{center}
    \singlespacing
    \vspace*{2em}
    \begin{center}
        \textbf{
        Miguel Correia,$^1$ Mathieu Giroux,$^1$
        Sebastian Mizera$^{2,3,4}$
        }
    \end{center}
    \vspace*{1em}
    \begin{center}
        \textsl{
        $^1$\ Department of Physics, McGill University \\
        Montr\'eal, QC H3A 2T8, Canada \\[\baselineskip]
        }
        \textsl{
        $^2$\ Department of Physics, Princeton University\\
        Princeton, NJ 08544, USA \\[\baselineskip]
        }
        \textsl{
        $^3$\ Princeton Center for Theoretical Science, Princeton University\\
        Princeton, NJ 08544, USA \\[\baselineskip]
        }
        \textsl{
        $^4$\ Institute for Advanced Study \\
         Einstein Drive, Princeton, NJ 08540, USA \\[\baselineskip]
        }
        \href{mailto:miguel.ribeirocorreia@mcgill.ca}{\small\texttt{miguel.ribeirocorreia@mcgill.ca}},
        \href{mailto:mathieu.giroux2@mail.mcgill.ca}{\small\texttt{mathieu.giroux2@mail.mcgill.ca}},
        \href{mailto:smizera@ias.edu}{\small\texttt{smizera@ias.edu}}
    \end{center}
    \vspace*{3em}
    \begin{abstract}
 \noindent We introduce {\SOFIA}, a \textsc{Mathematica} package  that automatizes the computation of singularities of Feynman integrals, based on new theoretical understanding of their analytic structure. Given a Feynman diagram, {\SOFIA} generates a list of potential singularities along with a candidate symbol alphabet. The package also provides a comprehensive set of
 tools for analyzing the analytic properties of Feynman integrals and related objects, such as cosmological and energy correlators. We showcase
 its capabilities by reproducing known results and predicting singularities and symbol alphabets of Feynman integrals at and beyond the high-precision frontier.
    \end{abstract}
    \vfill
    \centering{\version}
\end{titlingpage}
\newpage

\textsc{Program summary}

\emph{Program title}: {\SOFIA} (Singularities of Feynman Integrals Automatized)


\emph{Developer's repository link}: \href{https://github.com/StrangeQuark007/SOFIA}{\texttt{https://github.com/StrangeQuark007/SOFIA}} \cite{repo}

\emph{Licensing provisions}: MIT license

\emph{Programming language}: \textsc{Mathematica} 13 or higher

\emph{Supplementary material}: The example file \texttt{SOFIA\_examples.nb} in \cite{repo}.





\newpage
\tableofcontents 
\setcounter{page}{2}

\section{Introduction}

High-precision predictions for particle colliders rely more than ever on the computation of multi-loop Feynman integrals. For this reason, the study of Feynman integrals and their analytic structure has become a major area of research in theoretical particle physics (see, e.g., \cite{Weinzierl:2022eaz,Badger:2023eqz} for reviews). A central challenge in these computations is the reliable prediction of Feynman integral singularities as functions of kinematic parameters, without the need for a direct evaluation of the integral. In this paper, we tackle this problem by introducing {\SOFIA} (Singularities of Feynman Integrals Automatized), a \textsc{Mathematica} package designed for this purpose, which we release for public use:
\begin{equation}
\text{\href{https://github.com/StrangeQuark007/SOFIA}{\texttt{SOFIA GitHub repository}} \cite{repo}.}\nonumber
\end{equation}

\paragraph{Differential equations.}
The most common application is in computing Feynman integrals using the method of differential equations \cite{Kotikov:1990kg}. In the simplest cases, these equations take the form
\begin{equation}\label{eq:differential-equation}
\d \vec{I} = \sum_{i} \mathbf{\Omega}_{i}\, \d \log (W_i)\, \vec{I}\, ,
\end{equation}
where $\vec{I}$ is a vector of basis Feynman integrals with $\d$ refers to the differential in the kinematic variables. On the right-hand side, $\mathbf{\Omega}_{i}$ are matrices with entries rational in the kinematics and the spacetime dimension $\D$. In fact, one often looks for bases in which all $\mathbf{\Omega}_i$ are linear in the dimensional-regularization parameter $\varepsilon = \frac{4-\D}{2}$ \cite{Henn:2013pwa}. In this case, the resulting Feynman integrals $\vec{I}$ are often expressible in terms of multiple polylogarithms. The arguments of the logarithms, $W_i$, are known as the \emph{letters} \cite{Goncharov:2010jf}. Being able to predict these letters is currently one of the main bottlenecks in computations of polylogarithmic integrals: once $W_i$'s are known, one can fit the coefficients $\mathbf{\Omega}_i$ and consequently compute $\vec{I}$ using the differential equation together with some boundary conditions. 

Many caveats and extra complications can arise in this story, but the motivation remains the same: We want to determine the set of all possible $W_i$'s known as the \emph{alphabet}. This application is the main motivation for our work.

\begin{figure}
    \centering
    \begin{subfigure}{0.32\textwidth}
        \centering
        \includegraphics[scale=0.83,valign=c]{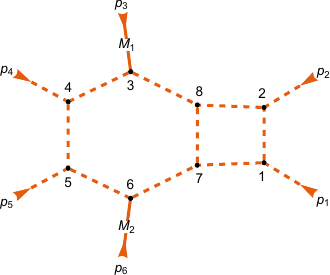}
        \caption{\centering\textbf{135 singularities} (10m)\\
        \textbf{369 letters} (84m)\\
        \texttt{SolverBound -> 1000}}\label{fig:diagrams0-a}
    \end{subfigure}
    \hfill
    \begin{subfigure}{0.32\textwidth}
        \centering
        \raisebox{1.83cm}{\includegraphics[scale=0.83,valign=c]{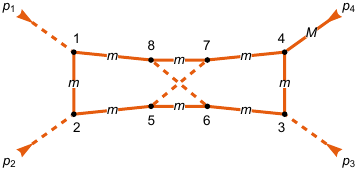}}
        \caption{\centering\textbf{125 singularities} (11m)\\
        \textbf{237 letters} (70m)\\
        \texttt{SolverBound -> 100}}\label{fig:diagrams0-b}
    \end{subfigure}
    \hfill
    \begin{subfigure}{0.32\textwidth}
        \centering
        \raisebox{1.89cm}{\includegraphics[scale=0.83,valign=c]{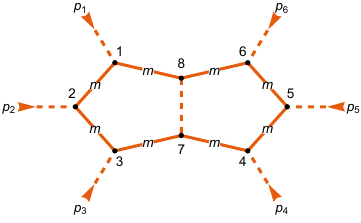}}
        \caption{\centering\textbf{275 singularities} (10m)\\
        \textbf{? letters} \\
        \texttt{SolverBound -> 500}}\label{fig:diagrams0-c}
    \end{subfigure}
    \caption{\label{fig:diagrams0}Examples of cutting‐edge diagrams with timings (in minutes) to obtained singularities and letters. See \cite{repo} for expressions. 
    Dashed lines denote massless particles and otherwise mass labels are displayed along the edges. The choice of \texttt{SolverBound} is determined by the maximum value for which the letters could be subsequently obtained. For diagram (c), although the singularity analysis finished in a reasonable time, the extraction of letters did not complete (hence the ``?'') for any reasonable choice of \texttt{SolverBound}. For diagram (a), we used a machine with two Intel Xeon Gold CPUs and 512GB memory, while for (b) and (c) we used a MacBook Pro with an M3 chip and 16GB memory. 
    }
\end{figure}

\paragraph{Local behavior near singularities.}
Since problems involving analytic structure are notoriously subtle, let us be more precise about the type of questions that we aim to address. A \emph{singularity} refers to a pole or a branch point in the (complexified) kinematic space. Let us collectively denote the kinematic variables as $\vec{s}\in\mathbb{C}^{\mathfrak{s}}$: they consist of the $\mathfrak{s}$ Mandelstam invariants and masses involved in the problem. While non-perturbatively scattering amplitudes might have a much richer singularity structure, based on examples in perturbation theory \cite{Bogner:2017xhp}, it is expected that singularities of the type
\begin{equation}\label{eq:local-behavior}
I_a \sim f_j^A\, \log^B f_j\, ,
\end{equation}
where $f_j = f_j(\vec{s})$ is a polynomial and $A$, $B$ are rational numbers. For example, $\sim \sqrt{f_j} \log f_j$ is one possibility. Note that this refers to a \emph{local} behavior of the amplitude near the locus $f_j = 0$. It is much simpler than the \emph{global} behavior which has to be compatible with \eqref{eq:local-behavior} for a list of all possible $f_j$'s. Indeed, it is known that Feynman integrals are expressed in terms of complicated functions such as multi-polylogarithms, elliptic functions, periods of higher-genus Riemann surfaces and Calabi--Yau manifolds, etc. \cite{Caola:2022ayt,Bourjaily:2022bwx} even though their local behavior \eqref{eq:local-behavior} remains simple. The goal of singularity analysis is to exploit this simplicity.

The connection between the local behavior near singularities \eqref{eq:local-behavior} and the global structure of differential equations \eqref{eq:differential-equation} is very simple. It must be that zeros, poles, and singularities of $W_i$'s are contained in the list of singularities $f_j$. Indeed, our goal will be to first compute a (possibly overcomplete) list of $f_j$'s. Then, we will reconstruct a list of possible $W_i$'s compatible with the aforementioned constraints.

\paragraph{Physical interpretation.}
In the literature, the aforementioned singularities $f_j = 0$ are also interchangeably known as \emph{Landau singularities} or \emph{anomalous thresholds} \cite{Bjorken:1959fd,Landau:1959fi,10.1143/PTP.22.128,Cutkosky:1960sp}. 
The equation $f_j = 0$ is a condition for a network of classical on-shell particles following their classical, though possibly complexified, paths. The simplest case corresponds to positive energies and real angles \cite{Coleman:1965xm}, i.e., the physical region. But the interpretation extends to the rest of the physical sheet, and in fact any sheet, after relaxing positivity and reality conditions; see, e.g., \cite{Caron-Huot:2023ikn}.

\paragraph{Recent progress.}
There has been an explosion of recent activity in the topic of determining singularities of Feynman integrals. Let us mention a couple of algorithms that have been implemented as parts of public packages and can handle state-of-the-art Feynman integrals. \texttt{HyperInt} implements a reduction algorithm that finds a superset of Landau singularities \cite{Panzer:2014caa}. \texttt{PLD.jl} implements an algorithm based on toric geometry that finds a computable subset of Landau singularities \cite{Fevola:2023kaw,Fevola:2023fzn}. Both of these approaches work in the Schwinger-like parametrizations of Feynman integrals. A database comparing the two codes can be found in \cite{MathRepo}. The package \texttt{Baikovletter} attempts to find symbol alphabets using Baikov representations \cite{Jiang:2024eaj}. 
We refer the readers to Sec.~\ref{sec:SingAnalysis} for more detailed comparison and the introduction of \cite{Fevola:2023fzn} for a more complete list of references.

Importantly, it was recently observed that applying singularity analysis to the elastic unitarity constraint (relating the imaginary part of a diagram to its ``cut'' sub-diagrams) yields a powerful recursion formula for the calculation of Landau singularities in the loop momentum space \cite{Caron-Huot:2024brh}. This method enables efficient computation of the leading Landau singularities in two-particle reducible diagrams, successfully reproducing known results from the database \cite{MathRepo} while also making new predictions, including all-loop examples such at the families of ladders and penta-ladders.

In this paper, we extend this approach to any Feynman diagram. Building on the aforementioned advanced computational techniques and additional theoretical understanding, we propose an algorithm that offers a systematic way to study the full set of Landau singularities in general Feynman diagrams.

\paragraph{New contributions.}
The main goal of {\SOFIA} is to streamline the analysis of singularities with minimal user input. The package comes with a companion \textsc{Mathematica} notebook containing 35 examples, where known results are reproduced and new predictions are given. This paper serves as a manual for {\SOFIA}.

The first main novelty comes from a new variant of \emph{Baikov representations} \cite{Baikov:1996iu,Frellesvig:2024ymq} adapted to minimize the number of integration variables and polynomials involved in the singularity analysis. {\SOFIA} implements an optimization technique that results in a minimal integral representation (according to a metric explained in Sec.~\ref{sec:Baikov}) without any user input 
besides the Feynman diagram under consideration. 
We believe it could become useful in other aspects of Feynman analysis that rely on Baikov representations, such as integration-by-parts reduction or analyzing Feynman geometries \cite{Frellesvig:2017aai,Bogner:2019lfa,Frellesvig:2019uqt,Chen:2022lzr,Giroux:2022wav,Pogel:2022vat,Delto:2023kqv,Giroux:2024yxu,Duhr:2024bzt,Frellesvig:2024zph}. We illustrate the latter application in Sec.~\ref{sec:example-geometries}. Likewise, we extend our analysis to eikonalized Feynman integrals, such as those in post-Minkowskianan expansions with an example in Sec.~\ref{sec:example-eikonal}.

The second key innovation is the integration of advanced polynomial system solving methods for singularity analysis into the optimized Baikov representation. In particular, we implement a variant of the \emph{Fubini reduction algorithm} \cite{Brown:2009ta} that results in an upper bound on the singularity locus. We also implement a \textsc{Mathematica} interface to \texttt{PLD.jl}, which allows one to seamlessly perform a numerical and symbolic analysis on a given diagram. We emphasize that this algorithm can be applied to any Euler integral \cite{Matsubara-Heo:2023ylc} and give a couple of examples of applications to cosmological and energy correlators in Sec.~\ref{sec:example-beyond}.

The third and final significant novelty is the implementation of graph‐theoretical tools to break down and speed up the computation of singularities. It has long been observed that the symmetries of a Feynman diagram lead to an identification between Schwinger parameters of the leading physical sheet singularity \cite{kolkunov1960singular,KORS} (see \cite{Correia:2021etg} for a modern application in the context of the S-matrix bootstrap). In this work, we extend this idea by relating different sets of subleading singularities through the construction and identification of subtopologies that are connected via symmetry transformations accounting for variations in their kinematics. This procedure effectively mitigates the factorial growth in the number of subtopologies that would otherwise need to be analyzed.

Finally, we use the recent implementation of the publicly available packge \texttt{Effortless} \cite{Effortlessxxx} to provide an end-to-end function that takes a diagram as the input and provides a candidate symbol alphabet as the output. By ``candidate'' we mean that the alphabet might be overcomplete. An example is provided in Sec.~\ref{sec:example-alphabet}. 

We believe that {\SOFIA} establishes substantial new milestones in the singularity analysis of Feynman integrals, producing results that were previously unattainable. The ancillary files \cite{repo} contain several new predictions for singularities and alphabets, including those presented in Fig.~\ref{fig:diagrams0}, which exemplify how {\SOFIA} can be applied to next-frontier examples, namely, $HZ+2$ jets, $H+$jet, and multi-jets production through light- (for (a)) or heavy-quark (for (b) and (c)) loop‐induced gluon fusion at the LHC.

We hope that {\SOFIA} will serve as a tool to explore new avenues and sheds light on the analytic structures of perturbative scattering amplitudes that were previously inaccessible.

\paragraph{Outline.}
We provide theoretical background in Sec.~\ref{sec:theoretical-background}, including the discussion of different variants of the Baikov representation, singularity analysis of integrals, and mapping from singularities to symbol alphabets. The manual of {\SOFIA} is given in Sec.~\ref{sec:manual} with an outline of all the functions and options. Sec.~\ref{sec:examples} focuses on examples and various applications. Finally, future directions are discussed in Sec.~\ref{sec:conclusion}.

\section{\label{sec:theoretical-background}Theoretical background}

The purpose of this section is to introduce three concepts: the Baikov representations, singularity analysis, and symbol alphabets. 

\subsection{\label{sec:Baikov}Baikov representations}

{\SOFIA} is based on the \emph{Baikov representation} of Feynman integrals.
In this section, we illustrate two variations of the Baikov representation: global and loop-by-loop. We will illustrate them directly on the simple example of the sunrise diagram. More complicated examples will be studied in Sec.~\ref{sec:examples}, and a detailed derivation of the Baikov representation can be found in App.~\ref{app:Baikov}. 
All the (loop-by-loop) Baikov representations given below can be generated with the {\SOFIA} package described in Sec.~\ref{sec:functions}.

\paragraph{Normalization.}
Throughout this paper and in the package, we use the following conventions for scalar Feynman integrals
\begin{equation}\label{eq:loopMomDef}
I_{\vec{\nu}}(\vec{s}) = \frac{1}{(i\pi^{\D/2})^L} \int \frac{\d^\D \ell_1 \, \d^\D \ell_2 \cdots \d^\D \ell_L}{(q_1^2 - m_1^2 )^{\nu_1} (q_2^2 - m_2^2 )^{\nu_2} \cdots (q_N^2 - m_N^2 )^{\nu_N}}\, ,
\end{equation}
where $\D$ is the spacetime dimension, $\ell_a$ denote the loop momenta, and $q_e$ and $m_e$ are the momenta and masses of the internal edges. The $\nu_i$'s are the (integer) powers of the propagators.
The list $\vec{s}$ collectively denotes the external kinematic variables such as Mandelstam invariants and masses.

\paragraph{General strategy.}
The idea of Baikov representation is to translate the integration over the loop momenta $\ell_a$ into that over Lorentz products of the form $\ell_a \cdot p_i$ and $\ell_a \cdot \ell_b$, where $p_i$ denote the external momenta in all-incoming conventions. Schematically, we want to convert the integration measure
\begin{equation}
\d^\D \ell_1 \, \d^\D \ell_2 \cdots \d^\D \ell_L
\quad\mapsto\quad
\mathcal{J}\, \d (\ell_1 \cdot p_1)\, \d(\ell_1 \cdot p_2) \cdots \d (\ell_1 \cdot \ell_1)\, \d (\ell_1 \cdot \ell_2) \cdots\, .
\end{equation}
The change of variables comes with a Jacobian which we called $\mathcal{J}$. We can go one step further after we notice that the new variables are related to inverse propagators $x_i = (\ell_a + \ldots + p_i + \ldots)^2 - m_i^2$, which are always quadratic in the loop momenta, by a linear transformation. Therefore, we can express
\begin{equation}\label{eq:change-of-variables}
\d^\D \ell_1 \, \d^\D \ell_2 \cdots \d^\D \ell_L
\quad\mapsto\quad \tilde{\mathcal{J}}\, \underbracket[0.4pt][2.7pt]{\d x_1 \, \d x_2\, \ldots \, \d x_E}_{\text{inverse propagators}} \, \underbracket[0.4pt][2pt]{\d x_{E+1} \, \ldots \d x_\mathcal{N}}_{\text{ISPs}}\, ,
\end{equation}
with yet another Jacobian $\tilde{\mathcal{J}}$.
The caveat is that the number of variables $\mathcal{N}$ is in general larger than the number of propagators $E$. Therefore, $x_1, \ldots x_E$ label inverse propagators, while $x_{E+1}, \ldots, x_\mathcal{N}$ are the so-called irreducible scalar products (ISPs).

\paragraph{Global vs. loop-by-loop approaches.} There are a few ways to implement the change of variables \eqref{eq:change-of-variables}. The general idea dates back to the S-matrix theory in the 1960s \cite{Mandelstam:1959bc, Gribov:1962ft, Correia:2020xtr}, although in the more recent literature on Feynman integrals parametrizations of this type are referred to as \text{Baikov representations} \cite{Baikov:1996iu}. In the \emph{global} (also called \emph{standard} or \emph{democratic} or \emph{original}) Baikov representation, one transforms all variables at the same time. This results in the representation of the form 
\begin{equation}\label{eq:global-Baikov}
I(\vec{s}) = c\, \tilde{\mathcal{G}}(\vec{s})^{\tilde{\gamma}} \int_{\Gamma'} \mathcal{G}(\vec{x};\vec{s})^{\gamma} \frac{\d^{N'} \vec{x}}{x_1^{\nu_1} x_2^{\nu_2} \cdots x_E^{\nu_E}}\, ,
\end{equation}
where $\vec{x} = (x_1, \ldots, x_E, \ldots, x_{N'})$ are the integration variables with
\begin{equation}\label{eq:Nprime}
\mathcal{N}\mapsto N'=\frac{L(L+1)}{2} + L \mathcal{E}\,,
\end{equation}
where $\mathcal{E}$ is the number of linearly-independent external momenta and $\vec{s}$ collectively denotes the external kinematic variables such as Mandelstam invariants and masses. The exponents $\gamma$, $\tilde\gamma$, as well as the prefactor $c$ are expressed in terms of the topology of the diagram (number of edges, etc). as well as the spacetime dimension $\D$. The derivation is given in App.~\ref{app:Baikov} and the explicit expressions for all the ingredients in \eqref{eq:global-Baikov} are spelled out in \eqref{eqs:Baikov-expressions}.

The central object in the global Baikov representation is $\mathcal{G}(\vec{x},\vec{s})$ which arises as the general Jacobian for the change of variables. It can be expressed as a Gram determinant of a certain matrix.
The integration contour $\Gamma'$ is determined by $\vec{x}$ such that $\mathcal{G}(\vec{x},\vec{s}) > 0$. Likewise, $\tilde{\mathcal{G}}(\vec{s})$ is another Gram determinant that depends only on external kinematics and hence acts as an overall normalization. The derivation in App.~\ref{app:Baikov} makes it clear that one can include different powers of propagators as well as non-trivial numerators with minimal modifications. For completeness, we review these aspects in App.~\ref{app:Baikov} (see also \cite{Lee:2009dh,Mastrolia:2018uzb}).

The alternatives proposed by Frellesvig and Papadopoulos are \emph{loop-by-loop Baikov representations} \cite{Frellesvig:2017aai}. The idea is simply to Baikov parametrize each loop of the diagram sequentially in a given order. For each loop, we look for a shift in the loop momentum $\ell_a$ so that fewer invariants---such as $\ell_a \cdot p_i$ and $\ell_a \cdot \ell_b$---appear before changing variables from the $\ell$'s to the $x$'s. This procedure has the effect of decreasing the total number of integration variables $N \leqslant N'$.  
Typically, the minimal/optimal $N$, is \cite{Frellesvig:2024ymq}  
\begin{equation}\label{eq:NnoPrime}  
    \mathcal{N}\mapsto N = L + \sum_{a=1}^{L} \mathcal{E}_a\,,  
\end{equation}  
where $ \mathcal{E}_a$ denotes the number of linearly independent external momenta associated with loop $ \ell_a $. Note that in \eqref{eq:NnoPrime}, $N$ scales \emph{linearly} with $L$, whereas in \eqref{eq:Nprime}, $N'$ exhibits a \emph{quadratic} dependence.

Loop-by-loop representations are not unique, as they depend on the exact change of variables. Therefore, one can distinguish between different \emph{variants}, depending on how such choices are being made. Below, we will describe our optimization procedure.
In any case, all the integral representations takes the general form:
\begin{equation}\label{eq:LBL-Baikov}
I(\vec{s}) = c' \int_{\Gamma'} \tilde{\mathcal{G}}_1^{\tilde\gamma_1} \mathcal{G}_1^{\gamma_1} \tilde{\mathcal{G}}_2^{\tilde\gamma_2} \mathcal{G}_2^{\gamma_2} \cdots \tilde{\mathcal{G}}_L^{\tilde\gamma_L} \mathcal{G}_L^{\gamma_L} \frac{\d^{N} \vec{x}}{x_1^{\nu_1} x_2^{\nu_2} \cdots x_E^{\nu_E}}\, .
\end{equation}
Structurally, the only difference is that it features $L$ pairs of Gram determinants $\tilde{\mathcal{G}}_a$, $\mathcal{G}_a$ for every loop $a = 1,2,\ldots,L$, instead of just one. Note that the order in which we label the loop momenta $\ell_1, \ell_2, \ldots, \ell_L$ plays a role. The integration contour is given by the intersection of regions defined by positivity conditions on internal-external Gram determinant ratios: $\Gamma' = \bigcap_{l=1}^{L}\{  \mathcal{G}_l / \tilde{\mathcal{G}}_l >0 \}$.
Details of the derivation are given in App.~\ref{app:Baikov}.  

\paragraph{Generalized cuts.}
Due to complicated integration contours, Baikov representations are typically not used in the explicit evaluation of Feynman integrals. However, they are extremely useful for integral reduction, integration-by-parts identities, singularity analysis, and other applications that rely on the simplicity of the integrand. The main advantage comes from the application of generalized unitarity. Generalized cuts can be taken by simply replacing the integration contour with residues around the relevant propagators (similarly, unitarity or Cutkosky cuts can be taken after an extra complex conjugation). For example, the \emph{maximal cut} is obtained by
\begin{equation}\label{eq:maxcut}
\int \d^\mathcal{N} \vec{x}
\quad\mapsto\quad
\frac{1}{(2\pi i)^E} \oint_{x_1 = 0} \oint_{x_2 = 0} \cdots \oint_{x_E = 0} \int \d^{\mathcal{N}-E} \vec{x}\,,
\end{equation}
in either the global or loop-by-loop Baikov approach (i.e., $\mathcal{N}=N\,, N'$). This replacement exists if the original integration domain, $\Gamma$ or $\Gamma'$, restricted to $x_1 = x_2 = \ldots = x_\mathcal{N} = 0$ is non-empty (more on that is discussed in Sec.~\ref{sec:functions}). Note that after taking the cut, $\mathcal{N}-E$ variables still have to be integrated over. 

\paragraph{Implementation.} As mentioned above, Baikov representations are \emph{not} unique. They depend on the order and labeling of the loop momenta $\ell_a$, as well as additional shifts.

We implement a search algorithm that optimizes for the most efficient loop-by-loop Baikov representation without any user input. It proceeds in two steps. First, when possible, the loops are ordered automatically so that the number of external momenta $\mathcal{E}_a$ (associated with each loop momentum $\ell_a$, for $1 \leq a \leq L$) increases from the smallest to the largest.\footnote{This strategy is inspired by the recursive elastic unitarity approach of \cite{Caron-Huot:2024brh}, where the selection and ordering of cuts play a crucial role in determining the complexity of the singularity analysis. In \cite{Caron-Huot:2024brh}, this selection was performed manually, whereas in the present work, the analogous loop-ordering process is fully automated.} For example, for the diagram in Fig.~\ref{fig:diagrams0-a}, the algorithm selects the box (with $\mathcal{E}_1=3$) as the first loop and the hexagon (with $\mathcal{E}_2=5$) as the second. Second, the translation invariance of the integration measure is exploited by shifting the loop momenta, i.e., $\ell_a \mapsto \ell_a' = \ell_a + \lambda_a$, to eliminate as many scalar products as possible. An explicit example of such shifts is provided in the sunrise example below.

\paragraph{Comparison with \texttt{BaikovPackage}.} Recently, a semi-automatized \textsc{Mathematica} package \texttt{BaikovPackage} for generating standard and loop-by-loop Baikov representations of Feynman integrals was released \cite{Frellesvig:2024ymq}. Unlike {\SOFIA}, which by default selects the loop-by-loop representation minimizing the number of integration variables $N$, \texttt{BaikovPackage} requires the user to make that choice and specify the order and labeling of the loop momenta $\ell_a$. Consequently, the outputs may differ depending on these selections.

\paragraph{Example.}
Let us illustrate the differences between the global and loop-by-loop approaches on (arguably) the simplest example: the sunrise diagram. To make expressions shorter, we take all internal masses equal $m$ and external momentum $p^2 = s$. In this case, we have only three propagators, $E=3$, with:
\begin{equation}
x_1 = \ell_1^2 - m^2 \, ,\qquad
x_2 = \ell_2^2 - m^2\, , \qquad
x_3 = (p - \ell_1 - \ell_2)^2 - m^2\, .
\end{equation}

As predicted by \eqref{eq:Nprime} with $L=2$ and $\mathcal{E}=1$, expanding these propagators out, the result depends on $N' = 5$ internal kinematic invariants: $\ell_1 \cdot p$, $\ell_2 \cdot p$, $\ell_1 \cdot \ell_2$, $\ell_1^2$, $\ell_2^2$, in addition to the external parameters $\vec{s} = (s, m^2)$. The $5$ internal variables can be traded for $\vec{x} = (x_1, x_2, x_3, x_4, x_5)$ by a linear change of variables, where $x_4$ and $x_5$ are arbitrarily-chosen ISPs, say
\begin{equation}
x_4 = \ell_1\cdot p\, , \qquad
x_5 = \ell_2\cdot p\, .
\end{equation}
This procedure results in the global Baikov parametrization \eqref{eq:global-Baikov} with
\begin{equation}
\hspace{-0.3cm}
    \tilde{\mathcal{G}}(s,m^2) = s\,,\quad
    \mathcal{G}(\vec{x},s,m^2) = \det \left[
\begin{array}{ccc}
 m^2{+}x_1 & *  & * \\
 \frac{1}{2} \left(x_3{-}x_1{-}x_2{-}m^2{-}s\right){+}x_4{+}x_5 & m^2{+}x_2 & * \\
 x_4 & x_5 & s \\
\end{array}
\right]\,,
\end{equation}
where the ``$*$'' denote symmetric entries.

In contrast, loop-by-loop representations first parametrize all invariants associated with the first loop momentum $\ell_1$. They are $\ell_1 \cdot (\ell_2 - p)$ and $\ell_1^2$. Notice that the former is counted as one variable: $\ell_1 \cdot \ell_2$ and $\ell_1 \cdot p$ appear in the expression for the Feynman integral only in the combination $\ell_1 \cdot (\ell_2 - p)$. Hence, integrating out the first loop momentum would give a function of $(\ell_2 - p)^2$, which is treated as external kinematics from this point of view. When it comes to parameterizing the second loop momentum $\ell_2$, we are thus left with only two kinematic invariants: $(\ell_2 - p)^2$ and $\ell_2^2$. Consequently, the total number of variables required is $N=4$. This is consistent with \eqref{eq:NnoPrime} for $L=2$ loops and $\mathcal{E}_1=\mathcal{E}_2=1$. Notably, this approach reduces the total number of variables compared to the global Baikov parametrization, which requires $N'=5$ variables.

Equivalently, we could have first relabeled $\ell'_2 = \ell_2 - p$, which turns the inverse propagators into
\begin{equation}
x_1 = \ell_1^2 - m^2\, , \qquad x_2 = (\ell_2' + p)^2 - m^2\,, \qquad x_3 = (\ell_1 + \ell_2')^2 - m^2\, .
\end{equation}
This makes it clear that only four invariants are needed: $\ell_1^2$, $\ell_2'^2$, $\ell_1 \cdot \ell_2'$, and $p \cdot \ell_2'$.
Taking $x_4=\ell_2\cdot p$ as the only ISP, the resulting representation has
\begin{align}\label{eq:sunriseLBL}
\tilde{\mathcal{G}}_1 &= m^2 + s + x_2 - 2x_4, & \mathcal{G}_1 &= \det \left[
\begin{array}{cc}
 m^2+x_1 & * \\
 \frac{x_3-x_1-x_2-s-m^2}{2}+x_4 & m^2+s+x_2-2x_4
\end{array}
\right], \nonumber\\
\tilde{\mathcal{G}}_2 &= s, & \mathcal{G}_2 &= \det \left[
\begin{array}{cc}
 m^2+x_2 & * \\
x_4 & s
\end{array}
\right].
\end{align}
We now proceed to explain how the singularity analysis is carried out in Baikov space.

\subsection{Singularity analysis: Fubini reduction algorithm}\label{sec:SingAnalysis}

As a mathematical problem, singularity analysis amounts to analyzing the topology of the union of hypersurfaces
\begin{equation}\label{eq:union}
\mathcal{U}(\vec{x},\vec{s}) \equiv \{ g_1(\vec{x}; \vec{s}) = 0 \}
\cup
\{ g_2(\vec{x}; \vec{s}) = 0 \}
\cup
\cdots 
\cup
\{ g_k(\vec{x}; \vec{s}) = 0 \}\,,
\end{equation}
in the complex space $\mathbb{C}^N$. Here, $\vec{x} \in \mathbb{C}^N$ denotes the integration variables and $\vec{s}$ denotes the collection of all the parameters, such as the kinematic invariants and the masses. Each of the $k$ polynomials $g_i(\vec{x}; \vec{s})$ describes a singularity of the integrand (such as a pole) or an integration boundary. A singularity can arise only if the topology of \eqref{eq:union} changes, e.g., when a mass shell degenerates into a lightcone or two mass shells pinch.

\paragraph{Example.}
For instance, loop-by-loop Baikov representations have
\begin{equation}\label{eq:LandauPoly}
(g_1, g_2, \ldots, g_{2L-1}, g_{2L}, g_{2L+1}, \ldots, g_{2L+E}) = (\G_1, \tilde \G_1, \ldots, \G_L, \tilde\G_L, x_1, \ldots, x_E)\equiv\mathcal{L}\, .
\end{equation}
The first $2L$ polynomials are the internal and external Gram determinants, while the last $E$ are the inverse propagators (and possibly ISPs). We do not need any extra polynomial for the integration boundaries, since they are already described in terms of the Gram determinants.

As another example, on the maximal cut, we need $E$ fewer polynomials:
\begin{equation}
(g_1, g_2, \ldots, g_{2L-1}, g_{2L}) = (\G_1, \tilde \G_1, \ldots, \G_L, \tilde\G_L) \Big|_{x_1 = x_2 = \ldots = x_E = 0}\, .
\end{equation}
In either case, one can add $x_{E+1}, \ldots, x_N$ to the list if one allows for negative powers of ISPs.

\paragraph{Singularity analysis.}
A conventional way to phrase this problem is to compute the Jacobian of the union of hypersurfaces:
\begin{equation}
J = \left[ \nabla g_1 \; \nabla g_2 \; \ldots \; \nabla g_k \right]\, ,
\end{equation}
where $\nabla = \left[ \partial_{x_1}\; \partial_{x_2}\; \ldots\; \partial_{x_N} \right]^\intercal$, and ask for what values of $\vec{s}$ the rank of $J$ changes; see, e.g., \cite[Ch.~9]{Cox2015}. In other words, we are asking for at least one null vector $\vec{\beta} = \left[ \beta_1, \beta_2, \ldots, \beta_k \right]$ of $J$. (When applied to the loop-momentum representation \eqref{eq:loopMomDef} with $g_i = q_i^2 - m_i^2$, $\beta$'s have the physical interpretation of Schwinger proper times). This requirement can be written as the following system of equations:
\begin{equation}\label{eq:null-vector-equations}
\beta_1 \nabla g_1(\vec{x},\vec{s}) + \beta_2 \nabla g_2(\vec{x},\vec{s}) + \ldots + \beta_k \nabla g_k(\vec{x},\vec{s}) = 0\, ,
\end{equation}
where $\vec{x} \in \mathcal{U}$, where $\mathcal{U}$ is defined in \eqref{eq:union}. We call the codimension-1 part of the solution set $\mathcal{S}(\vec{s})$ and write it as
\begin{equation}\label{eq:Sdef}
\mathcal{S}(\vec{s}) = \{ f_1(\vec{s}) = 0 \} \cup \{ f_2(\vec{s}) = 0 \} \cup \cdots \cup \{ f_K(\vec{s}) = 0 \}\, .
\end{equation}
The goal of singularity analysis is to convert $\mathcal{U}$ into $\mathcal{S}$.
We have essentially converted a topological problem into an algebraic one. It has a certain combinatorial aspect since we can classify the solutions based on the number of $\{ g_i = 0\}$ surfaces that intersect at $\vec{x}$. The problem becomes even more difficult when $\mathcal{U}$ has a complicated topology since one has to take into account its non-normal crossings through blow-ups.

Singularity analysis of this kind applied to Feynman integrals dates back to the pioneering work of Bjorken \cite{Bjorken:1959fd}, Landau \cite{Landau:1959fi}, and Nakanishi \cite{10.1143/PTP.22.128} who wrote systems of polynomial equations \eqref{eq:null-vector-equations} in various parametrizations. These are nowadays known as \emph{Landau equations} and serve as a prototype for the analysis of singularities. This analysis was made rigorous through the work of Pham, Thom, Whitney, and others with the application of algebraic geometry tools to finite Feynman integrals; see \cite{pham2011singularities} for a review. While mathematically rigorous, these methods tend to explode in computational complexity in anything but the simplest examples; see, e.g., \cite{Helmer:2024wax} for a recent implementation in \texttt{Macaulay2}.
Hence, recent resurgence of interest in this topic aims to make the singularity locus more computable.

\paragraph{Numerical approaches.}
Let us briefly outline the scope of numerical tools that are currently available for cutting-edge Feynman integrals. The \textsc{Julia} package \texttt{PLD.jl} \cite{Fevola:2023kaw,Fevola:2023fzn} uses techniques from toric and polyhedral geometry to give a computable subset of the singularity locus $\mathcal{S}$. It also implements the Euler characteristic test, which is another way to probe the topology of the space $\mathcal{U}$. If for a given set of parameters $\vec{s}_\ast$, the absolute value of the Euler characteristic of 
\begin{equation}
(\mathbb{C}^\ast)^N \setminus \mathcal{U}(\vec{x},\vec{s})\,,
\end{equation}
drops compared to its generic value, then $\vec{s}_\ast$ belongs to the singular locus. This test provides a quick way to determine whether a given $\vec{s}_\ast$ belongs to the singular locus $\mathcal{S}$. {\SOFIA} provides an interface to both functionalities of \texttt{PLD.jl}.

However, the package \texttt{PLD.jl} is based on toric geometry and hence uses $\mathbb{C}^\ast = \mathbb{C} \setminus \{0\}$ instead of $\mathbb{C}$. This is
appropriate for parametric representations of Feynman integrals, but in the Baikov representation it has the effect of introducing all $\{ x_i = 0\}$ in the hypersurface union $\mathcal{U}$. This is equivalent to putting ISPs with negative exponents.

\paragraph{Analytical approaches.}
Let us now outline the available analytical approaches that lie at the heart of {\SOFIA}. In contrast to the aforementioned numerical approach, the goal will be to  analytically compute a superset of the singularity locus $\mathcal{S}$.
In other words, we expect to encounter certain ``fictitious'' polynomials $f_i(\vec{s})$ that do not satisfy the original system of equations \eqref{eq:null-vector-equations}. However, as we explain below, our algorithm is designed to minimize their occurrence.

Let us describe the \textit{Fubini reduction algorithm}, which is a variation of \cite[Sec.~4.2]{Brown:2008um}. It circumvents the need to solve systems of equations such as \eqref{eq:null-vector-equations}. Instead, it proceeds by eliminating one variable $x_i$ at a time. We thus construct the sets of polynomials $\mathcal{L}$, $\mathcal{L}_{x_i}$, $\mathcal{L}_{x_i, x_j}$, etc. recursively, where the subscript refers to the variables we already eliminated. At each step, we define 
\begin{equation}\label{eq:L-xs}
[\mathcal{L}_{x_i, \ldots, x_j}]_{x_k} \equiv \bigcup_{g \,\in\, \mathcal{L}_{x_i, \ldots, x_j}} \!\!\!\!  \{ D_{x_k}(g),\, R_{x_k}(\infty, g) \}  \;\cup \bigcup_{g,\hat{g} \,\in\, \mathcal{L}_{x_i, \ldots, x_j}} \!\!\!\! \{ R_{x_k}(g,\hat{g}) \}\, .
\end{equation}

On the right-hand side, $D_{x_k}(g)$ refers to the \emph{discriminant} of $f$ with respect to $x_k$, and $R_{x_k}(g,\hat{g})$ is the \emph{resultant} of $g$ and $\hat{g}$ with respect to $x_k$. Writing the polynomial $g = c_0 + c_1 x_k + \ldots + c_d x_k^d$, the resultant $R_{x_k}(\infty, g) = c_d$ equals to the coefficient of the highest power. 
The set $[\mathcal{L}_{x_i, \ldots, x_j}]_{x_k}$ might consist of reducible polynomials or repeated entries. We hence define $\mathcal{L}_{x_i, \ldots, x_j, x_k} = \text{irred} [\mathcal{L}_{x_i, \ldots, x_j}]_{x_k}$ containing only the unique irreducible polynomials.

The three contributions in \eqref{eq:L-xs} have the following interpretations illustrated in Fig.~\ref{fig:fubini}. $D_{x_k}(g)$ corresponds to self-pinching of $g=0$ (star in Fig.~\ref{fig:fubini}), $R_{x_k}(\infty, g)$ gives the self-pinch at infinity (omitted in Fig.~\ref{fig:fubini}), while $R_{x_k}(g,\hat{g})$ is associated with a pinch between two curves $g=0$ and $\hat{g}=0$ (triangle in Fig.~\ref{fig:fubini}).

\begin{figure}
    \centering
 \begin{minipage}{0.45\textwidth}
        \centering
        \tikzset{every picture/.style={line width=0.75pt}}
   \tikzset{every picture/.style={line width=0.75pt}}
\begin{tikzpicture}[x=0.75pt,y=0.75pt,yscale=-1.4,xscale=1.4]
\draw[<-]    (190.5,50.5) -- (190.5,181.5) node[above, pos=0]{remaining $x$'s};
\draw[->]   (181,170) -- (350.5,170) node[right]{$x_k$};
\draw [color=Maroon  ,draw opacity=1 ]   (185.5,83.5) .. controls (193.5,113.5) and (279.5,125.5) .. (285.11,142.64) .. controls (290.71,159.77) and (241.5,165.5) .. (243.5,138.5) .. controls (245.5,111.5) and (302.5,63.5) .. (348.5,63.5) node[left,above]{$\hat{g}=0$};
\draw [color=RoyalBlue  ,draw opacity=1 ]   (184.5,58.5) .. controls (238,51) and (344.5,76.5) .. (354.5,113.5) node[left]{$g=0$};
\draw  [dash pattern={on 0.84pt off 2.51pt},gray!40]  (182,158) -- (351.5,157.5) ;
\draw  [dash pattern={on 0.84pt off 2.51pt},gray]  (182,146) -- (351.5,145.5) ;
\draw  [dash pattern={on 0.84pt off 2.51pt},gray!40]  (182,123) -- (351.5,122.5) ;
\draw  [dash pattern={on 0.84pt off 2.51pt},gray!40]  (182,75) -- (351.5,75) ;
\draw  [fill=black  ,fill opacity=1 ] (242.5,145.75) .. controls (242.5,144.78) and (243.28,144) .. (244.25,144) .. controls (245.22,144) and (246,144.78) .. (246,145.75) .. controls (246,146.72) and (245.22,147.5) .. (244.25,147.5) .. controls (243.28,147.5) and (242.5,146.72) .. (242.5,145.75) -- cycle ;
\draw  [fill=black  ,fill opacity=1 ] (283.36,145.39) .. controls (283.36,144.42) and (284.14,143.64) .. (285.11,143.64) .. controls (286.07,143.64) and (286.86,144.42) .. (286.86,145.39) .. controls (286.86,146.35) and (286.07,147.14) .. (285.11,147.14) .. controls (284.14,147.14) and (283.36,146.35) .. (283.36,145.39) -- cycle ;
\draw  [fill=black  ,fill opacity=1 ] (249.09,120) -- (249.68,121.61) -- (251.28,120.88) -- (250.52,122.43) -- (252.18,123) -- (250.52,123.57) -- (251.28,125.12) -- (249.68,124.39) -- (249.09,126) -- (248.5,124.39) -- (246.91,125.12) -- (247.66,123.57) -- (246,123) -- (247.66,122.43) -- (246.91,120.88) -- (248.5,121.61) -- cycle ;
\begin{scope}[yshift=-0.5]
    \draw  [fill=white  ,fill opacity=1 ] (263.25,157.75) .. controls (263.25,156.78) and (264.03,156) .. (265,156) .. controls (265.97,156) and (266.75,156.78) .. (266.75,157.75) .. controls (266.75,158.72) and (265.97,159.5) .. (265,159.5) .. controls (264.03,159.5) and (263.25,158.72) .. (263.25,157.75) -- cycle ;
\end{scope}
\draw  [fill=black  ,fill opacity=1 ] (304.09,72) -- (305.43,74.25) -- (306.77,76.5) -- (304.09,76.5) -- (301.41,76.5) -- (302.75,74.25) -- cycle ;
\draw[<->,gray]    (180,140) -- (180,152);
\draw  [draw opacity=0] (240.59,143.29) .. controls (240.02,141.22) and (239.92,138.94) .. (240.41,136.65) .. controls (240.7,135.29) and (241.17,134.02) .. (241.78,132.87) -- (252.21,139.15) -- cycle ; \draw[->,gray]   (240.59,143.29) .. controls (240.02,141.22) and (239.92,138.94) .. (240.41,136.65) .. controls (240.7,135.29) and (241.17,134.02) .. (241.78,132.87) ;  
\draw  [draw opacity=0] (245.3,154.32) .. controls (245.28,154.3) and (245.26,154.29) .. (245.24,154.27) .. controls (243.27,152.73) and (241.82,150.79) .. (240.94,148.69) -- (252.67,144.78) -- cycle ; \draw[<-,gray]   (245.3,154.32) .. controls (245.28,154.3) and (245.26,154.29) .. (245.24,154.27) .. controls (243.27,152.73) and (241.82,150.79) .. (240.94,148.69) ; 
\draw  [draw opacity=0] (278.46,132.02) .. controls (280.54,132.57) and (282.55,133.63) .. (284.29,135.2) .. controls (286.39,137.1) and (287.78,139.45) .. (288.45,141.92) -- (276.21,144.15) -- cycle ; \draw[<-,gray]   (278.46,132.02) .. controls (280.54,132.57) and (282.55,133.63) .. (284.29,135.2) .. controls (286.39,137.1) and (287.78,139.45) .. (288.45,141.92) ;  
\draw  [draw opacity=0] (289.3,147.58) .. controls (288.72,149.65) and (287.63,151.65) .. (286.03,153.37) .. controls (285.36,154.09) and (284.63,154.72) .. (283.86,155.27) -- (277.21,145.15) -- cycle ; \draw[->,gray]   (289.3,147.58) .. controls (288.72,149.65) and (287.63,151.65) .. (286.03,153.37) .. controls (285.36,154.09) and (284.63,154.72) .. (283.86,155.27) ;    
\end{tikzpicture}
    \end{minipage}%
    \hfill
    \begin{minipage}{0.52\textwidth}
        \centering
        \tikzset{every picture/.style={line width=0.75pt}}
   \tikzset{every picture/.style={line width=0.75pt}}
\begin{tikzpicture}[x=0.75pt,y=0.75pt,yscale=-1.4,xscale=1.4]
\draw[<-]    (190.5,50.5) -- (190.5,181.5) node[above, pos=0]{remaining $x$'s};
\draw[->]   (181,170) -- (350.5,170) node[right]{$x_k$};
\draw [color=Maroon  ,draw opacity=1 ]   (185.5,83.5) .. controls (193.5,113.5) and (279.5,125.5) .. (285.11,142.64) .. controls (290.71,159.77) and (241.5,165.5) .. (243.5,138.5) .. controls (245.5,111.5) and (302.5,63.5) .. (348.5,63.5) node[left,above]{$\hat{g}=0$};
\draw [color=RoyalBlue  ,draw opacity=1 ]   (184.5,58.5) .. controls (238,51) and (344.5,76.5) .. (354.5,113.5) node[left]{$g=0$};
\draw  [dash pattern={on 0.84pt off 2.51pt},gray!40]  (182,158) -- (351.5,157.5) ;
\draw  [dash pattern={on 0.84pt off 2.51pt},gray!90]  (243.5,180) -- (243.5,50) ;
\begin{scope}[yshift=-0.5]
    \draw  [fill=white  ,fill opacity=1 ] (263.25,157.75) .. controls (263.25,156.78) and (264.03,156) .. (265,156) .. controls (265.97,156) and (266.75,156.78) .. (266.75,157.75) .. controls (266.75,158.72) and (265.97,159.5) .. (265,159.5) .. controls (264.03,159.5) and (263.25,158.72) .. (263.25,157.75) -- cycle ;
\end{scope}
\begin{scope}[yshift=-11,xshift=-16]
    \draw  [fill=white  ,fill opacity=1 ] (263.25,157.75) .. controls (263.25,156.78) and (264.03,156) .. (265,156) .. controls (265.97,156) and (266.75,156.78) .. (266.75,157.75) .. controls (266.75,158.72) and (265.97,159.5) .. (265,159.5) .. controls (264.03,159.5) and (263.25,158.72) .. (263.25,157.75) -- cycle ;
\end{scope}
\end{tikzpicture}
    \end{minipage}
        \caption{(Left) Two-dimensional slicing of $\mathbb{C}^N$, with the current variable to eliminate placed along the horizontal axis and the remaining variables along the vertical axis. The curves represent the zero loci of two elements in $\mathcal{L}_{x_i,...,x_j}$ we call $g=0$ and $\hat{g}=0$. Geometrically, \eqref{eq:L-xs} means that it suffices to know the special points where the roots of these polynomials (black dots) collide, indicated by the star and the white circle (detected by the discriminant) as well as the triangle (detected by the resultant). (Right) The white circle is a fictitious singularity that arises solely from the choice of coordinates. For instance, horizontal and vertical slicings lead to different fictitious singularities.}
    \label{fig:fubini}
\end{figure}

One strategy in computing the singularity locus would be therefore to start with
\begin{equation}
\mathcal{L} = \{ g_1 , g_2, \ldots, g_k \}\, .
\end{equation}
and then perform the reduction in some order:
\begin{equation}
\mathcal{L} \mapsto \mathcal{L}_{x_1} \mapsto \mathcal{L}_{x_1,x_2} \mapsto \ldots \mapsto \mathcal{L}_{x_1,x_2,\ldots,x_N}\, .
\end{equation}
The result $\mathcal{L}_{x_1,x_2,\ldots,x_N}$ consists of polynomials that depend on $\vec{s}$ only. In general, it is a (large) superset of the singularity locus $\mathcal{S}$. However, the result of the computation depends on the specific order of the variables we chose in the reduction.

We refer to the difference between $\mathcal{L}_{x_1,x_2,\ldots,x_N}$ and $\mathcal{S}$ as \emph{fictitious} singularities. They are artifacts of the reduction procedure and have no physical meaning. In general, we are not concerned with catching a few fictitious singularities, since the goal is to find a superset of $\mathcal{S}$ anyway. However, it is beneficial to minimize their number. Such a fictitious point is illustrated by the circle in Fig.~\ref{fig:fubini}, which makes it clear that its location is potentially sensitive to the order in which the variables are eliminated (i.e., to how the slicing of $\mathbb{C}^N$ is performed).

The Fubini reduction algorithm instructs us to scan over all choices of $\mathcal{L}_{x_1,x_2,\ldots,x_N}$ and take the intersection of their results, that is
\begin{equation}\label{eq:SlowFubini}
\mathcal{L}_{\{x_1, x_2, \ldots, x_N\}} = \bigcap_{\sigma \in S_N} \mathcal{L}_{\sigma(x_1, x_2, \ldots, x_N)}\,,
\end{equation}
where $S_N$ denotes the permutation set of $N$ labels (two polynomials are considered equivalent for the purposes of intersection if their vanishing locus is the same). In practice, a lot of the intermediate results can be recycled between each permutation. Hence, the optimal recursion is
\begin{equation}\label{eq:Fubini}
\mathcal{L}_{\{x_1, \ldots, x_{k-1},x_k\}} = \bigcap_{i = 1}^{k}\, \text{irred} [\mathcal{L}_{\{x_1, \ldots, x_{k-1},x_k\} \setminus x_i}]_{x_i}\, ,
\end{equation}
where each $[\mathcal{L}_{\{x_1, \ldots, x_{k-1},x_k\} \setminus x_i}]_{x_i}$ is defined as in \eqref{eq:L-xs}. The central difference between \eqref{eq:SlowFubini} and \eqref{eq:Fubini} is that the former intersects $N!$ sets of proposed singularities at the very end, while the latter does so at the intermediate stages. This comes with enormous computational benefit and provides a slimmer superset of the singular locus $\mathcal{S}$, often coinciding with $\mathcal{S}$.
The equation \eqref{eq:Fubini} is implemented in {\SOFIA} as the default solver option \texttt{FastFubini}.

Brown \cite{Brown:2009ta} and Panzer \cite{Panzer:2015ida} have also proposed more sophisticated versions of the Fubini algorithm known as \emph{compatibility graph reduction}.
They are implemented in the \textsc{Maple} package \texttt{HyperInt} \cite{Panzer:2014caa}; see \cite[App.~A]{Fevola:2023fzn} for a review. We found that our implementation of this approach was slower than \eqref{eq:Fubini} in practical applications, and is therefore not shipped with {\version} of {\SOFIA}.

\paragraph{Other approaches.}
Other approaches to determining symbol alphabets based on Landau analysis exist; for example, utilizing momentum twistor techniques or Schubert analysis; see, e.g., \cite{Morales:2022csr,Lippstreu:2023oio,He:2024fij} for recent developments and more exhaustive lists of references.
Out of the algorithms implemented as computer codes, the \textsc{Mathematica} package \texttt{Baikovletter} \cite{Jiang:2023qnl,Jiang:2023oyq,Jiang:2024eaj} uses Baikov representations to attempt constructing symbol letters. Instead of a reduction algorithm, it aims to return a list of candidate symbol letters directly. There are instances in which the resulting alphabet is not complete; see, e.g., \cite{Abreu:2024yit,Henn:2025xrc}. 

\subsection{From singularities to letters}\label{SSec:Letters}

If a Feynman integral is sufficiently simple, it can admit a representation as a linear combination of multiple polylogarithms. In such cases, there is a great interest in knowing a priori the list of possible \emph{symbol letters} $W_i$, which translate to arguments of these polylogarithms; see, e.g., \cite{Goncharov:2010jf,Duhr:2011zq,Lee:2024kkm}. The collection of all independent letters is known as the \emph{symbol alphabet} $\mathbb{A}$. 

Before we move on, we note that even after identifying that a family of Feynman integrals includes integrals evaluating to non-polylogarithmic functions, knowledge of $\mathbb{A}$ remains valuable. This is because integrals in subsectors typically remain polylogarithmic (see \cite{Bogner:2019lfa,Muller:2022gec,Giroux:2024yxu} for explicit examples). 

\paragraph{Even letters.}
As reviewed in the introduction, letters are closely related to singularities. By definition, the zeros, poles, and singularities of letters are singularities of Feynman integrals. The simplest example is $\log(W_i)$, which has branch points at $W_i = 0$ and $W_i = \infty$. Hence the symbol alphabet must at least contain all the singularities themselves:
\begin{equation}\label{eq:WF}
(W_1, W_2, \ldots, W_K) = (f_1, f_2, \ldots, f_K) = \mathbb{A}^{\text{even}}\, .
\end{equation}
(One can also consider ratios of polynomials of the form $f_j/f_k$.)
These are sometimes called \emph{even} letters and certainly constitute a subset of the alphabet $\mathbb{A}^{\text{even}} \subseteq \mathbb{A}$.
The question therefore becomes how to leverage this information to recover the remaining part of $\mathbb{A}$.

\paragraph{Odd letters.}
We start by assuming the remaining letters, which we call \emph{odd}, take the general form
\begin{equation}\label{eq:odd-letter}
W_i = \frac{P_i + R_i\sqrt{Q_i}}{P_i - R_i\sqrt{Q_i}}\, ,
\end{equation}
where $P_i = P_i(\vec{s})$, $Q_i = Q_i(\vec{s})$, and $R_i = R_i(\vec{s})$ are some homogeneous polynomials in $\vec{s}$, such that the mass dimensions match. We assume that $Q_i$ is square-free, so that it does not have any factors outside the square root.
This is of course just an ansatz, and one can extend it to more complicated forms if necessary. We can now apply the same analysis as before. First of all, $W_i$ itself has a singularity (square-root branch point) at $Q_i = 0$. Therefore, we must have
\begin{equation}\label{eq:Q-polynomial}
Q_i = c_i \prod_{j=1}^{K} f_j^{e_{ij}}\, ,
\end{equation}
where $c_i \in \mathbb{Q}$ is a rational number and $e_{ij} \in \mathbb{Z}_{\geq 0}$ are non-negative integer powers. Next, we need to check that zero and poles are singularities, which leads to a similar constraint:
\begin{equation}\label{eq:P2-Q-polynomial}
(P_i + R_i \sqrt{Q_i})(P_i - R_i \sqrt{Q_i}) = P_i^2 - R_i^2 Q_i = c'_i \prod_{j=1}^{K} f_j^{e'_{ij}}\, ,
\end{equation}
for some new sets of constants $c_i' \in \mathbb{Q}$ and exponents $e_{ij}' \in \mathbb{Z}_{\geq 0}$.

\paragraph{Differential equation.}
As another perspective on the same problem, let us insert the odd letter ansatz \eqref{eq:odd-letter} into the differential equation \eqref{eq:differential-equation}. The relevant term is
\begin{equation}
\d \log (W_i) = \frac{P_i R_i \d Q_i - 2 (\d P_i R_i - P_i \d R_i) Q_i}{(P_i^2 - R_i^2 Q_i)\sqrt{Q_i}}\, .
\end{equation}
Hence, indeed, the only singular points of the differential equation are located at the positions of Landau singularities $f_j = 0$ for all possible $j$.

Therefore, knowing the symbol alphabet is not essential for constructing an ansatz for the differential equation \eqref{eq:differential-equation} based solely on the Landau singularities $f_j$. However, such an approach would lead to an unwieldy polynomial in the numerator. The symbol alphabet provides a practical way to reduce this ansatz into a more manageable form.

\paragraph{\texttt{Effortless} implementation.}

There are a few different strategies to finding odd letters. To our knowledge, they were first explored in \cite{Heller:2019gkq}.
We will follow \cite[Sec.~3.4]{Matijasic:2024too}, since this is the algorithm implemented in the publicly available package \texttt{Effortless} \cite{Effortlessxxx,repoEffortless}.
The package assumes that a list of all candidate polynomials $Q_i$, $R_i$ in \eqref{eq:odd-letter} and constants $c_i'$ from \eqref{eq:P2-Q-polynomial} are given.  It then attempts to construct all polynomials $R_i^2 Q_i + c_i' \prod_{j=1}^{K} f_j^{e_{ij}'}$ that are a perfect square $P_i^2$. This can be done by scanning over all choices of $e_{ij}'$. There is only a finite number of choices, since the set of singularities $f_j$ is finite and the exponents are limited because the whole expression has to have a uniform mass dimension. Repeating the same exercise for all choices of $Q_i$, $R_i$, and $c_i'$ results in the proposed odd letters. Further checking for linear dependence, one obtains a candidate odd alphabet $\mathbb{A}^{\text{odd}}$ and hence also $\mathbb{A} = \mathbb{A}^{\text{even}} \cup \mathbb{A}^{\text{odd}}$.

In practice, we consider all $Q_i$ of the form 
\begin{equation}\label{eq:doubleRoots}
\sqrt{Q_i} = \sqrt{g_a} \sqrt{h_b} \sqrt{f_k} \sqrt{f_l}\,,
\end{equation}
for every pair $k,l = 1,2,\ldots, K$ with $R_i = 1$ and $c_i' = \pm 4$ (the default option in \texttt{Effortless}), as well as a possible user-specified set of polynomials $(g_1, g_2,\ldots, g_A)$ and $(h_1, h_2,\ldots, h_B)$ with $a=1,2,\ldots,A$ and $b = 1,2,\ldots,B$, see Sec.~\ref{sec:SOFIA-command} for details. Hence, in total, there are $A \times B \times K^2$ possibilities checked. The odd letter ansatz in {\SOFIA} {\version} is formulated using the root structure defined in \eqref{eq:doubleRoots}.

There is a handful of examples of Feynman integrals in which letters more complicated than the ansatz \eqref{eq:odd-letter} exist, for example those with nested square roots \cite{FebresCordero:2023pww,Badger:2024fgb,Becchetti:2025oyb}. These are currently not implemented in \texttt{Effortless}.

\section{\label{sec:manual}Manual}

This section covers how to install {\SOFIA} and its dependencies, summarizes the key functions, and provides complete documentation.

\subsection{Installation}

As mentioned above, {\SOFIA} is hosted in a GitHub repository at \cite{repo}, where a compressed file with the latest version can be downloaded or cloned. The package requires a working installation of \textsc{Mathematica}. It was developed and tested using the \textsc{Mathematica} versions 13.2--14.0 and can be loaded by running the command:
\begin{lstlisting}[extendedchars=true,language=Mathematica]
Get["path/to/SOFIA.m"]
\end{lstlisting}
Alternatively, one may add the path to the package to the variable \texttt{\$Path}, in which case the loading of the package amounts to calling \texttt{Get["SOFIA.m"]}. Running this command by itself uses the default options:
\begin{lstlisting}[extendedchars=true,language=Mathematica]
SOFIAoptionFiniteFlow = False;
SOFIAoptionJulia = False;
\end{lstlisting}
When an Internet connection is available, the \texttt{Effortless} package is automatically loaded from GitHub \cite{repoEffortless}, ensuring that {\SOFIA} always uses the latest version. If needed, users can overwrite this path with \texttt{SOFIAoptionEffortlessPath = "path/to/Effortless.m"} to point to, say, a locally cloned copy. 
To speed up the alphabet reconstruction with \texttt{Effortless}, we recommend enabling \texttt{FiniteFlow} \cite{Peraro:2019svx} by setting \texttt{SOFIAoptionFiniteFlow = True} before running \texttt{Get["SOFIA.m"]}. Similarly, the \textsc{Julia} dependencies can be activated by setting \texttt{SOFIAoptionJulia = True}, although this option is typically unnecessary (more on that below). A short guide for setting up an optional link between \textsc{Mathematica} and \textsc{Julia} can be found in App.~\ref{app:julia}.

\subsection{Summary for people with busy schedules}

We illustrate how to use the main functionality of {\SOFIA}, which converts a diagram into a proposed list of singularities or candidate symbol letters depending on the options used.

\paragraph{Input.} The input diagram is specified through the list of its \texttt{edges} and \texttt{nodes}. The list \texttt{edges} consists of tuples \texttt{\{\{i,j\},m\}} specifying that a vertex \texttt{i} and \texttt{j} are connected through a propagator with mass \texttt{m}. The order does not matter. The \texttt{nodes} is a list of tuples \texttt{\{i,M\}} specifying that an external momentum $p_i$ with $p_i^2 = \mathtt{M}^2$ is attached to the vertex \texttt{i}. The order \emph{does} matter here since we call the external momenta $p_1, p_2, \ldots$ in the order of appearance.

As an example, the ``parachute'' diagram shown at the top of Fig.~\ref{fig:parSing} corresponds to 
\begin{lstlisting}[extendedchars=true,mathescape=true,language=Mathematica]
edges = {{{1, 2}, m}, {{1, 2}, m}, {{1, 3}, m}, {{2, 3}, m}};
nodes = {{1, $\mathtt{M_1}$}, {2, $\mathtt{M_2}$}, {3, $\mathtt{M_3}$}};
\end{lstlisting}
The code automatically assigns independent Mandelstam invariants of the form \texttt{sij\ldots} $= (p_i + p_j + \ldots)^2$ in the cyclic basis. For example, $6$-particle Mandelstam invariants are \texttt{s12}, \texttt{s23}, \texttt{s34}, \texttt{s45}, \texttt{s56}, \texttt{s16}, \texttt{s123}, \texttt{s234}, \texttt{s345} (we work in $\D$ dimensions).
Alternatively, one can run the command:
\begin{lstlisting}[extendedchars=true,language=Mathematica]
{edges, nodes} = FeynmanDraw
\end{lstlisting}
which opens a window in which one can draw the diagram directly. The output of this command is the pair \texttt{\{edges, nodes\}} with generic masses. 
One can visually confirm that the diagram is correctly specified by plotting it with:
\begin{lstlisting}[extendedchars=true,language=Mathematica]
FeynmanPlot[{edges, nodes}]
\end{lstlisting}
which reproduces the top of Fig.~\ref{fig:parSing}.

Note that the input masses in \texttt{\{edges, nodes\}} typically follow the format $ \mathtt{M} $, $ \mathtt{M}{}_{\mathtt{i}} $ (equivalent to \texttt{Subscript[M, i]}), $ \mathtt{m} $, or $ \mathtt{m}{}_{\mathtt{i}} $. However, as long as the \textsc{Julia} dependencies are not enabled, users can choose their own variable names. The only effect is that the mass variables will not be automatically homogenized in terms of mass squared in the output of the singularity analysis performed by the functions we now describe.

\paragraph{\texttt{SOFIA} command: Most important options and output format.} With this input, running {\SOFIA} amounts to:
\begin{lstlisting}[extendedchars=true,language=Mathematica]
SOFIA[{edges, nodes}]
\end{lstlisting}
This command produces a list of candidate singularities. For the parachute diagram with \texttt{edges} and \texttt{nodes} given above, this yields the list at the top of Fig.~\ref{fig:parGraph}.
The detailed structure of the output—including variations produced under different options—will be discussed later in this section.

The function \texttt{SOFIA[]} comes with many options that can customize the computation depending on the specific application. The most important options (and their default values) are: 
\begin{lstlisting}[extendedchars=true,language=Mathematica]
SOFIA[{edges, nodes}, FindLetters -> False
                    , MaxCut -> True
                    , IncludeSubtopologies -> True
                    , Symmetries -> True
                    , SolverBound -> 100]
\end{lstlisting}
The complete list of options is given in Sec.~\ref{sec:functions}.

\paragraph{Constructing candidate alphabets.} When the \texttt{FindLetters -> True} option is enabled, \texttt{SOFIA[]} first extracts the kinematic singularities and then uses the publicly available package \texttt{Effortless} \cite{repoEffortless} to generate all odd candidate letters, in addition to the even letters already found by \texttt{SOFIA[]}. The structure of the output is then a nested list, with independent even and odd letters in the first and second entries, and square roots in the third entry, respectively:
\begin{lstlisting}[extendedchars=true,language=Mathematica]
{{Even letters}, {Odd letters}, {Square roots}}
\end{lstlisting}

In contrast, with \texttt{FindLetters -> False}, the code simply returns the list of candidate singularities without generating any associated letters. As emphasized above, {\SOFIA} looks only for candidate singularities and a candidate symbol alphabet, meaning that it might find fictitious terms (in the sense of Sec.~\ref{sec:SingAnalysis}) that are not true singularities or true letters of a given Feynman integral. \decosix

\paragraph{Maximal cuts and subtopologies.} 
The option \texttt{MaxCut -> True}, enabled by default, computes the leading singularities, which we take to be the singularities on the maximal cut \eqref{eq:maxcut}. When \texttt{IncludeSubtopologies -> True} is also enabled, the calculation iterates over all subtopologies, therefore computing also subleading singularities.

To restrict the computation to leading singularities only, one can set \texttt{IncludeSubtopologies -> False} while keeping \texttt{MaxCut -> True}. Conversely, setting \texttt{MaxCut -> False} disables cuts entirely, computing both leading and subleading singularities using the top topology alone. However, this significantly increases runtime due to the much larger number of Baikov variables that must be eliminated by the polynomial solver. In this case, it is recommended to set \texttt{IncludeSubtopologies -> False}, as subleading singularities are already being analyzed. \decosix

 \paragraph{Symmetries.} When \texttt{IncludeSubtopologies -> True}, the role of the option \texttt{Symmetries -> True} is to
reduce the total number of subtopologies analyzed, by identifying those that are symmetric under a map between kinematic scales. These include transformations between internal and external masses, and Mandelstam invariants of the subtopologies. After the minimal set of subtopologies is analyzed, the singularities of the remaining ones are obtained via replacement rules. Thus, the option
\texttt{Symmetries -> True} should reproduce the results as if all diagrams were considered, but with significantly improved efficiency. 
This is particularly valuable in high-multiplicity cases (i.e., six or more external particles), where it generally speeds up the \texttt{SOFIA[]} command by \emph{at least} a factor of 2 to 10 in the examples considered. In general, its effect on reducing runtime is expected to increase with the number of external legs and loops.
The specifics of how the symmetry map between subtopologies is constructed using the \texttt{Symmetry[]} function is discussed later in Sec.~\ref{ssec:GF} below.

Enabling \texttt{Symmetries -> reflections} additionally accelerates the generation of subtopologies for highly symmetric graphs, as explained in more detail below. However, because the performance gain is graph-specific, the default value is set to \texttt{True} instead of \texttt{reflections}. \decosix

\paragraph{Bound in the number of monomials.} The option \texttt{SolverBound -> n} ensures that only intermediate polynomials $\mathcal{L}_{x_i, \ldots, x_j}$ in \eqref{eq:L-xs} with at most \texttt{n} monomials are considered in the elimination process. This option essentially controls the accuracy and speed of the computation. The higher \texttt{SolverBound} is, the more accurate the result becomes, but it will also take longer runtime. We recommend starting with a low value such as \texttt{SolverBound -> 100}. If the function returns results within a reasonable time, one should continue increasing \texttt{SolverBound}, as this generates a more complete list of singularities and letters if it is not already complete. The safest value is \texttt{SolverBound -> Infinity}. One of the few examples considered that required a high \texttt{SolverBound} is the ``rocket diagram'' from \cite[Fig.~3c]{Abreu:2024yit} covered in the notebook \texttt{SOFIA\_examples.nb} \cite{repo}. \decosix

\subsection{\label{sec:functions}Documentation of functions}

The purpose of this section is to provide a more comprehensive documentation of all the most important functions included in {\SOFIA} and their options. All these functions are memoized (that is, the results are stored and can be reused for future inputs, without being recomputed) to improve performance, and in Sec.~\ref{sec:examples} we provide concrete examples that illustrate their usage.

\subsubsection{\label{sec:SOFIA-command}\texttt{SOFIA} command}
The singularities and associated alphabets are computed by the central function \texttt{SOFIA[]}, whose high-level behavior and performance may depend on several user-specified options. Below, we list all its available options together with the default values: 
\begin{lstlisting}[extendedchars=true,language=Mathematica]
SOFIA[{edges, nodes}, FindLetters -> False
                    , IncludeSubtopologies -> True
                    , IncludeISPs -> False
                    , MaxCut -> True
                    , LoopEdges -> Automatic
                    , ShowPossiblyDegenerate -> False
                    , Symmetries -> True
                    , ShowHistoryGraph -> False
                    , UnclogTime -> 10^17
                    , StartAtSubtopology -> 1
                    , EndAtSubtopology -> -1
                    , Substitutions -> {}
                    , Solver -> FastFubini
                    , SolverBound -> 100
                    , FactorResult -> True
                    , PLDMethod -> sym
                    , PLDHomogeneous -> true
                    , PLDHighPrecision -> false
                    , PLDCodimStart -> -1
                    , PLDFaceStart -> 1
                    , PLDRunASingleFace -> false
                    , AddSing -> {}
                    , DoubleRoots -> {}
                    , AddLetters -> {}]
\end{lstlisting}

Lines 1--12 define options related to the Baikov representation, graph specifications, and scanning over subtopologies.

\paragraph{Baikov 
options.} \texttt{LoopEdges} specifies which internal edges are assigned $\ell_1, \dots, \ell_L$ before {\SOFIA} searches for both an ordering and linear shifts in the $\ell$'s minimizing the number $N$ of Baikov integration variables. For example, \texttt{LoopEdges -> \{3,7\}} means $\ell_1$ and $\ell_2$ are initially assigned to the $3^\text{rd}$ and $7^\text{th}$ edge in the list \texttt{edges}. If \texttt{LoopEdges -> Automatic}, this choice is automatically made by the package to minimize $N$. \decosix

\paragraph{ISPs.}
Negative powers of ISPs are included in \eqref{eq:LandauPoly} if \texttt{IncludeISPs -> True} and are excluded otherwise. In other words, \texttt{IncludeISPs -> True} places ISPs in the denominators, while \texttt{IncludeISPs -> False} places them in the numerators, where they do not contribute to singularities. \decosix

\paragraph{Skipping and targeting subtopologies.}
For examples that are still computationally expensive with \texttt{IncludeSubtopologies -> True} and \texttt{Symmetries -> True}, the user may want to set an upper bound on the time \texttt{T} (in seconds) taken by {\SOFIA} to perform the singularity analysis. This can be done using the options \texttt{UnclogTime -> T}. The default value is the age of the Universe. If the analysis of a (sub)topology exceeds time \texttt{T}, the code will record the position of the clogging components, allowing the user to inspect them individually later and optimize performance by adjusting other options.

The options \texttt{StartAtSubtopology -> n} and \texttt{EndAtSubtopology -> m} allow the user to start and end the singularity analysis on the \texttt{n}-th and \texttt{m}-th subtopologies, respectively. \decosix

\begin{figure}
    \centering
    \begin{subfigure}{1\textwidth}
        \centering
        \includegraphics[scale=1.1,valign=c]{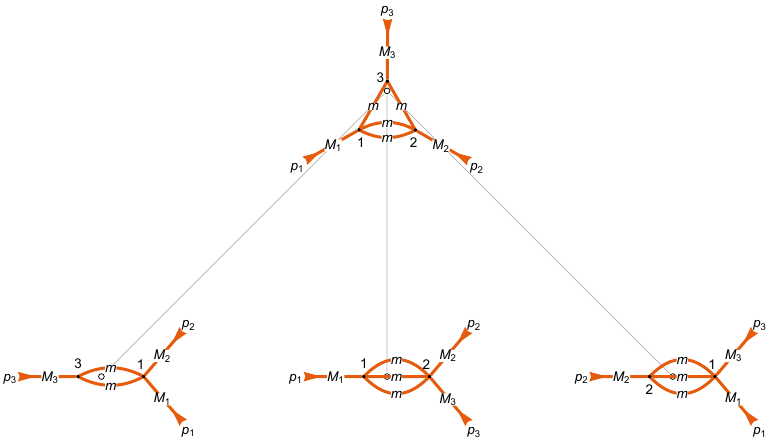}
        \caption{}\label{fig:parSing}
    \end{subfigure}
    \hfill
    \begin{subfigure}{1\textwidth}
        \centering
        \includegraphics[scale=1.1,valign=c]{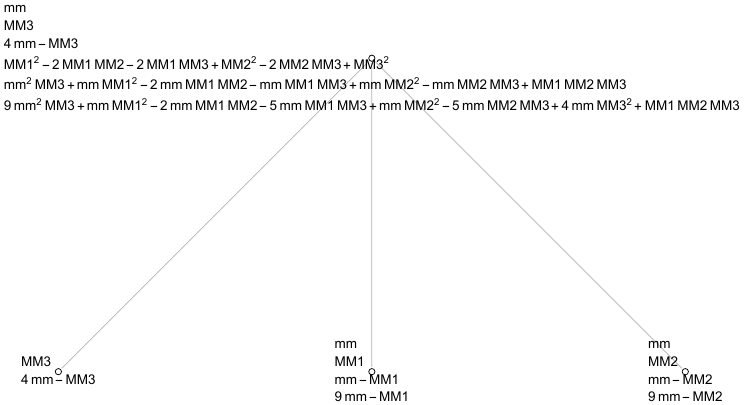}
        \caption{}\label{fig:parGraph}
    \end{subfigure}
    \caption{\label{fig:tree}Example output of \texttt{SOFIA[\{edges, nodes\}, ShowHistoryGraph -> True, SolverBound -> Infinity]} applied to the parachute diagram with \texttt{edges = \{\{\{1,2\},m\}, \{\{1,2\},m\}, \{\{2,3\},m\}, \{\{1,3\},m\}\}} and  
    \texttt{nodes = \{\{1,$\mathtt{M_1}$\}, \{2,$\mathtt{M_2}$\}, \{3,$\mathtt{M_3}$\}\}}. Panel (a) illustrates how diagrams are connected through edge contractions (with tadpoles removed), while panel (b) shows the corresponding candidate singularities. As explained in the main text, \texttt{SOFIA[]} uses the shorthands $\mathtt{mm = m^2}$ and $\mathtt{MMi = M_i^2}$ in its outputs. Note that the singularities in (b) are in one-to-one correspondence with those in the
    database \cite{MathRepo}---that is, there are no fictitious singularities in this example. 
    }
\end{figure}

\paragraph{Subtopology history graph.}
When \texttt{IncludeSubtopologies -> True}, the option \texttt{ShowHistoryGraph} 
can be used to print a top-bottom graph (laid out based on its spanning tree) whose nodes represent (sub)topologies, with edges connecting those related by edge contractions. If \texttt{ShowHistoryGraph -> True}, both the diagrams and the corresponding singularities are displayed at each node (see Figs.~\ref{fig:parSing} and \ref{fig:parGraph}, respectively); if \texttt{ShowHistoryGraph -> SingOnly}, only the singularities are shown.

\paragraph{Numerical substitutions.}
Finally, \texttt{Substitutions} allows the user to substitute some of the kinematic invariants (e.g., \texttt{Substitutions -> \{s12 -> 1\}}).  Substitutions are applied before the polynomial equations are solved. \decosix

\paragraph{Degenerate cuts.}
As mentioned in the discussion following \eqref{eq:maxcut}, computing the maximal cut of a valid loop-by-loop Baikov representation may result in a degenerate (e.g., measure zero) integration contour. \texttt{SOFIA[]} implements heuristics to detect such situations and prints a warning. Whenever this happens, an alternative choice of \texttt{LoopEdges} might be needed; see examples in \cite{repo}. In case of doubt, the warnings can be enabled by setting \texttt{ShowPossiblyDegenerate -> True}. \decosix 

Lines 13--21 in the code snippet above give options that control the solving strategy. 

\paragraph{Solver options.}
Within {\SOFIA}, there are two available solvers: \texttt{FastFubini} (which implements the analytic procedure outlined in Sec.~\ref{sec:SingAnalysis}) and \texttt{momentumPLD} (which instead uses the numerical approach of Sec.~\ref{sec:SingAnalysis}). The former relies solely on \textsc{Mathematica} functions, while the latter requires the working \textsc{Julia} dependencies detailed in App.~\ref{app:julia}. All the options \texttt{PLDname} relate specifically to \texttt{momentumPLD} (lines 13–18) are described under the same \texttt{name} as in \cite{MathRepoTutorial}. Finally, the option \texttt{FactorResult -> False} can be used to skip the factorization into irreducible components of $\mathcal{S}(\vec{s})$ in \eqref{eq:Sdef}, which, for large examples, can require both a long runtime and a large amount of memory.

We found that the native \textsc{Mathematica} solver outperforms the \textsc{Julia} solver on virtually all examples. Nevertheless, we keep the \textsc{Julia} solver available as an option to, e.g., cross-check the results obtained from the Fubini reduction in cases where it is practical. \decosix

Lines 22–24 in the code snippet above set options for \texttt{Effortless}, which are relevant only when \texttt{FindLetters -> True} is enabled.

\paragraph{\textbf{\texttt{Effortless}} options.}
At times, users might want to supplement the singularities computed by the \texttt{SOFIA[]} command with additional ones. For example, if singularities of subtopologies are already known or anticipated and \texttt{IncludeSubtopologies -> False} is used. To do so, one can use the option \texttt{AddSing -> \{p1,p2,...\}} to append the specified polynomials to the singularities found by \texttt{SOFIA[]}.

By default (\texttt{DoubleRoots -> \{\}}), \texttt{SOFIA[]} uses $g_a=h_b=1$ in \eqref{eq:doubleRoots} to construct the candidate odd alphabet $\mathbb{A}^{\text{odd}}$ from the ansatz in \eqref{eq:odd-letter}.  The \texttt{DoubleRoots} option allows more flexibility. Specifying \texttt{DoubleRoots -> \{\{fm,..., fn\},\{fp,..., fq\}\}} forces \texttt{SOFIA[]} to consider in \eqref{eq:doubleRoots} the polynomials provided by the user (yet typically belonging to \eqref{eq:WF}): \texttt{\{fm,..., fn\}} for $g_a$ and \texttt{\{fp,..., fq\}} for $h_b$. For instance, setting \texttt{DoubleRoots -> \{\{f1, f2\},\{f1,f3,f4\}\}} ensures that $g_a = \mathtt{f1}, \mathtt{f2}$ and $h_b=\mathtt{f1}, \mathtt{f3}, \mathtt{f4}, \mathtt{f4}$ are considered in building the candidate $\mathbb{A}^{\text{odd}}$. Selecting \texttt{DoubleRoots -> \{\{fm, ..., fn\}, All\}} instructs \texttt{SOFIA[]} to assign the list \texttt{\{fm,..., fn\}} to $g_a$ while using all $f_i$ from \eqref{eq:WF} (which are internally constructed by \texttt{SOFIA[]}) for $h_b$. 

In contrast, choosing \texttt{DoubleRoots -> All} directs \texttt{SOFIA[]} to apply all $f_i$ from \eqref{eq:WF} to both $g_a$ and $h_b$.  In most cases, using the ``sledgehammer'' option \texttt{DoubleRoots -> All}  comes at the cost of using significantly more computational resources. An example using the \texttt{DoubleRoots} option is discussed in Sec.~\ref{sec:example-alphabet}.

The option \texttt{AddLetters -> \{\{Even letters\},\{Odd letters\}\}} allows the user to append additional even and odd letters to the list constructed by \texttt{Effortless}. The default value is the empty list \texttt{\{\}}. \decosix

\subsubsection{\texttt{SOFIADecomposeAlphabet} command}
It is often useful to be able to check if two alphabets are equivalent. The command:
\begin{lstlisting}[extendedchars=true,language=Mathematica]
SOFIADecomposeAlphabet[alphabet1, alphabet2]
\end{lstlisting}
will check if (the differential of) each letter in the list \texttt{alphabet1} can be decomposed linearly in terms of that of \texttt{alphabet2}. The input format for the first entry \texttt{alphabet1} is a list of logarithms (for an explicit example, see \texttt{knownAlphabet} in Sec.~\ref{sec:example-alphabet}). The second entry \texttt{alphabet2} should have the same format as the output of the \texttt{SOFIA[]} command when \texttt{FindLetters -> True}, that is \texttt{alphabet2 = \{\{Even letters\}, \{Odd letters\}, \{Square roots\}\}}. Consequently, one can directly set $\mathtt{alphabet2 = SOFIA}$\texttt{[...]}. This is useful to check whether the output of \texttt{SOFIA[]} spans an alphabet in the case that it is already known.

\subsubsection{\texttt{SOFIABaikov} command}
A loop-by-loop Baikov representation minimizing $N'$ can be generated using:
\begin{lstlisting}[extendedchars=true,language=Mathematica]
SOFIABaikov[{edges, nodes}, LoopEdges -> Automatic
                          , Dimension -> dim
                          , MaxCut -> False
                          , ShowXs -> False
                          , ShowDetailedDiagram -> True]
\end{lstlisting}

The options in lines 1 and 3 work the same way as the options with the same names in the \texttt{SOFIA[]} command. To set the spacetime dimension, one uses \texttt{Dimension -> dim}. When restricting to an integer dimension (e.g., $\mathtt{dim} = 4$), then either \eqref{eq:LBL-Baikov} or its maximal cut \eqref{eq:maxcut} might vanish. This can happen because the prefactors include $\Gamma$-functions that may diverge; see, e.g., \eqref{eq:generalBaikov}, which could indicate a potential $\frac{\varepsilon}{\varepsilon}$ cancellation. In such cases, {\SOFIA} will automatically display a warning and suggest to use dimensional regularization (e.g., $\mathtt{dim} = 4 - 2\varepsilon$).

The Baikov variables, chosen internally by the code, can be printed in terms of loop and external momenta by setting \texttt{ShowXs -> True}. Enabling \texttt{ShowDetailedDiagram -> True} additionally displays these variables on the diagram specified by \texttt{\{edges, nodes\}}, along with the chosen internal edge momenta.

\subsubsection{\label{ssec:GF}Graphical functions}
{\SOFIA} offers several functions that leverage graph-theoretical techniques to streamline and accelerate singularity analysis, which may also be useful for other purposes. In particular:
\begin{lstlisting}[extendedchars=true,language=Mathematica]
FeynmanDraw
FeynmanPlot[{edges,nodes}, EdgeLabels -> {},
                         , HighlightedEdges -> {}]
Subtopologies[{edges,nodes}, Reflections -> False]
Symmetry[{edges1,nodes1},{edges2,nodes2}]
SymmetryQuotient[ListOfDiagrams]
\end{lstlisting}

\paragraph{Drawing diagrams.} The function \texttt{FeynmanDraw} opens a window that lets the user draw the desired Feynman diagram by clicking and dragging the pointer. It outputs a list of edges and nodes in the format \texttt{\{edges,nodes\}} with generic internal and external masses, which can be later adjusted by the user. By convention, the external momenta are all incoming and are assigned according to the ordering in \texttt{nodes}. The list \texttt{\{edges,nodes\}} completely specifies a Feynman diagram associated to \eqref{eq:loopMomDef}, and is the main input of most functions in {\SOFIA}. \decosix

\paragraph{Plotting diagrams.} A tuple of the form \texttt{\{edges, nodes\}} can be verified to represent the correct Feynman diagram by running \texttt{FeynmanPlot[\{edges, nodes\}]}. The options \texttt{EdgeLabels} and \texttt{HighlightedEdges} allow labeling each element of \texttt{edges} (for instance, with edge momenta or Baikov variables) and highlighting specific edges by specifying their positions in \texttt{edges}. By default, if no options are provided, internal edges are decorated with the masses from \texttt{edges} and external edges with the masses from \texttt{nodes}.

\paragraph{Generating subtopologies.}
Given a Feynman diagram represented as \texttt{\{edges, nodes\}}, the command \texttt{Subtopologies[\{edges, nodes\}]} generates a list of edges and nodes for each non-trivial subtopology of the original graph. It automatically removes contact diagrams, one-vertex reducible diagrams and tadpoles from the list of diagrams produced. For example, applying this command to any two-loop diagram will return a list of its one- and two-loop subtopologies, where the former arise from tadpole excisions—i.e., by removing internal edges of the form \texttt{\{\{i,i\},\_\}} (see for example Fig.~\ref{fig:parSing}).

The \texttt{Reflections -> True} option leverages \texttt{Symmetry[]} (discussed below) to concurrently eliminate subtopologies related by reflection transformations (i.e., symmetries that do not involve any kinematic replacement rules). This option is particularly effective when most subtopologies are reflection-symmetric. For example, for the triple-box diagram in \texttt{SOFIA\_examples.nb} \cite{repo}, this option is shown to improve the computation speed by approximately an order of magnitude. Note that behind the scenes, this is what the option \texttt{Symmetries -> reflections} is doing in the \texttt{SOFIA[]} command. \decosix

\paragraph{Symmetries.} Given two Feynman diagrams represented as \texttt{\{edges1, nodes1\}} and \texttt{\{edges2, nodes2\}}, the command \texttt{Symmetry[\{edges1,nodes1\},\{edges2,nodes2\}]} identifies symmetry transformations relating them, accounting for masses and external momenta. If no symmetry exists, it returns an empty list: \texttt{\{\}}. Otherwise, it outputs \texttt{\{\{edges,nodes\},rule\}} where \texttt{\{edges, nodes\}} is one of the input diagrams and \texttt{rule} is the minimal replacement rule on the masses and Mandelstam variables required to relate it to the other diagram. In the above context, a reflection refers to a symmetry where \texttt{rule = \{\}}. 

\begin{figure}
    \centering
    \begin{subfigure}{0.20\textwidth}
        \centering
        \includegraphics[scale=0.71]{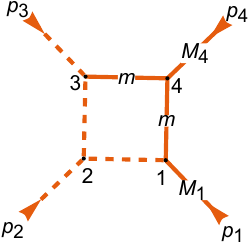}
        \caption{}\label{fig:sym-box1}
    \end{subfigure}
    \hfill
    \begin{subfigure}{0.20\textwidth}
        \centering
        \includegraphics[scale=0.71]{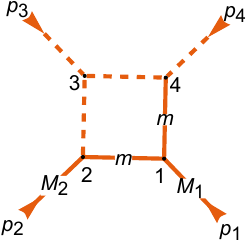}
        \caption{}\label{fig:sym-box2}
    \end{subfigure}
    \hfill
    \begin{subfigure}{0.26\textwidth}
        \centering
        \includegraphics[scale=0.71]{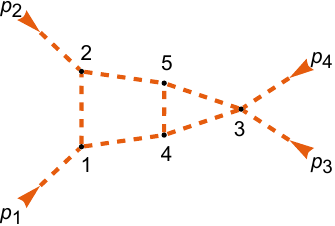}
        \caption{}\label{fig:sym-rep1}
    \end{subfigure}
     \hfill
    \begin{subfigure}{0.20\textwidth}
        \centering
        \includegraphics[scale=0.71]{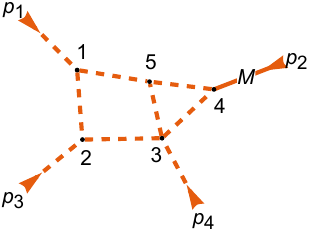}
        \caption{}\label{fig:sym-rep2}
    \end{subfigure}
    \caption{\label{fig:sym} Diagrams related by the \texttt{Symmetry[]} function as described in the text. These diagrams were all generated by the \texttt{FeynmanPlot[]} function. The dashed lines denote  massless particles and otherwise the mass label is displayed on the edges.}
\end{figure}

For example, Fig.~\ref{fig:sym} displays two pairs of diagrams (a-b and c-d) related by symmetry. Running \texttt{Symmetry[(a),(b)]}---with \texttt{(a)} and \texttt{(b)} representing the lists of edges and nodes of the diagrams of the same names---gives \texttt{\{(a),\{$\mathtt{s_{1,2}\!\to\!s_{2,3}, s_{2,3}\!\to\! s_{1,2}, M_1^2\!\to \!M_2^2, M_4^2\!\to\!M_1^2}$\}\}}, which corresponds to a clockwise rotation of $90^\circ$ relating diagram (a) to diagram (b).

Diagrams (c) and (d) have a less obvious relation. Running \texttt{Symmetry[(c),(d)]} outputs \texttt{\{(d),\{$\mathtt{s_{1,2}\!\to\!0, s_{2,3}\!\to\!0, M^2 \to s_{1,2}}$\}\}} which effectively amounts to setting the momentum $p_4$ of diagram (d) to zero, $p_4 \to 0$. Indeed, in terms of Mandelstam variables, we have $\,p_4 \cdot p_1 = \, p_2 \cdot p_3 \propto s_{23} \to 0$ and $\,p_4 \cdot p_3 = \,p_1 \cdot p_2 \propto s_{12} \to 0$. In this case, \texttt{Symmetry[]} also correctly recognizes that the mass scale $M^2$ in diagram (d) plays the role of the Mandelstam invariant $s_{34} =s_{12}$ in diagram (c).

Note that the symmetry relation between diagrams is not necessarily bi-directional, as the example (d) $\to$ (c) demonstrates. We observe that \texttt{Symmetry[]} typically returns the diagram with the larger number of kinematic scales among the input diagrams. Furthermore, \texttt{Symmetry[]} can relate diagrams with different numbers of external edges. This includes cases where external momenta meet at a common vertex or where an external leg effectively ``detaches'' by setting its momentum exactly to zero.

In such cases, kinematic replacements can be subtle because the operations of specializing kinematics and computing singularities generally do not commute (a phenomenon analogous to that encountered when specializing kinematics and evaluating Feynman integrals; see \cite{Mizera:2021icv,Dlapa:2023cvx,Fevola:2023kaw}). To address this, we introduce a small parameter $\delta$ and extract the leading non-vanishing coefficient for each component of the singular locus $\mathcal{S}$ as $\delta \to 0$. See App.~\ref{app:subtle} for an explicit demonstration.

Before moving on, we introduce the higher-level command \texttt{SymmetryQuotient[L]}. This command takes a list of diagrams formatted as \texttt{L = \{\{edges1, nodes1\}, \{edges2, nodes2\}, \{edges3, nodes3\}, ...\}} and returns a minimal set of representative diagrams, along with the corresponding replacement rules for kinematic scales needed to reproduce all the diagrams in the input list. In the main command \texttt{SOFIA[]}, enabling the option \texttt{Symmetries -> True} internally applies \texttt{SymmetryQuotient} to the subtopologies generated by \texttt{Subtopologies}. \decosix

\subsubsection{Solvers}
All the documented functions so far have been specifically designed for applications to Feynman integrals. However, the solvers used behind the scenes can handle a variety of other problems, some of which are explored in Sec.~\ref{sec:examples}. They can be used independently of the earlier commands via:
\begin{lstlisting}[extendedchars=true,language=Mathematica]
SolvePolynomialSystem[polynomials, variables, Solver -> FastFubini
                                            , SolverBound -> 100
                                            , FactorResult -> True
                                            , PLDMethod -> sym
                                            , PLDHomogeneous -> true
                                            , PLDHighPrecision -> false
                                            , PLDCodimStart -> -1
                                            , PLDFaceStart -> 1
                                            , PLDRunASingleFace -> false]
\end{lstlisting}
Here, instead of the diagram information, the inputs consist of a set of \texttt{polynomials} and a set of \texttt{variables} to eliminate. The available options remain the same as above. In the case where \texttt{Solver -> momentumPLD}, the user must ensure that the variables to be eliminated are given as a text string in the form \texttt{variables = \{x1, x2, x3, ..., xn\}}. These variables are used to parametrize \texttt{polynomials} and are the default variables employed by the {\SOFIA} wrapper for \textsc{Julia}, i.e., they are predefined in the \textsc{Julia} external evaluator and not using them would cause failure. While the \texttt{SOFIA} function ensures this automatically when called, this step must be handled manually when using \texttt{SolvePolynomialSystem}. 

\section{\label{sec:examples} Detailed
examples}

\begin{figure}
    \centering
    \begin{subfigure}{0.32\textwidth}
        \centering
        \includegraphics[scale=0.83]{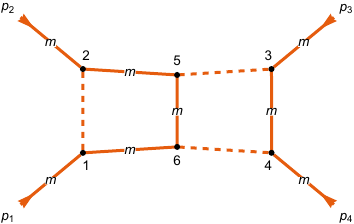}
        \caption{}\label{fig:diagrams-a}
    \end{subfigure}
    \hfill
    \begin{subfigure}{0.32\textwidth}
        \centering
        \includegraphics[scale=0.83]{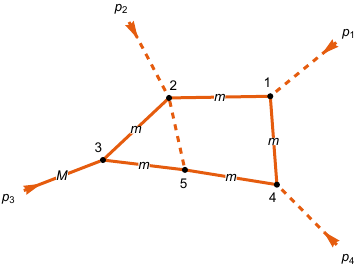}
        \caption{}\label{fig:diagrams-b}
    \end{subfigure}
    \hfill
    \begin{subfigure}{0.32\textwidth}
        \centering
        \includegraphics[scale=0.83]{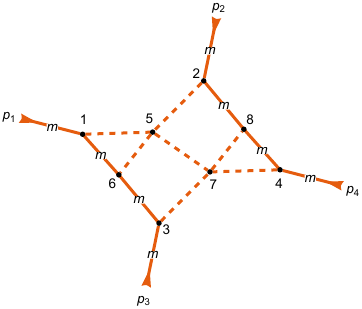}
        \caption{}\label{fig:diagrams-c}
    \end{subfigure}
    \caption{\label{fig:diagrams}Example diagrams considered in Sec.~\ref{sec:examples}. These diagrams were all generated by the \texttt{FeynmanPlot} function. The dashed lines denote  massless particles and otherwise the mass label is displayed on the edges.}
\end{figure}

We now demonstrate the explicit use of the functions introduced above with concrete examples that are not too large to fit within the text. Collectively, these examples took less than 30 seconds to run on a conventional laptop. A list of 30 additional examples, including many multi-scale and multi-loop ones, can be found in the \textsc{Mathematica} notebook \texttt{SOFIA\_examples.nb} \cite{repo}.

\subsection{\label{sec:example-alphabet} Symbol alphabet}

In this example, we illustrate how {\SOFIA}, combined with the publicly available package \texttt{Effortless} \cite{Effortlessxxx}, can be used to determine a (super)set of logarithmic letters, both even and odd, for a given Feynman integral. 

We consider the diagram in Fig.~\ref{fig:diagrams-a}, which contributes to the $t$-channel planar Møller and Bhabha scattering amplitudes at $\mathcal{O}(e^6)$ in QED. The massive electrons and/or positrons are represented by solid lines, while the massless photon exchanges are depicted by dashed lines. This diagram was computed explicitly using differential equations in \cite{Duhr:2021fhk} and thus serves as an excellent crosscheck to the {\SOFIA} implementation.

A candidate alphabet can be generated by defining the graph in terms of its nodes and edges and then running the \texttt{SOFIA[]} command as follows: 
\begin{lstlisting}[extendedchars=true,language=Mathematica]
edges = {{{1,2},0}, {{2,5},m}, {{3,5},0}, {{3,4},m},
         {{4,6},0}, {{1,6},m}, {{5,6},m}};
nodes = {{1,m}, {2,m}, {3,m}, {4,m}};
extraRoots = {-s12, -s23, -s12-s23}; 
SOFIAalphabet = SOFIA[{edges, nodes}, FindLetters -> True
                                    , SolverBound -> Infinity
                                    , DoubleRoots -> {extraRoots, All}]
\end{lstlisting}

The output is a (super)set of 10 even and 42 odd candidate letters, which we collect in the alphabet \texttt{SOFIAalphabet[[1;;2]]}\,$=\{\{\tilde{W}_1,\ldots,\tilde{W}_{10}\},\{\tilde{W}_{11},\ldots,\tilde{W}_{52}\}\}$. We use the notation $\tilde W_i = \log(W_i)$. For example: 
\begin{lstlisting}[extendedchars=true,mathescape=true,language=Mathematica]
$\tilde{W}_{31}$ =$\,$ Log[$\frac{\mathtt{s12\,(4 mm^2 + mm\,s12 - s12\,s23)+r[[5]]\,r[[10]]}}{\mathtt{s12\,(4 mm^2 + mm\,s12 - s12\,s23)-r[[5]]\,r[[10]]}}$];
$\tilde{W}_{32}$ =$\,$ Log[$\frac{\mathtt{s12\,(4 mm - s12)(3 mm - s23)+r[[5]]\,r[[10]]}}{\mathtt{s12\,(4 mm - s12)(3 mm - s23)-r[[5]]\,r[[10]]}}$];
(* 50 more odd and even candidate letters *)
\end{lstlisting}
where the $\mathtt{r[[i]]}$'s are square roots, which will be discussed below. The $\tilde{W}_i$ depend on $\mathfrak{s}=3$ variables: $\mathtt{s12}$, $\mathtt{s23}$, and $\mathtt{mm}$. We introduce $\mathtt{mm} = \mathtt{m}^2$ here because  \texttt{SOFIA} internally homogenizes the output polynomials and thus automatically replaces the mass variables $\mathtt{m}$ with the square root of their squares $\sqrt{\mathtt{mm}}$. In doing so, all the variables in the letters have the same mass dimension. In general, and as we will demonstrate explicitly below, not all of the letters in $\mathtt{SOFIAalphabet}$ are expected to appear in the differential equation \eqref{eq:differential-equation}. 

It should also be noted that, in the example shown in Fig.~\ref{fig:diagrams-a}, if the odd letters are assumed to be constructible only from the square roots of individual candidate singularities---as is the case with the default value \texttt{DoubleRoots -> \{\}} turned on---the resulting alphabet would be incomplete. 

To be more explicit, let us clearly distinguish the effect of working with the option \texttt{DoubleRoots -> \{extraRoots,All\}} instead of the default one. Running:
\begin{lstlisting}[extendedchars=true,language=Mathematica]
SOFIA[{edges, nodes}, (* same options as before *), DoubleRoots -> {}]
\end{lstlisting}
returns a set of odd letters that may depend on any of the 10 candidate square roots:
\begin{lstlisting}[extendedchars=true,mathescape=true,language=Mathematica]
r0s = {Sqrt[mm],
       Sqrt[4 mm^3 - mm^2 s12 + 2 mm s12 s23 - s12 s23^2],
       (* 8 more single square roots *)};
\end{lstlisting}
found by \texttt{SOFIA}, but where each odd letter contains \emph{at most} one square root. Now, keeping the same settings but with the replacement \texttt{DoubleRoots -> \{extraRoots, All\}}, \texttt{SOFIA} will (internally) add to this initial list of 10 allowed roots $10 \times 3 = 30$ double roots:
\begin{lstlisting}[extendedchars=true,mathescape=true,language=Mathematica]
r = Flatten[{Map[Sqrt[extraRoots] #&, r0s], r0s}];
\end{lstlisting}
for a total of 40 (single \emph{and} double) square roots 
to use in the construction of $\mathbb{A}^{\text{odd}}$:
\begin{lstlisting}[extendedchars=true,mathescape=true,language=Mathematica]
r[[5]] = Sqrt[-s12] Sqrt[4mm-s12];
r[[10]] = Sqrt[-s12] Sqrt[4 mm^3 - mm^2 s12 + 2 mm s12 s23 - s12 s23^2];
(* 38 more single and double square roots *)
\end{lstlisting}
As we shall show momentarily, $\mathtt{r}$ is sufficient for the \texttt{SOFIA} command to produce a complete basis of odd letters. (Note that the signs chosen in the \texttt{DoubleRoots} option above are purely conventional and ensure that the comparison below is performed using the root conventions of \cite{Duhr:2021fhk}).

At this stage, we can check the validity of \texttt{SOFIAalphabet} by comparing it against the alphabet \texttt{knownAlphabet} in \cite[Eq.~(2.8)]{Duhr:2021fhk}. This is done by running:
\begin{lstlisting}[extendedchars=true,mathescape=true,language=Mathematica]
knownAlphabet = Log[{$\frac{\mathtt{s12}}{\mathtt{mm}}$, $\frac{\mathtt{s12\,s23+r[[5]]\,r[[16]]}}{\mathtt{s12\,s23-r[[5]]\,r[[16]]}}$, (* 14 more letters *)}];
SOFIADecomposeAlphabet[knownAlphabet, SOFIAalphabet]
\end{lstlisting}

In line 1 above, the syntax reflects that the roots defined in \cite[Eq.~(2.6)]{Duhr:2021fhk} were manually re-expressed, for comparison purposes, in terms of the roots $\mathtt{r[[i]]}$.
The \texttt{SOFIADecomposeAlphabet} command then checks whether it is possible to decompose each element of \texttt{knownAlphabet} in terms of \texttt{SOFIAalphabet} and returns the decomposition if it is. For example, the most complicated letter in \texttt{knownAlphabet} decomposes according to:
\begin{equation}
\hspace{-0.3cm}
     \d\log \frac{r_{5} r_{10} \left(m^2{-}s_{12}\right){-}4 m^6 s_{12}{-}3 m^4 s_{12}^2{+}m^2 s_{12}^3{+}3 m^2 s_{12}^2 s_{23}{-}s_{12}^3 s_{23}}{r_{5} r_{10} \left(m^2{-}s_{12}\right){+}4 m^6 s_{12}{+}3 m^4 s_{12}^2{-}m^2 s_{12}^3{-}3 m^2 s_{12}^2 s_{23}{+}s_{12}^3 s_{23}}=\d\tilde{W}_{31}{-}2\d\tilde{W}_{32}\,.
\end{equation}
See the \textsc{Mathematica} notebook \texttt{SOFIA\_examples.nb}~\cite{repo} for all the details of the complete decomposition, as well as how the roots $r_{i}=\mathtt{r[[i]]}$ in \texttt{r} and the letters $\tilde{W}_{j}$ in \texttt{SOFIAalphabet} are defined.  In the same notebook, we show that the same procedure successfully reproduces other non-trivial alphabets, including those arising in the two-loop massless and one-mass pentagon functions \cite{Chicherin:2017dob,Abreu:2020jxa,Chicherin:2021dyp} as well as the genuine two-loop massless six-particle letters \cite{Abreu:2024fei,Henn:2025xrc}.

\subsection{\label{sec:example-geometries} Feynman geometries}

In the following example, we examine the diagram shown in Fig.~\ref{fig:diagrams-b} with a top-quark loop of mass $m$. This diagram is particularly interesting because it contributes, through the double-box containing it, to the $\mathcal{O}(g_s^4 m)$ amplitude for producing a Higgs boson of mass $M$ and a jet via gluon fusion at the LHC.

An important early step in computing this amplitude is to determine the types of functions to which the contributing diagrams are expected to evaluate, i.e., the function space. This information is encoded in the underlying geometry of the associated master integrals. To reveal it, it suffices to carefully analyze the maximal cut of each master integral in four dimensions ($\varepsilon=0$), which is conveniently done in the Baikov representation thanks to \eqref{eq:maxcut}. This approach is reliable because the underlying geometry is characterized by the solution of the homogeneous part of the differential equation satisfied by the uncut integrals, which coincides with the differential equation governing the maximal cut integral \cite{Primo:2016ebd,Bosma:2017ens}.

The maximal cut procedure \emph{always} reduces Feynman integrals to periods over an algebraic variety. The geometry of this variety dictates the class of functions that appear after integration. For instance, a genus-0 (rational) curve gives rise to polylogarithms, a genus-1 (elliptic) curve to elliptic functions, and more intricate cases may involve higher-genus Riemann surfaces (see, e.g.,~\cite{Marzucca:2023gto}) or even higher-dimensional Calabi--Yau manifolds (see Sec.,\ref{sec:example-eikonal}), pointing to increasingly rich—and still not fully understood—mathematical structures. Strictly speaking, this way, one can only place an upper bound on the complexity of a given Feynman geometry; verifying that it is not degenerate generally requires further analysis. (To our knowledge, there is currently no automated algorithm to \emph{systematically} determine the precise function space of a Feynman integral, e.g., whether it is polylogarithmic, elliptic, or beyond. Even less is known at the amplitude level, where full cancellations among non-polylogarithmic sectors may occur; see, e.g., \cite{Bern:2024adl}).

For the diagram in Fig.~\ref{fig:diagrams-b}, we can use {\SOFIA} to learn about its geometry. Computing its maximal cut amounts to running:
\begin{lstlisting}[extendedchars=true,language=Mathematica]
edges = {{{1,2},m}, {{2,3},m}, {{3,5},m}, {{4,5},m}, {{1,4},m}, {{2,5},0}};
nodes = {{1,0}, {2,0}, {3,M}, {4,0}};
SOFIABaikov[{edges, nodes}, MaxCut -> True, Dimension -> 4]
\end{lstlisting}
The result, obtained in a fraction of a second, is 
\begin{equation}\label{eq:ex2MC}
    \frac{-1}{256\pi^9}\int\frac{\d x_7}{\sqrt{\prod_{i=1}^{4}(x_7-r_i)}}\,, \quad \text{where}\quad 
    \begin{sqcases}
    r_{1,2}= (m\pm M)^2\,\\
    r_{3,4}= m^2+s_{12}\pm \frac{2m\sqrt{s_{12}s_{23}(s_{12}+s_{23}-M^2)}}{s_{23}}\,
    \end{sqcases}\,.
\end{equation}
The zero locus of the \emph{irreducible} quartic in $x_7$, $P_4(x_7)=\prod_{i=1}^{4}(x_7-r_i)$, in the denominator \emph{defines} the elliptic curve controlling the function space of this integral. 

From \eqref{eq:ex2MC}, it is also easy to read off the leading Landau singularity (where all propagators are on-shell) of the diagram in Fig.~\ref{fig:diagrams-b}. Indeed, these are given by the zeroes of the discriminant of $P_4(x_7)$ in $x_7$, i.e.,
\begin{equation}
\begin{split}
D_{x_7}P_4=m^2M^2s_{12}s_{23}&(M^2{-}s_{12})(M^2{-}s_{12}{-}s_{23})[16 m^4 (s_{12}{+}s_{23})^2{+}s_{23}^2 (M^2{-}s_{12})^2\\&{-}8 m^2 s_{23} (M^2 (s_{23}{-}s_{12}){+}s_{12}(s_{12}{+}s_{23}))]\,.
\end{split}
\end{equation}
These singularities can, of course, also be found automatically by running:
\begin{lstlisting}[extendedchars=true,language=Mathematica]
SOFIA[{edges, nodes}, MaxCut -> True]
\end{lstlisting}
One can verify that all the singularities in the $\varepsilon$-factorized differential equation from \cite{Gorges:2023zgv} are included in the resulting list available in \cite{repo}.

\subsection{\label{sec:example-eikonal} Eikonal/Regge/heavy-mass physics}

This example considers the eikonal/Regge scattering of two heavy masses---e.g., black holes or neutron stars, treated as point particles---via graviton exchange. In particular, we focus on the diagram shown in Fig.~\ref{fig:diagrams-c}, which contributes to the process at $\mathcal{O}(G_N^5)$. This example demonstrates two key points: first, how to take the eikonal limit of the Baikov representation; and second, how the latter turns out sufficient to uncover the underlying Calabi--Yau threefold geometry \cite{Frellesvig:2023bbf,Frellesvig:2024rea}.

In order to take the eikonal/Regge/heavy-mass expansion and keep the leading order term which captures the relevant classical physics \cite{Brandhuber:2021eyq}, we first parametrize the momenta in Fig.~\ref{fig:diagrams-c} such that
\begin{equation}
    \begin{split}
        p_1=\overline{p}_1 - \frac{q}{2}, \qquad p_3 = -\overline{p}_1 -\frac{q}{2}, \qquad p_2 = \overline{p}_2 + \frac{q}{2} , \qquad p_4 = -\overline{p}_2 + \frac{q}{2}\,.
    \end{split}
\end{equation}
Recall that we use all-incoming conventions for the momenta $p_i$.
The spacelike momentum transfer $q$ satisfies the transverse condition $q\cdot\overline{p}_i\equiv0$ for all $i$ as and is normalized so that $q^2\equiv-1$. Setting $\overline{p}_i=\overline{m} u_i$, with $u_i\cdot u_i\equiv1$ and $u_1\cdot u_2\equiv y$, the loop-by-loop representation can be obtained in two steps.

First, we run the \texttt{SOFIABaikov} command on the diagram with the original kinematics. This can be done with:
\begin{lstlisting}[extendedchars=true,language=Mathematica]
edges = {{{1,5},0}, {{2,5},0}, {{5,6},0}, {{5,7},0}, {{7,8},0}, {{3,7},0},
         {{4,7},0}, {{1,6},m}, {{3,6},m}, {{2,8},m}, {{4,8},m}};
nodes = {{1,m}, {2,m}, {3,m}, {4,m}};
maxCut = SOFIABaikov[{edges, nodes}, MaxCut -> True
                                   , Dimension -> 4
                                   , ShowXs -> True]
\end{lstlisting}
The output \texttt{maxCut} is an expression for the maximal cut of this diagram. The option \texttt{ShowXs} prints also the definition of all the $x_i$'s, including the leftover integration variables $x_{12}$, $x_{13}$, $x_{14}$ we get after cutting all the $x_1, x_2, \ldots, x_{11}$. In particular, we find
\begin{equation}
x_{12} = \ell_4^2\, ,\qquad
x_{13} = \ell_3^2\, ,\qquad
x_{14} = \ell_4 \cdot p_1 = \overline{m} \underbracket[0.4pt]{\ell_4 \cdot u_1}_{\hat{x}_{14}} + \ldots\, .
\end{equation}
The assignment of the loop momenta was automatic, but the precise knowledge of which momentum is which does not affect the analysis. We find:
\begin{equation}\label{eq:gravityInt}
\mathtt{maxCut} = \frac{1}{2^{15} \pi^{17}} \int\frac{\ed x_{12}\, \ed x_{13}\, \ed x_{14}}{\sqrt{x_{12}(4m^2 - x_{12})}
\sqrt{x_{13}(4m^2 - x_{13})}
\sqrt{P_2}\sqrt{P_4}} \,,
\end{equation}
where $P_2$ and $P_4$ are two polynomials with degrees $2$ and $4$ in the $x_i$'s respectively.

Implementing the eikonal limit is a matter of substituting the kinematic variables and expanding in large $\overline{m} = \mathtt{mb}$. It can be implemented with:
\begin{lstlisting}[extendedchars=true,mathescape=true,language=Mathematica]
eikonal = {s${}_{1,2}$ -> 2 (1 + y) mb^2,
           s${}_{2,3}$ -> 2 (1 - y) mb^2,
           m -> Sqrt[mb^2 - 1/4],
           x14 -> x14h mb};
maxCut = Series[Applyd[maxCut/.eikonal, {x12, x13, x14h}]
                                      , {mb, $\infty$, 4}, Assumptions -> mb > 0]
\end{lstlisting}
The list \texttt{eikonal} defines the substitutions, and the {\SOFIA} function \texttt{Applyd} implements the Jacobian for the change of variables to $(x_{12}, x_{13}, \hat{x}_{14}) = (\mathtt{x12}, \mathtt{x13}, \mathtt{x14h})$.
The leading order of this expansion gives:
\begin{equation}
\mathtt{maxCut} = -\frac{1}{2^{19} \pi^{17}\overline{m}^4} \int \frac{\d x_{12}\, \d x_{13}\, \d \hat{x}_{14}}{\sqrt{x_{12}} \sqrt{x_{13}} \sqrt{\hat P_2} \sqrt{\hat P_4}} + \mathcal{O}(\overline{m}^{-5})\, ,
\end{equation}
where the polynomials are given by
\begin{align}
\hat P_2 &= 4 \hat{x}_{14}^2 + (y^2 - 1)(1 + x_{12})^2\, ,\\
\hat P_4 &= \hat{x}_{14}^2 (1 + x_{13})^2 - x_{12} x_{13} (1 + x_{12} + x_{13})\, .
\end{align}
The curve $\hat P_4(x_{12}, x_{13}, \hat x_{14}) = 0$ defines a Calabi--Yau threefold (as a quartic polynomial in $\mathbb{CP}^3$ \cite{Hubsch:1992nu}), in agreement with the results of \cite{Frellesvig:2023bbf,Frellesvig:2024rea}.

The singularities of the resulting integral are simple to analyze. In {\SOFIA}, it boils down to running:
\begin{lstlisting}[extendedchars=true,mathescape=true,language=Mathematica]
SolvePolynomialSystem[{x12, x13, $\hat{P}_2$, $\hat{P}_4$}, {x12, x13, x14h}
                                        , SolverBound -> Infinity]
\end{lstlisting}
The result gives singularities at $y=0$ and $y = \pm 1$, corresponding physically to the high-energy and static limits, respectively.

\subsection{\label{sec:example-beyond} Beyond Feynman integrals}

In this subsection, we illustrate how {\SOFIA} can be used to study singularities of other classes of Feynman-like integrals.

\subsubsection{Energy-energy correlators}

As an example, we focus on the type of integrals arising in the computation of the four-point energy-energy correlator (EEC) in the collinear limit within $\mathcal{N}=4$ super Yang-Mills theory \cite{Chicherin:2024ifn}. Physically, the EEC captures meaningful correlations of energy flux after the emission of four final-state particles from an initial energetic parton and is now an active topic of research in both theoretical and experimental high-energy physics, see, e.g.\cite{Hofman:2008ar,Larkoski:2017jix}.

For four-point EEC, the relevant integrals to compute originate from a phase-space integral over a squared $1\to4$ amplitude \cite[Eqs.\,(2-4)]{Chicherin:2024ifn}. For our discussion, it is sufficient to quote the two families of integrals considered in \cite{{Chicherin:2024ifn}}:
\begin{subequations}\label{eq:EECfam}
    \begin{align}
        A_{a_1\dots a_7| a_8 a_9 a_{10}}&\equiv \int_0^\infty\d x_1\d x_2\d x_3\d x_4\frac{\delta[1-(x_1+x_2+x_3+x_4)]}{s_{1234}^{a_1}s_{123}^{a_2}s_{234}^{a_3}x_{1234}^{a_4}x_{234}^{a_5}x_{123}^{a_6}x_{34}^{a_7}s_{12}^{a_8}s_{23}^{a_9}s_{34}^{a_{10}}}\,,
        \\
        B_{a_1\dots a_4|a_{5}a_{6}}&\equiv\int_0^\infty\d x_1\d x_2\d x_3\frac{\delta[1-(x_1+x_2+x_3)]}{s_{123}^{a_1}x_{12}^{a_2}x_{23}^{a_3}x_{123}^{a_4}s_{12}^{a_5}s_{23}^{a_6}}\,,\label{eq:EEC-B}
    \end{align}
\end{subequations}
where $a_i\in\mathbb{Z}$ and the variables entering \eqref{eq:EECfam} are
\begin{subequations}\label{eq:EECvar}
    \begin{align}
        s_{l_1 l_2 \cdots l_n} \equiv \sum_{1 \leq i < j \leq n} x_{l_i}\, x_{l_j}\, (z_{l_i} - z_{l_j})\, (\bar{z}_{l_i} - \bar{z}_{l_j})\,, \qquad  x_{l_1 l_2 \cdots l_n}\equiv\sum_{i=1}^n x_{l_i}\,,\\
        \text{with}\qquad
        ( z_1, \overline{z}_1, z_2, \overline{z}_2, z_3, \overline{z}_3, z_4, \overline{z}_4) = \left( 0,0, z, \overline{z}, 1 , 1, \frac{z-w}{1-w}, \frac{\overline{z}-\overline{w}}{1-\overline{w}} \right)\, .
    \end{align}
\end{subequations}
We treat $z$, $\bar z$, $w$, $\bar w$ as independent variables.
The parametrization of these families is overcomplete since, e.g., $a_2$ and $a_3$ in \eqref{eq:EECfam} cannot be simultaneously positive. This would introduce an overlapping pole $\frac{1}{x_{12} x_{23}}$ which is forbidden by a version of Steinmann relations. We thus find it simpler to perform the singularity analysis at the level of master integrals introduced in \cite[App. B]{Chicherin:2024ifn} (in the same way we typically do the analysis at the level of individual Feynman diagram topologies, as opposed to all integrals at a given loop order at the same time).
 
The complete analysis is given in \cite{repo}.
Below, we illustrate the procedure on a simple master integral
\begin{equation}\label{eq:exampleEEC}
    B_1\equiv|z_1-z_2|^2\int_0^\infty\d x_1\d x_2\d x_3\frac{x_1 \delta[1-(x_1+x_2+x_3)]}{s_{123}\, x_{123}}\,.
\end{equation}
As explained in Sec.~\ref{sec:SingAnalysis}, to find the singular locus of this master integral, we can simply apply the Fubini reduction algorithm to the system of polynomials
\begin{equation}
\mathcal{L} = \{s_{123},x_{123},x_1,x_2,x_3\}\, ,
\end{equation}
which explicitly can be inputted as:
\begin{lstlisting}[extendedchars=true,language=Mathematica]
L = {z zb (x1 x2 + x2 x3) + x2 x3 (1 - z - zb) + x1 x3,
     x1 + x2 + x3,
     x1,
     x2,
     x3};
\end{lstlisting}
Note that $x_1\,,x_2$ and $x_3$ are included in $\mathcal{L}$ since $x_1=x_2=x_3=0$ appear in the integration boundaries of the $B$ family in \eqref{eq:EEC-B}. Within {\SOFIA}, the singularities are computed by running:
\begin{lstlisting}[extendedchars=true,language=Mathematica]
SolvePolynomialSystem[L, {x1, x2, x3}, SolverBound -> Infinity]
\end{lstlisting}
The output is 
\begin{equation}
    \{z,\;
    \overline{z},\;
    1-z,\;
    1-\overline{z},\;
    z-\overline{z},\;
    1- z \bar{z},\;
    z+\bar{z} - z \bar{z},\;
    z + \bar{z} - 2 z \bar{z}\}\,.
\end{equation}
Looping this command over all the master integrals in \cite[App. B]{Chicherin:2024ifn} immediately reproduces the polynomial part of the alphabet quoted in that reference, namely
\begin{equation}
    \begin{split}
        \{z\,,& \overline{z}\,, 1-z\,, 1-\overline{z}\,, z-\overline{z}\,,w\overline{z}-\overline{w}z\,,w\overline{z}-\overline{w}\,,z\overline{w}-w\,,1-w-\overline{w}+z\overline{w}\,,\\&
        1-w-\overline{w}+w\overline{z}\,,|z|^2-|w|^2\,,z-|w|^2\,,\overline{z}-|w|^2\}\cup (z\leftrightarrow w)\,. 
    \end{split}
\end{equation}
Notably, the analysis over the master integrals
\begin{equation}\label{eq:exampleEEC2}
    \{A_{22},A_{23}\}\equiv|z_2-z_3|^4\int_0^\infty\d x_1\d x_2\d x_3\d x_4\frac{\{x_2,x_3\} \delta[1-(x_1+x_2+x_3+x_4)]}{s_{123}\,s_{234}\,x_{1234}}\,,
\end{equation}
returns a distinctly complicated letter (see \texttt{SOFIA\_examples.nb} \cite{repo} for details):
\begin{equation}\label{eq:compactedEEC}
L =    w^4 \overline{w}^2 z^2+2 w^3 \overline{w}^3 z^2-2 w^3 \overline{w}^2 z^2+w^2 \overline{w}^4 z^2-2 w^2 \overline{w}^3 z^2-2 w^2 \overline{w}^2 z^3+(\text{151 terms})\,.
\end{equation}

 It has an interesting geometric interpretation. As explained in \cite{Chicherin:2024ifn}, the solutions to the maximal cut conditions of the integrals in \eqref{eq:exampleEEC2} (i.e., $s_{123}=s_{234}=x_{1234}=0$) define a singular cubic curve after projection onto the $(x_2,x_3)$-plane. Its roots $a\,,b\,,c$, which give rise to the \emph{non-integer} polynomial part of the alphabet listed in \cite[Eq. (12)]{Chicherin:2024ifn}, satisfy
\begin{equation}\label{eq:rootsEEC}
F=\Big\{abc=-|z|^2|w|^2\,,\, a+b+c=1-z-\overline{z}-w-\overline{w}\,,\, \frac{1}{a}+\frac{1}{b}+\frac{1}{c}=1-\frac{1}{z}-\frac{1}{\overline{z}}-\frac{1}{w}-\frac{1}{\overline{w}}\Big\}\,.
\end{equation}
The discriminant $D_X V$ of the cubic polynomial $V$ formed by these roots (given by Vièta's formula)
\begin{equation}\label{eq:EECcubic}
V = X^3-F_2 X^2+ F_1 F_3 X-F_1=0\,,
\end{equation}
is precisely \eqref{eq:compactedEEC}:
\begin{equation}
D_X V = [(a{-}b)(b{-}c)(c{-}a)]^2= L \,.
\end{equation}

\subsubsection{Cosmological correlators}

Similar singularity analysis can also be applied to cosmological correlators. Let us consider an example of a triangle diagram (also known as the one-loop three-site graph) for the conformally coupled scalar in Friedmann–Lemaître–Robertson–Walker cosmology. Following the notation of \cite{Benincasa:2024lxe}, the corresponding correlator expressed in Baikov representation is proportional to
\begin{equation}\label{eq:cosmo}
\int \frac{  y_{12} y_{23} y_{31} \d y_{12}\, \d y_{23}\, \d y_{31} }{q_{\mathfrak{g}_1} q_{\mathfrak{g}_2} q_{\mathfrak{g}_3}} \kappa^\chi \left[ \frac{1}{q_{\mathcal{G}_{12}}} \left( \frac{1}{q_{\mathfrak{g}_{23}}} + \frac{1}{q_{\mathfrak{g}_{31}}} \right) + \text{cyclic} \right] \, ,
\end{equation}
where 
\begin{subequations}
    \begin{align}
q_{\mathfrak{g}_j} &= x_j + y_{j-1,j} + y_{j,j+1}\,, \\
q_{\mathfrak{g}_{j,j+1}} &= x_j + x_{j+1} + y_{j-1,j} + y_{j+1,j+2}\,, \\
q_{\mathcal{G}_{j,j+1}} &= x_1 + x_2 + x_3 + y_{j,j+1}\,,
\end{align}
\end{subequations}
with the indices defined modulo $3$.
In addition, $\kappa$ is proportional to the Cayley--Menger determinant
\begin{equation}
\kappa \propto \det \begin{bmatrix}
0 & 1 & 1 & 1 & 1\\
1 & 0 & y_{12}^2 & y_{23}^2 & y_{31}^2 \\
1 & y_{12}^2 & 0 & x_2^2 & x_1^2 \\
1 & y_{23}^2 & x_2^2 & 0 & x_3^2 \\
1 & y_{31}^2 & x_1^2 & x_3^2 & 0
\end{bmatrix}\,,
\end{equation}
and $\chi$ is a generic parameter related to the type of cosmology.
Physically, the $x_i$'s are variables measuring energies of the external particles.

The singularity analysis of this integral is straightforward. We obtain it by considering all $6$ terms in the square brackets of \eqref{eq:cosmo} separately. We find 37 possible singularities, for example:
\begin{equation}
    L=x_1^4+2 x_1^3 x_2-x_1^2 x_3 (2 x_2+x_3)-2 x_1 (x_2-x_3) (x_2+x_3)^2+2 x_3 (x_2+x_3)^3\,,
\end{equation}
and record the rest in \texttt{SOFIA\_examples.nb} \cite{repo}.

To our knowledge, the singularity set for the triangle integral was never obtained before. Partial results exist \cite{Benincasa:2024ptf} and we have checked that our singularity set contains them.

\section{\label{sec:conclusion}Future directions}

In this work, we introduced the package {\SOFIA}, whose main purpose is to automate the computations of singularities of Feynman integrals. It combines a new theoretical understanding of Baikov representations with advanced polynomial reduction techniques and graph-theoretical tools.

This first implementation already pushes beyond the current state of the art in many cases. In particular, the ancillary file \cite{repo} includes several new predictions, including those of Fig.~\ref{fig:diagrams0}. These results highlight both the strength and versatility of {\SOFIA}, leading to results that were previously beyond reach.

This progress also opens up a number of future directions in automatizing the determination of the analytic structure of scattering amplitudes and other objects of interest, such as cosmological or energy correlators. 

\paragraph{Nature of singularities.} One of the most pressing questions is the algorithmic determination of the nature of singularities, i.e., the exponents $A$ and $B$ in the local behavior $f_j^A \log^B f_j$ as $f_j \to 0$. These exponents depend on the spacetime dimension and the geometry of the pinch manifold. In the case where the masses are sufficiently generic, this geometry is simple to describe, and consequently a closed-form formula for the exponents $A$ and $B$ is known \cite{Landau:1959fi}; see, e.g., \cite{pham2011singularities,Hannesdottir:2022bmo} for reviews. In ``non-generic'' situations, which are often the ones of physical interest, more work has to be done in analyzing the pinch manifold. Performing this analysis algorithmically would provide a significant improvement over the capabilities of {\SOFIA}, especially given that it would allow one to distinguish between square-root and logarithmic letters, thus optimizing the size of the proposed symbol alphabet.

\paragraph{Sheet structure.}
A related question concerns the Riemann sheet structure of Feynman integrals and scattering amplitudes. The simplest version of this question is to determine which singularities lie on the \emph{physical sheet} or in the \emph{physical region}. This question is usually addressed by either determining whether the loop momenta along the pinch manifold are physical, or whether the associated Schwinger parameters are non-negative. We have not attempted to address these questions since the Fubini reduction algorithm circumvents the need to solve the Landau equations directly. Therefore, it does not directly reconstruct the pinch manifold. Similarly, the discontinuity (monodromy) structure of Feynman integrals is more difficult to study in Baikov-like representations, as has been recognized for a long time \cite{Gribov:1962ft,islam1965analytic} (see \cite{Tourkine:2023xtu} for a recent study), since the integration contour itself depends on the kinematics. One possible direction is to keep track of the genealogy of singularities in the reduction algorithm, see \cite[Sec.~6]{Brown:2009ta}, which is closely related to the hierarchical structure; see, e.g., \cite{Berghoff:2022mqu}.

The questions about the sheet structure have their avatar in the symbol properties for polylogarithmic integrals. For example, presence on the physical sheet corresponds to the first-entry condition and the discontinuity structure is encoded in symbol adjacency. See \cite{Hannesdottir:2024cnn,Hannesdottir:2024hke} for recent work on the connection to Landau analysis. 
We believe that the same information is encoded in the commutation relations of the matrices $\mathbf{\Omega}_i$ in \eqref{eq:differential-equation} and boundary conditions even for non-polylogarithmic integrals, which means being able to predict it independently would put constraints on the $\mathbf{\Omega}_i$'s. See \cite{Abreu:2020jxa,Abreu:2021smk} for explicit examples in the context of extended Steinmann relations.

\paragraph{Finding the alphabet.}
To find the symbol alphabet we wrapped in {\SOFIA} the publicly available package \texttt{Effortless} \cite{Effortlessxxx}. 
More work is needed in the algorithmic determination of alphabets, which is currently the main bottleneck in {\SOFIA}  as can be seen in the timings indicated in Fig.~\ref{fig:diagrams0}.
We have already mentioned that this procedure would benefit from determining algorithmically which letters are square-root vs. logarithmic, though this information can be provided through the option \texttt{AddLetters}, as explained in Sec.~\ref{sec:SOFIA-command}. In addition, more conceptual work is needed to understand how general letters can be, including the maximum multiplicity of square roots that can be expected and which singularities are allowed to appear as root arguments. Here, we committed to the ansatz \eqref{eq:doubleRoots} with double square roots. More general ans\"atze can be implemented as needed. 

\paragraph{Feynman geometries.} We found the optimized loop-by-loop Baikov representation to be efficient at bounding the complexity of the geometries associated with Feynman integrals, e.g., whether they are expressible in terms of periods of elliptic curves or Calabi--Yau manifolds. In addition, {\SOFIA} allows one to find the loci of parameters for which they become singular. One may imagine extending {\SOFIA} to also analyze these geometries completely automatically.

\paragraph{Spacetime dimension.}
In the current implementation, {\SOFIA} assumes that the spacetime dimension $\D$ is large enough so that all external momenta remain independent up to momentum conservation.
It also assumes that $\D$ is generic, which is what is needed for dimensional regularization. This is why we only provide candidate singularities. Whenever one specializes the dimension $\D$, the actual list of singularities might, and in general will, reduce. Therefore, another improvement on {\SOFIA} would be to analyze such cases separately. Imposing a constraint on the external momenta is cumbersome because one has to work with constrained variables, or come up with a different set of kinematical data such as spinors. On the other hand, imposing a bound on the dimensionality of the loop momenta could potentially reduce the number of integration variables in Baikov representations.

Specializing the spacetime dimension may prove useful in other contexts, such as the S-matrix bootstrap \cite{Kruczenski:2022lot}, where understanding the singularity structure can be essential to make use of tools such as dispersion relations, particularly at higher points \cite{Caron-Huot:2023ikn,Guerrieri:2024ckc} and beyond lightest-particle scattering \cite{Homrich:2019cbt,Correia:2022dcu}. See also \cite{Vergu:2023rqz,Bargiela:2024hgt} for recent progress on computation of Feynman integrals via dispersion relations. Another relevant application is the study of two-dimensional integrable models, in which anomalous thresholds manifest as Coleman--Thun poles \cite{Coleman:1978kk,Braden:1990wx}. For example, an automated tool like {\SOFIA} could help extend recent studies on perturbative integrability beyond one-loop \cite{Dorey:2022fvs,Fabri:2024xmi}.

\paragraph{Bottleneck diagrams.} We observe that the performance of {\SOFIA} could be further optimized for Feynman diagrams that exhibit maximal connectivity---that is, diagrams for which the minimum number of vertex removals required to disconnect the graph into trees is maximized. This pattern becomes evident when monitoring the \texttt{SOFIA[]} command on sufficiently complicated diagrams with \texttt{IncludeSubtopologies -> True}: the (sub)topologies taking the highest runtime are those that are maximally, or nearly maximally, connected.

One main drawback of the loop-by-loop Baikov approach in these cases is that the resulting $N$ in \eqref{eq:NnoPrime} typically remains high, because each $\mathcal{E}_a$ is as large as possible for maximally connected diagrams, regardless of our minimizing strategy for the choice of \texttt{LoopEdges}. This poses a computational challenge starting at three-loops, because maximally connected diagrams lead to systems with a large number ($\ge5$) of variables to eliminate, even on the maximal cut. An example (with $N=5$ on the maximal cut) is the symmetric ``envelope diagram'' from \cite[Fig.~1e]{Mizera:2021icv}, whose singularities were previously calculated with \texttt{PLD.jl} in Schwinger parameter space for different masses assignments. Although the symbolic extraction of singularities for this diagram currently requires an impractical runtime in {\SOFIA}, we were able to use it to successfully reproduce numerically all of its known singularities. 

We emphasize that these diagrams remain equally challenging to solve with either \texttt{Solver -> FastFubini} or \texttt{Solver -> momentumPLD} in \texttt{SOFIA[]}. For the envelope, the difficulty in the latter arises because, on the max cut, one must still eliminate five Baikov variables (the $x$'s) \emph{and} five $\beta$'s from \eqref{eq:null-vector-equations} (with one $\beta$ fixed to 1). In total, this requires eliminating nine variables in Baikov space, compared to only five in Schwinger space.

Fortunately, the stringent criteria of maximal vertex connectivity ensure that, at a given loop order and multiplicity, the removal of only a few vertices results in most diagrams disconnecting into trees. Consequently, the number of maximally connected diagrams remains combinatorially small, especially for processes of phenomenological relevance, e.g., in the Standard Model. 

Finally, because the connectivity requirement naturally forces a uniformity in the distribution of vertices, maximally connected diagrams are generally highly symmetric. Fully exploiting this property, as demonstrated in \cite{KORS,Correia:2021etg} for the physical sheet singularity, could plausibly lead to further improvements of {\SOFIA}'s performance on these graphs.

\paragraph{Acknowledgments.} We thank Samuel Abreu, Giacomo Brunello, Simon Caron-Huot, Hjalte Frellesvig, Hofie Hannesdottir, Antonela Matijašić, Andrew McLeod, Ian Moult, Erik Panzer, Andrzej Pokraka, Simon Telen, and Simone Zoia for useful comments and discussions. In particular, we thank Antonela Matijašić and Julian Miczajka for sharing pre-release versions of \texttt{Effortless} with us. The work of M.C. and M.G. is supported by the National Science and Engineering Council of Canada (NSERC) and the Canada Research Chair program, reference number CRC-2022-00421. M.G. thanks the Institute for Advanced Study for
its hospitality during a crucial stage of this work. 

\appendix

\section{\label{app:Baikov}Derivation of Baikov representations}

\paragraph{Setup.} We begin by expressing a generic Feynman integral $I$ in a form that emphasizes the relevant structure for the derivation, particularly the integration measure:
\begin{equation}\label{eq:startBaikov}
    I = \int \d^\D \ell_1 \, \d^\D \ell_2 \cdots \d^\D \ell_L \, \mathcal{I}(\ell_1, \ell_2, \dots, \ell_L)\,,
\end{equation}
where the integrand $\mathcal{I}$ is a rational function of the loop and external momenta, $\ell_a$ and $p_i$ respectively, as well as possible other variables such as masses.

Let us use the label $q_i$ to collectively refer to the independent internal and external momenta, that is
\begin{equation}
(q_1, q_2, \ldots, q_{M}) = (p_1, p_2, \ldots, p_{\mathcal{E}}, \ell_L , \ldots, \ell_2, \ell_1)\, .
\end{equation}
We only include $\mathcal{E}$ external momenta, where
\begin{equation}
\mathcal{E} = \min(\D_0, n-1)\, .
\end{equation}
This is because there are at most $n-1$ independent $p_i$'s due to momentum conservation. In addition, the number of independent $p_i$'s cannot exceed the spacetime dimension $\D_0$ (in dimensional regularization, $\D = \D_0 - 2\varepsilon$).
The total number of such independent momenta $q_i$ is therefore $M = L+\mathcal{E}$.
They define a natural coordinate system, where we work in terms of the Lorentz-invariant scalar products: $X_{ij} = q_i \cdot q_j$. 

Let us consider the integration measure $\d^\D\ell_a$ in the order $a=1,2,\ldots,L$. We can assume that all the momenta $\ell_1, \ell_2, \ldots, \ell_{a-1}$ have been handled and now we consider $\ell_a$. The first key observation is that $\ell_a$ can be decomposed into two components:
\begin{equation}
    \ell_a = \ell_{a\parallel} + \ell_{a\perp}\, \qquad (1\le a\le L)\,,
\end{equation}
where $\ell_{a\parallel}$ lies in the space spanned by $(q_1, \dots, q_{M-a}) = (p_1, \ldots, p_{\mathcal{E}}, \ell_L, \ldots , \ell_{a+1})$, and $\ell_{a\perp}$ is perpendicular to this space. The integration measure then factorizes as
\begin{equation}
    \d^\D \ell_a = \d^{M-a} \ell_{a\parallel} \, \d^{\D-M+a} \ell_{a\perp}\, \qquad (1\le a\le L)\,.
\end{equation}
Our next goal is to change the variables in the parallel components $\ell_{a\parallel}$ of the loop momenta to the scalar products $X_{ij}$. In order to do so, we first derive the following general result.

\paragraph{Key equation.} Consider any vector $v$ in an $n$-dimensional space and express it in terms of a (not necessarily orthonormal) basis $(q_1,q_2,\dots,q_n)$:
\begin{equation}
v = \lambda_1\,q_1 + \lambda_2\,q_2 + \cdots + \lambda_n\,q_n\,.
\end{equation}
In components, this reads
\begin{equation}
v^i = \sum_{a=1}^{n}\lambda_a\,q_a^i\,.
\end{equation}
The differential volume element in the $n$-dimensional subspace is then given by
\begin{equation}\label{eq:dvn1}
\d^n v = \Bigl|\det\!\frac{\partial v^i}{\partial \lambda_a}\Bigr|\,\d^n \lambda\,.
\end{equation}
Since $\frac{\partial v^i}{\partial \lambda_a} = q_a^i$,
the corresponding Jacobian matrix $J$ has entries $J_{ia} = q_a^i$. Although the basis $(q_1,...,q_n)$ is not assumed orthonormal, we can compute
\begin{equation}
(J^T J)_{ab} =  q_a\cdot q_b \equiv G(q_1,q_2,\dots,q_n)_{ab}\,,
\end{equation}
where $G=G(q_1,q_2,\dots,q_n)$ is called the Gram matrix. Noting that $\det(J^T J) = (\det J)^2$, we deduce $(\det J)^2 = \det G$ and hence: $\det J = \pm \sqrt{\det G}$.
Thus, the change of variables yields
\begin{equation}\label{eq:genRes1}
\d^n v = \sqrt{\det G}\;\d^n \lambda \,.
\end{equation}
Note that the factor $\sqrt{\det G}$ is nothing but the volume of the parallelepiped spanned by the vectors $q_1,\dots,q_n$. This factor accounts for the distortion of the unit hypercube in the $\lambda$-space when mapped into the momentum space by the basis $\{q_a\}$. For an orthonormal basis, $\det G = 1$, and the volume is unchanged. For a non-orthonormal basis, the hypercube is transformed into a parallelepiped with volume $\sqrt{\det G}$, which corrects the integration measure accordingly. Equipped with \eqref{eq:genRes1}, we proceed with the derivation.

\textbf{Parallel components.} 
The parallel components are, by definition, linear combinations of the basis vectors $ q_1, q_2, \dots, q_{M-a} $:  
\begin{equation}\label{eq:lperpDef}
    \ell_{a\parallel} = \sum_{b=1}^{M-a} \lambda_{ab} q_b\, \qquad (1\le a\le L)\,.
\end{equation}  
Under the projection onto this $ q $-basis, the volume form transforms according to \eqref{eq:genRes1}:
\begin{equation}
\begin{split}
        \d^{M-a} \ell_{a\parallel} 
        &=\sqrt{\det G} \prod_{b=1}^{M-a} \d\lambda_{ab}\,,
\end{split}
\end{equation}
where now $G = G(q_1, q_2, \dots, q_{M-a})$.
Next, using \eqref{eq:lperpDef}, it is easy to see that the $X$'s and the $\lambda$'s are linearly related
\begin{equation}
    X_{ak}=q_k\cdot \ell_{a\parallel}=\sum_{b=1}^{M-a} \lambda_{ab} q_b\cdot q_k\iff X=\lambda\cdot G\,,
\end{equation}
where the right-hand side is a matrix equation. Thus, 
\begin{equation}\label{eq:parResult}
\begin{split}
        \d^{M-a} \ell_{a\parallel} 
        &=\sqrt{\det G} \Bigl|\det\!\frac{\partial X}{\partial \lambda}\Bigr|^{-1} \prod_{b=1}^{M-a} \d X_{ab}=\frac{\prod_{b=1}^{M-a} \d X_{ab}}{\sqrt{\det G}}\, \qquad (1\le a\le L)\,.
\end{split}
\end{equation}

\textbf{Perpendicular components.} The perpendicular components $\ell_{a\perp}$ are integrated in spherical coordinates, where the volume element in an $(\D-M+a)$-dimensional space is
\begin{equation}
    \d^{\D-M+a} \ell_{a\perp} = \ell_{a\perp}^{\D-M+a-1} \d \ell_{a\perp} \d\Omega_{\D-M+a}\,.
\end{equation}
By integrating out the angular part, we obtain
\begin{equation}\label{eq:measure000}
    \int\limits_{S_{\D-M+a}}\d^{\D-M+a} \ell_{a\perp} = \frac{1}{2} \Omega_{\D-M+a} \,\ell_{a\perp}^{\D-M+a-2} \, \d \ell_{a\perp}^2\,,
\end{equation}
where
\begin{equation}
\Omega_n = \frac{2\pi^{n/2}}{\Gamma(n/2)}\,,
\end{equation}
is the full $n$-dimensional solid angle. Choosing the $(q_1, \dots, q_{M-a})$ such that $\ell_{a\perp}^2 = X_{aa}$, we substitute $\d \ell_{a\perp}^2 = \d X_{aa}$, leading to
\begin{equation}\label{eq:perpResult}
    \hspace{-0.2cm}\int\limits_{S_{\D-M+a}}\hspace{-0.3cm}\d^{\D-M+a} \ell_{a\perp} =  \left[\frac{\det G(\ell_a, \ell_{a+1}, \dots, p_\mathcal{E})}{\det G(\ell_{a+1}, \dots, p_\mathcal{E})} \right]^{\frac{\D-M+a-2}{2}} \frac{\Omega_{\D-M+a}\,\d X_{aa}}{2}\,,
\end{equation}
for $1\le a\le L$. This formula has an insightful geometric interpretation. The loop momentum $ \ell_a $ is decomposed into two parts: one lying within the subspace spanned by $ (\ell_{a+1}, \dots, \ell_L, p_1, \dots, p_\mathcal{E}) $ (base space) and the other orthogonal to it (ruled hypersurface). The Cartesian product of these two subspaces forms a parallelepiped. Just as the height of a three-dimensional cylinder can be described as the ratio of two volumes, the height/radius of the ruled hypersurface can similarly be expressed in terms of the volume ratio:
\begin{equation}\label{eq:height0}
    |\ell_{a\bot}| \equiv \sqrt{X_{aa}}= \sqrt{\frac{\det G(\ell_a,\ell_{a+1},\dots,\ell_L,p_1,\dots,p_\mathcal{E})}{\det G(\ell_{a+1},\dots,\ell_L,p_1,\dots,p_\mathcal{E})}}\,.
\end{equation}
As in \eqref{eq:genRes1}, the Gram determinant $\det G(q_1,\dots,q_{\D-M+a}) $ gives the square of the volume of the parallelepiped spanned by the vectors $(q_i)_{i=1}^{\D-M+a}$. Hence, the numerator in \eqref{eq:height0} represents the volume of the entire parallelepiped formed by $(\ell_a,\ell_{a+1},\dots,\ell_L,p_1,\dots,p_\mathcal{E})$ while the denominator represents the volume of the base spanned by $(\ell_{a+1},\dots,\ell_L,p_1,\dots,p_\mathcal{E})$. 

Finally, since the integration region for the radial coordinate $\d \ell_{a\perp}^2$ in \eqref{eq:measure000} is $\ell_{a\perp}^2>0$, the integration region for the scalar measure in \eqref{eq:perpResult} is simply given by
\begin{equation}
    \Gamma_a \equiv \left\{ \ell_{a\perp}^2=\frac{\det G(\ell_a, \dots, \ell_L, p_1, \dots, p_\mathcal{E})}{\det G(\ell_{a+1}, \dots, \ell_L, p_1, \dots, p_\mathcal{E})} > 0 \right\}\, \qquad (1\le a\le L)\,.
    \label{eq:seedContour}
\end{equation}

\textbf{Final expression.} After substituting the parallel (see \eqref{eq:parResult}) and perpendicular  (see \eqref{eq:perpResult}) volume elements for each loop momentum, all but the $a=1$ and $a=L$ Gram determinants cancel, resulting in the Baikov representation of \eqref{eq:startBaikov} 
\begin{equation}\label{eq:generalBaikov}
\begin{split}
        \int \d^\D \ell_1 \d^\D \ell_2 \dots \d^\D &\ell_L\, \mathcal{I}(\ell_1, \dots, \ell_L)=   \frac{\pi^{\frac{L(1+2(\D+\mathcal{E})-L)}{4}}}{\prod_{l=1}^{L} \Gamma\left( \frac{\D-M+l}{2} \right) } \det G(p_1, \dots, p_\mathcal{E})^{(-\D+\mathcal{E}+1)/2}\\&    \times \int_\Gamma \prod_{i=1}^L\prod_{j=i}^M \d X_{ij} \det G(\ell_1, \dots, \ell_L, p_1, \dots, p_\mathcal{E})^{(\D-M-1)/2} f(X_{ij})\,,
\end{split}
\end{equation}
where the multiple factors of $i\pi^{\D/2}$ in \eqref{eq:loopMomDef} were absorbed in the integrand $f$. Moreover, the integration domain $\Gamma$ is given by imposing $L$ conditions given in \eqref{eq:seedContour}, such that $\Gamma \equiv \Gamma_1 \cap \Gamma_2 \cap \cdots \cap \Gamma_L=\{\det G(\ell_1, \dots, \ell_L, p_1, \dots, p_\mathcal{E})>0\}$.%
\footnote{The last equality follows from the observation that, since $\det G(q_1,q_2,\dots,q_n) > 0$, the biggest Gram matrix is positive definite. A key property of positive definite matrices is that every principal submatrix is also positive definite. In particular, Sylvester’s criterion \cite{Horn_Johnson_1985} tells us that all leading principal minors are positive, and in fact, this extends to any principal submatrix. Therefore, the submatrix corresponding to $q_2, q_3, \dots, q_n$ is positive definite, ensuring that $\det G(q_2,q_3,\dots,q_n) > 0$ and so on. In contrast, in the loop-by-loop case discussed in the main text, the multiple pairs of internal-external Gram matrices typically are not principal submatrices of one another. This occurs when they satisfy \eqref{eq:generalBaikov}, but with Gram matrices of different sizes (see, e.g., \eqref{eq:sunriseLBL}). In such cases, the integration contour remains the complete intersection of the corresponding $\Gamma_i$ regions.}

Note that the number of integration variables $ X_{ij} $ in the measure product corresponds to the number of unique pairs $ (i,j) $, where $ i $ runs from $ 1 $ to $ L $ and $ j $ runs from $ i $ to $ M $. This count evaluates to  
\begin{equation}
\sum_{i=1}^{L} (M - i + 1) = \frac{L(L+1)}{2} + L \mathcal{E} \equiv N'\,,
\end{equation}
when $ M = L + \mathcal{E}$. In the main text, we relabeled the $ N' $ integration variables as
\begin{equation}
    (x_1, \dots, x_{N'}) = \text{linear combinations of }\bigcup_{i=1}^{L} \{ X_{ij} \mid i \leq j \leq M \}\,,
\end{equation}
where the first $E$ $x$'s are propagators.  Given this notation, let us finally note that, in Sec.~\ref{sec:Baikov}, we denoted
\begin{subequations}\label{eqs:Baikov-expressions}
\begin{align}
c&= \frac{2^{L-N}(-i)^L \pi^{(L-N)/2}}{\prod_{l=1}^{L} \Gamma\left( \frac{\D-M+l}{2} \right)}\, ,\\
\gamma &=(\D-M-1)/2\, ,\\
\tilde\gamma &=(\mathcal{E}-\D+1)/2 \, ,\\
\mathcal{G}(\vec{x}; \vec{s}) &= \det G(\ell_1, \dots, \ell_L, p_1, \dots, p_\mathcal{E})\, ,\\
\tilde{\mathcal{G}}(\vec{s}) &= \det G(p_1, \dots, p_\mathcal{E})\, .
\end{align}
\end{subequations}
This completes the derivation of the global Baikov representation in \eqref{eq:global-Baikov}. 

\textbf{Loop-by-loop variants.} In contrast, the loop-by-loop Baikov representations \eqref{eq:LBL-Baikov} are derived from the global Baikov formula \eqref{eq:generalBaikov} by first choosing an ordering for the loops in the diagram considered. For each loop, one identifies the set of linearly independent external momenta (these include momenta from loops to be integrated later). Then, the one-loop Baikov representation (i.e., \eqref{eq:generalBaikov} for $L=1$) is applied to that loop momentum. Repeating this process for every remaining loop yields a representation that generally involves fewer integration variables than the global Baikov representation (see, e.g., the sunrise example discussed in Sec.~\ref{sec:Baikov}). Note that this procedure is not unique—the chosen loop ordering and the selection of independent momenta can influence both the final number of integration variables $N$ and the form of the integrand. The package {\SOFIA} automatically looks for a choice of loop shifts and ordering that minimizes $N$.

\section{\label{app:julia}Interface with \textsc{Julia}}

As mentioned in Sec.~\ref{sec:manual}, {\SOFIA} can optionally use the \textsc{Julia} package \texttt{PLD.jl} to extract the singular locus in \eqref{eq:Sdef}. A working installation of this package can be linked to {\SOFIA} with the option
\begin{lstlisting}[extendedchars=true,language=Mathematica]
SOFIAoptionJulia = True;
\end{lstlisting}
Running this command successfully for the first time requires a few steps. First, the user must ensure that \texttt{PLD.jl} and all its dependencies are correctly installed and running on their \textsc{Julia} interface. The installation procedure can be found in \cite{MathRepoTutorial}.

Next, the user needs to register the version of \textsc{Julia} in which \texttt{PLD.jl} runs (i.e., \texttt{v1.8.5}) as an external evaluator in \textsc{Mathematica} by running
\begin{lstlisting}[extendedchars=true,language=Mathematica]
RegisterExternalEvaluator["Julia", "/path/to/julia"]
\end{lstlisting}
where ``\texttt{/path/to/julia}'' is replaced by the actual path to the \textsc{Julia} binary. The user can then verify that \textsc{Mathematica} is communicating with \textsc{Julia} correctly by running, e.g.,
\texttt{ExternalEvaluate["Julia", "1 + 1"]}.
Only once this is done, one should be able to run \texttt{Get["SOFIA.m"]} with \texttt{SOFIAoptionJulia = True};
if the script is loaded successfully, a ``\texttt{Checking status}'' log should appear while loading without warnings or errors on the \textsc{Mathematica} interface.

\section{Subtleties with kinematic replacements}\label{app:subtle}

As described in Sec.~\ref{sec:manual}, in many cases the command \texttt{Symmetry[]} relates two diagrams with different numbers of scales. In such situations, some singularities might be missed if the kinematic replacement is not taken carefully.

\begin{figure}
    \centering
    \begin{subfigure}{0.49\textwidth}
        \centering
        \includegraphics[scale=1.1]{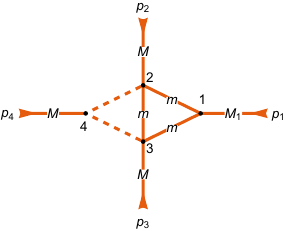}
        \caption{}\label{fig:acnd1}
    \end{subfigure}
    \hfill
    \begin{subfigure}{0.49\textwidth}
        \centering
        \includegraphics[scale=1.1]{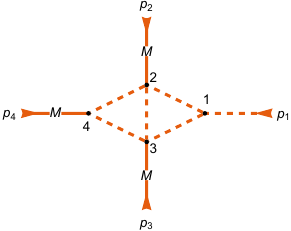}
        \caption{}\label{fig:acnd2}
    \end{subfigure}
    \caption{\label{fig:acnodes} Diagrams related by the command \texttt{Symmetry[]} and whose kinematic replacement rule must be taken with care, as described in the text. These diagrams were generated by the \texttt{FeynmanPlot[]} function.}
\end{figure}

To illustrate this point, we consider the two acnode diagrams \cite{Eden:1966dnq} defined by:
\begin{lstlisting}[extendedchars=true,mathescape=true,language=Mathematica]
(* Acnode diagram #1 *)
edges1 = {{{1,2},m}, {{2,3},m}, {{3,4},0}, {{2,4},0}, {{1,3},m}};
nodes1 = {{1,$\mathtt{M_1}$}, {2,$\mathtt{M}$}, {3,$\mathtt{M}$}, {4,$\mathtt{M}$}};
(* Acnode diagram #2 *)
edges2 = {{{1,2},0}, {{2,3},0}, {{3,4},0}, {{2,4},0}, {{1,3},0}};
nodes2 = {{1,0}, {2,$\mathtt{M}$}, {3,$\mathtt{M}$}, {4,$\mathtt{M}$}};
\end{lstlisting}
We can plot them via:
\begin{lstlisting}
FeynmanPlot/@{{edges1,nodes1},{edges2,nodes2}}
\end{lstlisting}
which outputs the diagrams in Fig.~\ref{fig:acnodes}, with diagram (a) corresponding to \texttt{\{edges1,nodes1\}}, and diagram (b) corresponding to \texttt{\{edges2,nodes2\}}. 

To find the symmetry relation we run:
\begin{lstlisting}[extendedchars=true,mathescape=true,language=Mathematica]
Symmetry[{edges1,nodes1},{edges2,nodes2}]
\end{lstlisting}
confirming that diagram \texttt{\{edges1,nodes1\}} is related to \texttt{\{edges2,nodes2\}} via the kinematic replacement $r_0=$\texttt{\{}$\mathtt{M_1^2}\to \mathtt{0}, \mathtt{m^2}\to \mathtt{0}$\texttt{\}}. 

Let us now see how this replacement rule applied naively is insufficient to reproduce all singularities. It is enough to perform the \emph{leading} singularity analysis with \texttt{IncludeSubtopologies -> False} on both diagrams:
\begin{lstlisting}[extendedchars=true,mathescape=true,language=Mathematica]
diags = {{edges1,nodes1},{edges2,nodes2}};
Do[singularitiesAcn[i] = SOFIA[diags[[i]], SolverBound -> 600
                                         , IncludeSubtopologies -> False]
                                         , {i,2}];
\end{lstlisting}
Applying the naive replacement rule $r_0$ directly to the eleven components of $\mathcal{S}$ obtained in \texttt{singularitiesAcn[1]} yields all eight components of $\mathcal{S}$ found for \texttt{singularitiesAcn[2]}, \emph{except for one}:
\begin{equation}\label{eq:missingSing}
    \mathtt{MM}^2-3\,\mathtt{MM}\,\mathtt{s12} + \mathtt{s12}^2+\mathtt{s12}\,\mathtt{s23}\,,
\end{equation}
where, we recall,  $\mathtt{MM}=\mathtt{M}^2$, $\mathtt{MM1}=\mathtt{M}_1^2$ and $\mathtt{sij}=\mathtt{s}_{ij}$.

However, by replacing $r_0$ with its ``blown-up'' version, $r_\delta = $ \texttt{\{}$ \mathtt{M_1^2}\to \mathtt{\delta}, \mathtt{m^2}\to \mathtt{\delta}$\texttt{\}}, expanding to leading order in $\mathtt{\delta}$, and factoring the result into irreducible components, we recover all the singularities in \texttt{singularitiesAcn[2]}, including \eqref{eq:missingSing}. In particular, from the $11^{\text{th}}$ component of $\mathcal{S}$ obtained for \texttt{singularitiesAcn[1]}, we have
\begin{equation}
   3\,\mathtt{MM}^{12}\,\mathtt{MM1}\,\mathtt{mm}+ (\text{1048 terms}) = 0\, \raisebox{0.4ex}{\mylongmapsto{$r_\delta$}}\,\delta^2\,\times\,
   \eqref{eq:missingSing}
   \times (\text{2 terms})\,.
\end{equation}
For all the details of this example, see the notebook \texttt{SOFIA\_examples.nb} in \cite{repo}. The $\delta$-procedure outlined above is implemented and executed automatically by {\SOFIA}.

\bibliographystyle{JHEP}
\bibliography{refs}

@article{Caron-Huot:2024brh,
    author = "Caron-Huot, Simon and Correia, Miguel and Giroux, Mathieu",
    title = "{Recursive Landau Analysis}",
    eprint = "2406.05241",
    archivePrefix = "arXiv",
    primaryClass = "hep-th",
    month = "6",
    year = "2024"
}

@article{Fevola:2023kaw,
    author = "Fevola, Claudia and Mizera, Sebastian and Telen, Simon",
    title = "{Landau Singularities Revisited: Computational Algebraic Geometry for Feynman Integrals}",
    eprint = "2311.14669",
    archivePrefix = "arXiv",
    primaryClass = "hep-th",
    doi = "10.1103/PhysRevLett.132.101601",
    journal = "Phys. Rev. Lett.",
    volume = "132",
    number = "10",
    pages = "101601",
    year = "2024"
}

@article{Frellesvig:2024ymq,
    author = "Frellesvig, Hjalte",
    title = "{The Loop-by-Loop Baikov Representation -- Strategies and Implementation}",
    eprint = "2412.01804",
    archivePrefix = "arXiv",
    primaryClass = "hep-th",
    month = "12",
    year = "2024"
}

@article{Frellesvig:2017aai,
    author = "Frellesvig, Hjalte and Papadopoulos, Costas G.",
    title = "{Cuts of Feynman Integrals in Baikov representation}",
    eprint = "1701.07356",
    archivePrefix = "arXiv",
    primaryClass = "hep-ph",
    doi = "10.1007/JHEP04(2017)083",
    journal = "JHEP",
    volume = "04",
    pages = "083",
    year = "2017"
}

@article{Brown:2009ta,
    author = "Brown, Francis C. S.",
    title = "{On the periods of some Feynman integrals}",
    eprint = "0910.0114",
    archivePrefix = "arXiv",
    primaryClass = "math.AG",
    month = "10",
    year = "2009"
}

@article{Panzer:2014caa,
    author = "Panzer, Erik",
    title = "{Algorithms for the symbolic integration of hyperlogarithms with applications to Feynman integrals}",
    eprint = "1403.3385",
    archivePrefix = "arXiv",
    primaryClass = "hep-th",
    doi = "10.1016/j.cpc.2014.10.019",
    journal = "Comput. Phys. Commun.",
    volume = "188",
    pages = "148--166",
    year = "2015"
}

@article{Frellesvig:2023bbf,
    author = "Frellesvig, Hjalte and Morales, Roger and Wilhelm, Matthias",
    title = "{Calabi-Yau Meets Gravity: A Calabi-Yau Threefold at Fifth Post-Minkowskian Order}",
    eprint = "2312.11371",
    archivePrefix = "arXiv",
    primaryClass = "hep-th",
    doi = "10.1103/PhysRevLett.132.201602",
    journal = "Phys. Rev. Lett.",
    volume = "132",
    number = "20",
    pages = "201602",
    year = "2024"
}

@article{Frellesvig:2024rea,
    author = {Frellesvig, Hjalte and Morales, Roger and P\"ogel, Sebastian and Weinzierl, Stefan and Wilhelm, Matthias},
    title = "{Calabi-Yau Feynman integrals in gravity: $\varepsilon$-factorized form for apparent singularities}",
    eprint = "2412.12057",
    archivePrefix = "arXiv",
    primaryClass = "hep-th",
    month = "12",
    year = "2024"
}

@article{Giroux:2024yxu,
    author = "Giroux, Mathieu and Pokraka, Andrzej and Porkert, Franziska and Sohnle, Yoann",
    title = "{The soaring kite: a tale of two punctured tori}",
    eprint = "2401.14307",
    archivePrefix = "arXiv",
    primaryClass = "hep-th",
    reportNumber = "UUITP-03/24, BONN-TH-2024-01",
    doi = "10.1007/JHEP05(2024)239",
    journal = "JHEP",
    volume = "05",
    pages = "239",
    year = "2024"
}

@article{Bogner:2019lfa,
    author = {Bogner, Christian and M\"uller-Stach, Stefan and Weinzierl, Stefan},
    title = "{The unequal mass sunrise integral expressed through iterated integrals on $\overline{\mathcal M}_{1,3}$}",
    eprint = "1907.01251",
    archivePrefix = "arXiv",
    primaryClass = "hep-th",
    doi = "10.1016/j.nuclphysb.2020.114991",
    journal = "Nucl. Phys. B",
    volume = "954",
    pages = "114991",
    year = "2020"
}

@inproceedings{Caola:2022ayt,
    author = "Caola, Fabrizio and Chen, Wen and Duhr, Claude and Liu, Xiaohui and Mistlberger, Bernhard and Petriello, Frank and Vita, Gherardo and Weinzierl, Stefan",
    title = "{The Path forward to N$^3$LO}",
    booktitle = "{Snowmass 2021}",
    eprint = "2203.06730",
    archivePrefix = "arXiv",
    primaryClass = "hep-ph",
    reportNumber = "SLAC-PUB-17658, BONN-TH-2022-06, MITP-22-021",
    month = "3",
    year = "2022"
}

@article{Mastrolia:2018uzb,
    author = "Mastrolia, Pierpaolo and Mizera, Sebastian",
    title = "{Feynman Integrals and Intersection Theory}",
    eprint = "1810.03818",
    archivePrefix = "arXiv",
    primaryClass = "hep-th",
    doi = "10.1007/JHEP02(2019)139",
    journal = "JHEP",
    volume = "02",
    pages = "139",
    year = "2019"
}

@article{Frellesvig:2019uqt,
    author = "Frellesvig, Hjalte and Gasparotto, Federico and Mandal, Manoj K. and Mastrolia, Pierpaolo and Mattiazzi, Luca and Mizera, Sebastian",
    title = "{Vector Space of Feynman Integrals and Multivariate Intersection Numbers}",
    eprint = "1907.02000",
    archivePrefix = "arXiv",
    primaryClass = "hep-th",
    doi = "10.1103/PhysRevLett.123.201602",
    journal = "Phys. Rev. Lett.",
    volume = "123",
    number = "20",
    pages = "201602",
    year = "2019"
}

@article{Landau:1959fi,
	author = "Landau, L.D.",
	title = "{On analytic properties of vertex parts in quantum field theory}",
	doi = "10.1016/B978-0-08-010586-4.50103-6",
	journal = "Nucl. Phys.",
	volume = "13",
	number = "1",
	pages = "181--192",
	year = "1960"
}

@article{10.1143/PTP.22.128,
	author = {Nakanishi, Noboru},
	title = "{Ordinary and Anomalous Thresholds in Perturbation Theory}",
	journal = {Prog. Theor. Phys.},
	fjournal = {Progress of Theoretical Physics},
	volume = {22},
	number = {1},
	pages = {128-144},
	year = {1959},
	month = {07},
	issn = {0033-068X},
	doi = {10.1143/PTP.22.128},
}

@book{Eden:1966dnq,
    author = "Eden, Richard John and Landshoff, Peter V. and Olive, David I. and Polkinghorne, John Charlton",
    title = "{The analytic S-matrix}",
    publisher = "Cambridge Univ. Press",
    address = "Cambridge",
    year = "1966"
}

@phdthesis{Bjorken:1959fd,
	author = "Bjorken, James Daniel",
	title = "{Experimental tests of Quantum electrodynamics and spectral representations of Green's functions in perturbation theory}",
	reportNumber = "RX-1037",
	school = "Stanford U.",
	year = "1959"
}

@article{Coleman:1965xm,
	author = "Coleman, S. and Norton, R.E.",
	title = "{Singularities in the physical region}",
	doi = "10.1007/BF02750472",
	journal = "Nuovo Cim.",
	volume = "38",
	pages = "438--442",
	year = "1965"
}

@book{Hannesdottir:2022bmo,
    author = "Hannesdottir, Holmfridur Sigridar and Mizera, Sebastian",
    title = "{What is the i\ensuremath{\varepsilon} for the S-matrix?}",
    eprint = "2204.02988",
    archivePrefix = "arXiv",
    primaryClass = "hep-th",
    doi = "10.1007/978-3-031-18258-7",
    isbn = "978-3-031-18257-0, 978-3-031-18258-7",
    publisher = "Springer",
    series = "SpringerBriefs in Physics",
    month = "1",
    year = "2023"
}

@article{Mizera:2021icv,
    author = "Mizera, Sebastian and Telen, Simon",
    title = "{Landau discriminants}",
    eprint = "2109.08036",
    archivePrefix = "arXiv",
    primaryClass = "math-ph",
    doi = "10.1007/JHEP08(2022)200",
    journal = "JHEP",
    volume = "08",
    pages = "200",
    year = "2022"
}

@book{pham2011singularities,
  title="{Singularities of integrals: Homology, hyperfunctions and microlocal analysis}",
  author={Pham, F.},
  isbn={9780857296023},
  series={Universitext},
  year={2011},
  publisher={Springer London}
}

@article{Berghoff:2022mqu,
    author = "Berghoff, Marko and Panzer, Erik",
    title = "{Hierarchies in relative Picard-Lefschetz theory}",
    eprint = "2212.06661",
    archivePrefix = "arXiv",
    primaryClass = "math-ph",
    month = "12",
    year = "2022"
}

@article{Correia:2021etg,
    author = "Correia, Miguel and Sever, Amit and Zhiboedov, Alexander",
    title = "{Probing multi-particle unitarity with the Landau equations}",
    eprint = "2111.12100",
    archivePrefix = "arXiv",
    primaryClass = "hep-th",
    reportNumber = "CERN-TH-2021-197",
    doi = "10.21468/SciPostPhys.13.3.062",
    journal = "SciPost Phys.",
    volume = "13",
    number = "3",
    pages = "062",
    year = "2022"
}

@book{Weinzierl:2022eaz,
    author = "Weinzierl, Stefan",
    title = "{Feynman Integrals: A Comprehensive Treatment for Students and Researchers}",
    eprint = "2201.03593",
    archivePrefix = "arXiv",
    primaryClass = "hep-th",
    reportNumber = "MITP/22-001",
    doi = "10.1007/978-3-030-99558-4",
    isbn = "978-3-030-99557-7, 978-3-030-99560-7, 978-3-030-99558-4",
    publisher = "Springer",
    series = "UNITEXT for Physics",
    year = "2022"
}

@article{Frellesvig:2024zph,
    author = "Frellesvig, Hjalte and Morales, Roger and Wilhelm, Matthias",
    title = "{Classifying post-Minkowskian geometries for gravitational waves via loop-by-loop Baikov}",
    eprint = "2405.17255",
    archivePrefix = "arXiv",
    primaryClass = "hep-th",
    doi = "10.1007/JHEP08(2024)243",
    journal = "JHEP",
    volume = "08",
    pages = "243",
    year = "2024"
}

@article{Giroux:2022wav,
    author = "Giroux, Mathieu and Pokraka, Andrzej",
    title = "{Loop-by-loop differential equations for dual (elliptic) Feynman integrals}",
    eprint = "2210.09898",
    archivePrefix = "arXiv",
    primaryClass = "hep-th",
    doi = "10.1007/JHEP03(2023)155",
    journal = "JHEP",
    volume = "03",
    pages = "155",
    year = "2023"
}

@article{Delto:2023kqv,
    author = "Delto, Maximilian and Duhr, Claude and Tancredi, Lorenzo and Zhu, Yu Jiao",
    title = "{Two-Loop QED Corrections to the Scattering of Four Massive Leptons}",
    eprint = "2311.06385",
    archivePrefix = "arXiv",
    primaryClass = "hep-ph",
    reportNumber = "BONN-TH-2023-13, TUM-HEP-1479/23",
    doi = "10.1103/PhysRevLett.132.231904",
    journal = "Phys. Rev. Lett.",
    volume = "132",
    number = "23",
    pages = "231904",
    year = "2024"
}

@article{Duhr:2024bzt,
    author = "Duhr, Claude and Gasparotto, Federico and Nega, Christoph and Tancredi, Lorenzo and Weinzierl, Stefan",
    title = "{On the electron self-energy to three loops in QED}",
    eprint = "2408.05154",
    archivePrefix = "arXiv",
    primaryClass = "hep-th",
    reportNumber = "BONN-TH-2024-12, MITP/24-065, TUM-HEP-1518/24",
    doi = "10.1007/JHEP11(2024)020",
    journal = "JHEP",
    volume = "11",
    pages = "020",
    year = "2024"
}

@article{Jiang:2023oyq,
    author = "Jiang, Xuhang and Lian, Ming and Yang, Li Lin",
    title = "{Recursive structure of Baikov representations: The top-down reduction with intersection theory}",
    eprint = "2312.03453",
    archivePrefix = "arXiv",
    primaryClass = "hep-ph",
    doi = "10.1103/PhysRevD.109.076020",
    journal = "Phys. Rev. D",
    volume = "109",
    number = "7",
    pages = "076020",
    year = "2024"
}

@article{Chen:2022lzr,
    author = "Chen, Jiaqi and Jiang, Xuhang and Ma, Chichuan and Xu, Xiaofeng and Yang, Li Lin",
    title = "{Baikov representations, intersection theory, and canonical Feynman integrals}",
    eprint = "2202.08127",
    archivePrefix = "arXiv",
    primaryClass = "hep-th",
    doi = "10.1007/JHEP07(2022)066",
    journal = "JHEP",
    volume = "07",
    pages = "066",
    year = "2022"
}

@article{Heller:2019gkq,
    author = "Heller, Matthias and von Manteuffel, Andreas and Schabinger, Robert M.",
    title = "{Multiple polylogarithms with algebraic arguments and the two-loop EW-QCD Drell-Yan master integrals}",
    eprint = "1907.00491",
    archivePrefix = "arXiv",
    primaryClass = "hep-th",
    reportNumber = "MITP/19-043, MSUHEP-19-012",
    doi = "10.1103/PhysRevD.102.016025",
    journal = "Phys. Rev. D",
    volume = "102",
    number = "1",
    pages = "016025",
    year = "2020"
}

@article{Henn:2013pwa,
    author = "Henn, Johannes M.",
    title = "{Multiloop integrals in dimensional regularization made simple}",
    eprint = "1304.1806",
    archivePrefix = "arXiv",
    primaryClass = "hep-th",
    doi = "10.1103/PhysRevLett.110.251601",
    journal = "Phys. Rev. Lett.",
    volume = "110",
    pages = "251601",
    year = "2013"
}

@article{Dlapa:2023cvx,
    author = "Dlapa, Christoph and Helmer, Martin and Papathanasiou, Georgios and Tellander, Felix",
    title = "{Symbol alphabets from the Landau singular locus}",
    eprint = "2304.02629",
    archivePrefix = "arXiv",
    primaryClass = "hep-th",
    reportNumber = "DESY-23-048",
    doi = "10.1007/JHEP10(2023)161",
    journal = "JHEP",
    volume = "10",
    pages = "161",
    year = "2023"
}

@article{Cutkosky:1960sp,
	author = "Cutkosky, R. E.",
	title = "{Singularities and discontinuities of Feynman amplitudes}",
	doi = "10.1063/1.1703676",
	journal = "J. Math. Phys.",
	volume = "1",
	pages = "429--433",
	year = "1960"
}

@article{Badger:2023eqz,
    author = "Badger, Simon and Henn, Johannes and Plefka, Jan Christoph and Zoia, Simone",
    title = "{Scattering Amplitudes in Quantum Field Theory}",
    eprint = "2306.05976",
    archivePrefix = "arXiv",
    primaryClass = "hep-th",
    doi = "10.1007/978-3-031-46987-9",
    journal = "Lect. Notes Phys.",
    volume = "1021",
    pages = "pp.",
    year = "2024"
}

@article{Goncharov:2010jf,
    author = "Goncharov, Alexander B. and Spradlin, Marcus and Vergu, C. and Volovich, Anastasia",
    title = "{Classical Polylogarithms for Amplitudes and Wilson Loops}",
    eprint = "1006.5703",
    archivePrefix = "arXiv",
    primaryClass = "hep-th",
    reportNumber = "BROWN-HET-1602",
    doi = "10.1103/PhysRevLett.105.151605",
    journal = "Phys. Rev. Lett.",
    volume = "105",
    pages = "151605",
    year = "2010"
}

@inproceedings{Bourjaily:2022bwx,
    author = "Bourjaily, Jacob L. and others",
    title = "{Functions Beyond Multiple Polylogarithms for Precision Collider Physics}",
    booktitle = "{Snowmass 2021}",
    eprint = "2203.07088",
    archivePrefix = "arXiv",
    primaryClass = "hep-ph",
    reportNumber = "BONN-TH-2022-05, UUITP-11/22, CERN-TH-2022-029, TUM-HEP-1391/22,
  HU-EP-22/08, MITP-22-022",
    month = "3",
    year = "2022"
}

@article{Lippstreu:2023oio,
    author = "Lippstreu, Luke and Spradlin, Marcus and Yelleshpur Srikant, Akshay and Volovich, Anastasia",
    title = "{Landau Singularities of the 7-Point Ziggurat II}",
    eprint = "2305.17069",
    archivePrefix = "arXiv",
    primaryClass = "hep-th",
    month = "5",
    year = "2023"
}

@article{Caron-Huot:2023ikn,
    author = "Caron-Huot, Simon and Giroux, Mathieu and Hannesdottir, Holmfridur S. and Mizera, Sebastian",
    title = "{Crossing beyond scattering amplitudes}",
    eprint = "2310.12199",
    archivePrefix = "arXiv",
    primaryClass = "hep-th",
    doi = "10.1007/JHEP04(2024)060",
    journal = "JHEP",
    volume = "04",
    pages = "060",
    year = "2024"
}

@article{Helmer:2024wax,
    author = "Helmer, Martin and Papathanasiou, Georgios and Tellander, Felix",
    title = "{Landau Singularities from Whitney Stratifications}",
    eprint = "2402.14787",
    archivePrefix = "arXiv",
    primaryClass = "hep-th",
    reportNumber = "DESY-24-023",
    month = "2",
    year = "2024"
}

@article{Fevola:2023fzn,
    author = "Fevola, Claudia and Mizera, Sebastian and Telen, Simon",
    title = "{Principal Landau determinants}",
    eprint = "2311.16219",
    archivePrefix = "arXiv",
    primaryClass = "math-ph",
    doi = "10.1016/j.cpc.2024.109278",
    journal = "Comput. Phys. Commun.",
    volume = "303",
    pages = "109278",
    year = "2024"
}

@article{Brown:2008um,
    author = "Brown, Francis",
    title = "{The Massless higher-loop two-point function}",
    eprint = "0804.1660",
    archivePrefix = "arXiv",
    primaryClass = "math.AG",
    doi = "10.1007/s00220-009-0740-5",
    journal = "Commun. Math. Phys.",
    volume = "287",
    pages = "925--958",
    year = "2009"
}

@phdthesis{Panzer:2015ida,
    author = "Panzer, Erik",
    title = "{Feynman integrals and hyperlogarithms}",
    eprint = "1506.07243",
    archivePrefix = "arXiv",
    primaryClass = "math-ph",
    doi = "10.18452/17157",
    school = "Humboldt U.",
    year = "2015"
}

@misc{repo,
  author       = {},
  title        = {},
  howpublished = {\href{https://github.com/StrangeQuark007/SOFIA}{SOFIA GitHub repository}},
  note         = {GitHub repository, accessed: 2025-02-05},
  year         = {2025}
}

@misc{repoEffortless,
  author       = {},
  title        = {},
  howpublished = {\href{https://github.com/antonela-matijasic/Effortless/tree/main}{Effortless GitHub repository}},
  note         = {GitHub repository, accessed: 2025-02-24},
  year         = {2025}
}

@misc{MathRepo,
  title        = {},
  howpublished = {\href{https://mathrepo.mis.mpg.de/PLD/}{Principal Landau Determinants MathRepo repository}},
  note         = {},
  year         = {2023}
}

@misc{MathRepoTutorial,
  author       = {C. Fevola and S. Mizera and S. Telen},
  title        = {\text{PLD} tutorial},
  howpublished = {\href{https://mathrepo.mis.mpg.de/PLD/PLDTutorial.html}{MathRepo}},
  note         = {},
  year         = {2023}
}

@article{Lee:2009dh,
    author = "Lee, R. N.",
    title = "{Space-time dimensionality D as complex variable: Calculating loop integrals using dimensional recurrence relation and analytical properties with respect to D}",
    eprint = "0911.0252",
    archivePrefix = "arXiv",
    primaryClass = "hep-ph",
    doi = "10.1016/j.nuclphysb.2009.12.025",
    journal = "Nucl. Phys. B",
    volume = "830",
    pages = "474--492",
    year = "2010"
}

@article{Baikov:1996iu,
    author = "Baikov, P. A.",
    editor = "Werlen, M. and Perret-Gallix, D.",
    title = "{Explicit solutions of the multiloop integral recurrence relations and its application}",
    eprint = "hep-ph/9611449",
    archivePrefix = "arXiv",
    reportNumber = "INP-96-42-449",
    doi = "10.1016/S0168-9002(97)00126-5",
    journal = "Nucl. Instrum. Meth. A",
    volume = "389",
    pages = "347--349",
    year = "1997"
}

@phdthesis{Matijasic:2024too,
    author = "Matija\v{s}i\'c, Antonela",
    title = "{Singularity structure of Feynman integrals with applications to six-particle scattering processes}",
    doi = "10.5282/edoc.34154",
    school = "Munich U.",
    year = "2024"
}

@article{Chicherin:2024ifn,
    author = "Chicherin, Dmitry and Moult, Ian and Sokatchev, Emery and Yan, Kai and Zhu, Yunyue",
    title = "{Collinear limit of the four-point energy correlator in $\mathcal{N}=4$ supersymmetric Yang-Mills theory}",
    eprint = "2401.06463",
    archivePrefix = "arXiv",
    primaryClass = "hep-th",
    reportNumber = "LAPTH-004/24",
    doi = "10.1103/PhysRevD.110.L091901",
    journal = "Phys. Rev. D",
    volume = "110",
    number = "9",
    pages = "L091901",
    year = "2024"
}

@article{Marzucca:2023gto,
    author = {Marzucca, Robin and McLeod, Andrew J. and Page, Ben and P\"ogel, Sebastian and Weinzierl, Stefan},
    title = "{Genus drop in hyperelliptic Feynman integrals}",
    eprint = "2307.11497",
    archivePrefix = "arXiv",
    primaryClass = "hep-th",
    reportNumber = "CERN-TH-2023-133, MITP-23-033, ZU-TH 33/23",
    doi = "10.1103/PhysRevD.109.L031901",
    journal = "Phys. Rev. D",
    volume = "109",
    number = "3",
    pages = "L031901",
    year = "2024"
}

@article{Effortlessxxx,
    author = "Matijašić, Antonela and Miczajka, Julia",
    title = "{\texttt{Effortless}}: Efficient generation of odd letters with multiple
roots as leading singularities",
    eprint = "25xx.xxxxx",
    archivePrefix = "arXiv",
    primaryClass = "hep-th",
    year = "2025",
    page = "to appear"
}

@article{Duhr:2021fhk,
    author = "Duhr, Claude and Smirnov, Vladimir A. and Tancredi, Lorenzo",
    title = "{Analytic results for two-loop planar master integrals for Bhabha scattering}",
    eprint = "2108.03828",
    archivePrefix = "arXiv",
    primaryClass = "hep-ph",
    reportNumber = "BONN-TH-2021-06, OUTP-21-19P",
    doi = "10.1007/JHEP09(2021)120",
    journal = "JHEP",
    volume = "09",
    pages = "120",
    year = "2021"
}

@article{Chicherin:2017dob,
    author = "Chicherin, Dmitry and Henn, Johannes and Mitev, Vladimir",
    title = "{Bootstrapping pentagon functions}",
    eprint = "1712.09610",
    archivePrefix = "arXiv",
    primaryClass = "hep-th",
    doi = "10.1007/JHEP05(2018)164",
    journal = "JHEP",
    volume = "05",
    pages = "164",
    year = "2018"
}

@article{Chicherin:2021dyp,
    author = "Chicherin, Dmitry and Sotnikov, Vasily and Zoia, Simone",
    title = "{Pentagon functions for one-mass planar scattering amplitudes}",
    eprint = "2110.10111",
    archivePrefix = "arXiv",
    primaryClass = "hep-ph",
    reportNumber = "LAPTH-041/21, MPP-2021-182",
    doi = "10.1007/JHEP01(2022)096",
    journal = "JHEP",
    volume = "01",
    pages = "096",
    year = "2022"
}

@article{Abreu:2024yit,
    author = "Abreu, Samuel and Chicherin, Dmitry and Sotnikov, Vasily and Zoia, Simone",
    title = "{Two-loop five-point two-mass planar integrals and double Lagrangian insertions in a Wilson loop}",
    eprint = "2408.05201",
    archivePrefix = "arXiv",
    primaryClass = "hep-th",
    reportNumber = "CERN-TH-2024-136, LAPTH-043/24, ZU-TH 40/24",
    doi = "10.1007/JHEP10(2024)167",
    journal = "JHEP",
    volume = "10",
    pages = "167",
    year = "2024"
}

@article{Brandhuber:2021eyq,
    author = "Brandhuber, Andreas and Chen, Gang and Travaglini, Gabriele and Wen, Congkao",
    title = "{Classical gravitational scattering from a gauge-invariant double copy}",
    eprint = "2108.04216",
    archivePrefix = "arXiv",
    primaryClass = "hep-th",
    reportNumber = "QMUL-PH-21-18, SAGEX-21-07",
    doi = "10.1007/JHEP10(2021)118",
    journal = "JHEP",
    volume = "10",
    pages = "118",
    year = "2021"
}

@article{Henn:2025xrc,
    author = "Henn, Johannes and Matija\v{s}i\'c, Antonela and Miczajka, Julian and Peraro, Tiziano and Xu, Yingxuan and Zhang, Yang",
    title = "{Complete function space for planar two-loop six-particle scattering amplitudes}",
    eprint = "2501.01847",
    archivePrefix = "arXiv",
    primaryClass = "hep-ph",
    reportNumber = "HU-EP-25/01-RTG, MITP-25-001, MPP-2025-2, USTC-ICTS/PCFT-25-01",
    month = "1",
    year = "2025"
}

@article{Hofman:2008ar,
    author = "Hofman, Diego M. and Maldacena, Juan",
    title = "{Conformal collider physics: Energy and charge correlations}",
    eprint = "0803.1467",
    archivePrefix = "arXiv",
    primaryClass = "hep-th",
    doi = "10.1088/1126-6708/2008/05/012",
    journal = "JHEP",
    volume = "05",
    pages = "012",
    year = "2008"
}

@article{Larkoski:2017jix,
    author = "Larkoski, Andrew J. and Moult, Ian and Nachman, Benjamin",
    title = "{Jet Substructure at the Large Hadron Collider: A Review of Recent Advances in Theory and Machine Learning}",
    eprint = "1709.04464",
    archivePrefix = "arXiv",
    primaryClass = "hep-ph",
    doi = "10.1016/j.physrep.2019.11.001",
    journal = "Phys. Rept.",
    volume = "841",
    pages = "1--63",
    year = "2020"
}

@article{Benincasa:2024lxe,
    author = "Benincasa, Paolo and Vaz\~ao, Francisco",
    title = "{The Asymptotic Structure of Cosmological Integrals}",
    eprint = "2402.06558",
    archivePrefix = "arXiv",
    primaryClass = "hep-th",
    month = "2",
    year = "2024"
}

@article{Benincasa:2024ptf,
    author = "Benincasa, Paolo and Brunello, Giacomo and Mandal, Manoj K. and Mastrolia, Pierpaolo and Vaz\~ao, Francisco",
    title = "{On one-loop corrections to the Bunch-Davies wavefunction of the universe}",
    eprint = "2408.16386",
    archivePrefix = "arXiv",
    primaryClass = "hep-th",
    month = "8",
    year = "2024"
}

@article{Bosma:2017ens,
    author = "Bosma, Jorrit and Sogaard, Mads and Zhang, Yang",
    title = "{Maximal Cuts in Arbitrary Dimension}",
    eprint = "1704.04255",
    archivePrefix = "arXiv",
    primaryClass = "hep-th",
    doi = "10.1007/JHEP08(2017)051",
    journal = "JHEP",
    volume = "08",
    pages = "051",
    year = "2017"
}

@article{Primo:2016ebd,
    author = "Primo, Amedeo and Tancredi, Lorenzo",
    title = "{On the maximal cut of Feynman integrals and the solution of their differential equations}",
    eprint = "1610.08397",
    archivePrefix = "arXiv",
    primaryClass = "hep-ph",
    reportNumber = "TTP16-046",
    doi = "10.1016/j.nuclphysb.2016.12.021",
    journal = "Nucl. Phys. B",
    volume = "916",
    pages = "94--116",
    year = "2017"
}

@article{Kotikov:1990kg,
    author = "Kotikov, A. V.",
    title = "{Differential equations method: New technique for massive Feynman diagrams calculation}",
    reportNumber = "ITF-90-31E",
    doi = "10.1016/0370-2693(91)90413-K",
    journal = "Phys. Lett. B",
    volume = "254",
    pages = "158--164",
    year = "1991"
}

@article{Gorges:2023zgv,
    author = {G\"orges, Lennard and Nega, Christoph and Tancredi, Lorenzo and Wagner, Fabian J.},
    title = "{On a procedure to derive \ensuremath{\varepsilon}-factorised differential equations beyond polylogarithms}",
    eprint = "2305.14090",
    archivePrefix = "arXiv",
    primaryClass = "hep-th",
    doi = "10.1007/JHEP07(2023)206",
    journal = "JHEP",
    volume = "07",
    pages = "206",
    year = "2023"
}

@Inbook{Cox2015,
author="Cox, David A.
and Little, John
and O'Shea, Donal",
title="The Dimension of a Variety",
bookTitle="Ideals, Varieties, and Algorithms: An Introduction to Computational Algebraic Geometry and Commutative Algebra",
year="2015",
publisher="Springer International Publishing",
address="Cham",
pages="469--538",
isbn="978-3-319-16721-3",
doi="10.1007/978-3-319-16721-3_9",
url="https://doi.org/10.1007/978-3-319-16721-3_9"
}

@article{Hannesdottir:2024hke,
    author = "Hannesdottir, Holmfridur and McLeod, Andrew and Schwartz, Matthew D. and Vergu, Cristian",
    title = "{The Landau Bootstrap}",
    eprint = "2410.02424",
    archivePrefix = "arXiv",
    primaryClass = "hep-ph",
    month = "10",
    year = "2024"
}

@article{Peraro:2019svx,
    author = "Peraro, Tiziano",
    title = "{\texttt{FiniteFlow}: multivariate functional reconstruction using finite fields and dataflow graphs}",
    eprint = "1905.08019",
    archivePrefix = "arXiv",
    primaryClass = "hep-ph",
    reportNumber = "ZU-TH 24/19",
    doi = "10.1007/JHEP07(2019)031",
    journal = "JHEP",
    volume = "07",
    pages = "031",
    year = "2019"
}

@article{Jiang:2023qnl,
    author = "Jiang, Xuhang and Yang, Li Lin",
    title = "{Recursive structure of Baikov representations: Generics and application to symbology}",
    eprint = "2303.11657",
    archivePrefix = "arXiv",
    primaryClass = "hep-ph",
    doi = "10.1103/PhysRevD.108.076004",
    journal = "Phys. Rev. D",
    volume = "108",
    number = "7",
    pages = "076004",
    year = "2023"
}

@article{Jiang:2024eaj,
    author = "Jiang, Xuhang and Liu, Jiahao and Xu, Xiaofeng and Yang, Li Lin",
    title = "{Symbol letters of Feynman integrals from Gram determinants}",
    eprint = "2401.07632",
    archivePrefix = "arXiv",
    primaryClass = "hep-ph",
    month = "1",
    year = "2024"
}

@article{Hannesdottir:2024cnn,
    author = "Hannesdottir, Holmfridur S. and Lippstreu, Luke and McLeod, Andrew J. and Polackova, Maria",
    title = "{Minimal Cuts and Genealogical Constraints on Feynman Integrals}",
    eprint = "2406.05943",
    archivePrefix = "arXiv",
    primaryClass = "hep-th",
    month = "6",
    year = "2024"
}

@article{Bern:2024adl,
    author = "Bern, Zvi and Herrmann, Enrico and Roiban, Radu and Ruf, Michael S. and Smirnov, Alexander V. and Smirnov, Vladimir A. and Zeng, Mao",
    title = "{Amplitudes, supersymmetric black hole scattering at $ \mathcal{O}\left({G}^5\right) $, and loop integration}",
    eprint = "2406.01554",
    archivePrefix = "arXiv",
    primaryClass = "hep-th",
    doi = "10.1007/JHEP10(2024)023",
    journal = "JHEP",
    volume = "10",
    pages = "023",
    year = "2024"
}

@article{Muller:2022gec,
    author = {M\"uller, Hildegard and Weinzierl, Stefan},
    title = "{A Feynman integral depending on two elliptic curves}",
    eprint = "2205.04818",
    archivePrefix = "arXiv",
    primaryClass = "hep-th",
    reportNumber = "MITP/22-037",
    doi = "10.1007/JHEP07(2022)101",
    journal = "JHEP",
    volume = "07",
    pages = "101",
    year = "2022"
}

@article{Pogel:2022vat,
    author = {P\"ogel, Sebastian and Wang, Xing and Weinzierl, Stefan},
    title = "{Bananas of equal mass: any loop, any order in the dimensional regularisation parameter}",
    eprint = "2212.08908",
    archivePrefix = "arXiv",
    primaryClass = "hep-th",
    doi = "10.1007/JHEP04(2023)117",
    journal = "JHEP",
    volume = "04",
    pages = "117",
    year = "2023"
}

@book{Hubsch:1992nu,
    author = "Hubsch, Tristan",
    title = "{Calabi-Yau manifolds: A Bestiary for physicists}",
    isbn = "978-981-02-1927-7",
    publisher = "World Scientific",
    address = "Singapore",
    year = "1994"
}

@article{Abreu:2020jxa,
    author = "Abreu, Samuel and Ita, Harald and Moriello, Francesco and Page, Ben and Tschernow, Wladimir and Zeng, Mao",
    title = "{Two-Loop Integrals for Planar Five-Point One-Mass Processes}",
    eprint = "2005.04195",
    archivePrefix = "arXiv",
    primaryClass = "hep-ph",
    doi = "10.1007/JHEP11(2020)117",
    journal = "JHEP",
    volume = "11",
    pages = "117",
    year = "2020"
}

@article{Abreu:2021smk,
    author = "Abreu, Samuel and Ita, Harald and Page, Ben and Tschernow, Wladimir",
    title = "{Two-loop hexa-box integrals for non-planar five-point one-mass processes}",
    eprint = "2107.14180",
    archivePrefix = "arXiv",
    primaryClass = "hep-ph",
    doi = "10.1007/JHEP03(2022)182",
    journal = "JHEP",
    volume = "03",
    pages = "182",
    year = "2022"
}

@article{Duhr:2011zq,
    author = "Duhr, Claude and Gangl, Herbert and Rhodes, John R.",
    title = "{From polygons and symbols to polylogarithmic functions}",
    eprint = "1110.0458",
    archivePrefix = "arXiv",
    primaryClass = "math-ph",
    reportNumber = "IPPP-11-56, DCPT-11-112",
    doi = "10.1007/JHEP10(2012)075",
    journal = "JHEP",
    volume = "10",
    pages = "075",
    year = "2012"
}

@article{Lee:2024kkm,
    author = "Lee, Roman N.",
    title = "{Polylogarithmic functions with prescribed branching locus and linear relations between them}",
    eprint = "2407.12503",
    archivePrefix = "arXiv",
    primaryClass = "hep-th",
    month = "7",
    year = "2024"
}

@article{FebresCordero:2023pww,
    author = "Febres Cordero, F. and Figueiredo, G. and Kraus, M. and Page, B. and Reina, L.",
    title = "{Two-loop master integrals for leading-color $ pp\to t\overline{t}H $ amplitudes with a light-quark loop}",
    eprint = "2312.08131",
    archivePrefix = "arXiv",
    primaryClass = "hep-ph",
    reportNumber = "CERN-TH-2023-240",
    doi = "10.1007/JHEP07(2024)084",
    journal = "JHEP",
    volume = "07",
    pages = "084",
    year = "2024"
}

@article{Badger:2024fgb,
    author = "Badger, Simon and Becchetti, Matteo and Giraudo, Nicol\`o and Zoia, Simone",
    title = "{Two-loop integrals for $ t\overline{t} $+jet production at hadron colliders in the leading colour approximation}",
    eprint = "2404.12325",
    archivePrefix = "arXiv",
    primaryClass = "hep-ph",
    reportNumber = "CERN-TH-2024-048, ZU-TH 22/24",
    doi = "10.1007/JHEP07(2024)073",
    journal = "JHEP",
    volume = "07",
    pages = "073",
    year = "2024"
}

@article{Becchetti:2025oyb,
    author = "Becchetti, Matteo and Dlapa, Christoph and Zoia, Simone",
    title = "{Canonical differential equations for the elliptic two-loop five-point integral family relevant to $t\bar t +$jet production at leading colour}",
    eprint = "2503.03603",
    archivePrefix = "arXiv",
    primaryClass = "hep-th",
    reportNumber = "DESY 25-029, ZU-TH 13/25",
    month = "3",
    year = "2025"
}

@article{Abreu:2024fei,
    author = "Abreu, Samuel and Monni, Pier Francesco and Page, Ben and Usovitsch, Johann",
    title = "{Planar Six-Point Feynman Integrals for Four-Dimensional Gauge Theories}",
    eprint = "2412.19884",
    archivePrefix = "arXiv",
    primaryClass = "hep-ph",
    reportNumber = "CERN-TH-2024-221, HU-EP-24/40-RTG",
    month = "12",
    year = "2024"
}

@article{He:2024fij,
    author = "He, Song and Jiang, Xuhang and Liu, Jiahao and Yang, Qinglin",
    title = "{Landau-based Schubert analysis}",
    eprint = "2410.11423",
    archivePrefix = "arXiv",
    primaryClass = "hep-th",
    month = "10",
    year = "2024"
}

@article{Morales:2022csr,
    author = "Morales, Roger and Spiering, Anne and Wilhelm, Matthias and Yang, Qinglin and Zhang, Chi",
    title = "{Bootstrapping Elliptic Feynman Integrals Using Schubert Analysis}",
    eprint = "2212.09762",
    archivePrefix = "arXiv",
    primaryClass = "hep-th",
    doi = "10.1103/PhysRevLett.131.041601",
    journal = "Phys. Rev. Lett.",
    volume = "131",
    number = "4",
    pages = "041601",
    year = "2023"
}

@article{Mandelstam:1959bc,
    author = "Mandelstam, Stanley",
    title = "{Analytic properties of transition amplitudes in perturbation theory}",
    doi = "10.1103/PhysRev.115.1741",
    journal = "Phys. Rev.",
    volume = "115",
    pages = "1741--1751",
    year = "1959"
}

@article{Gribov:1962ft,
    author = "Gribov, V. N. and Dyatlov, I. T.",
    title = "{Analytic continuation of the three-particle unitarity condition. Simplest diagrams}",
    journal = "Sov. Phys. JETP",
    volume = "15",
    pages = "140",
    year = "1962"
}

@article{islam1965analytic,
  title={Analytic Property of Three-Body Unitarity Integral},
  author={Islam, Jamal N and Kim, YS},
  journal={Physical Review},
  volume={138},
  number={5B},
  pages={B1222},
  year={1965},
  publisher={APS}
}

@article{Guerrieri:2024ckc,
    author = "Guerrieri, Andrea and Homrich, Alexandre and Vieira, Pedro",
    title = "{Multiparticle Flux-Tube S-matrix Bootstrap}",
    eprint = "2404.10812",
    archivePrefix = "arXiv",
    primaryClass = "hep-th",
    doi = "10.1103/PhysRevLett.134.041601",
    journal = "Phys. Rev. Lett.",
    volume = "134",
    number = "4",
    pages = "041601",
    year = "2025"
}

@article{Correia:2022dcu,
    author = "Correia, Miguel",
    title = "{Nonperturbative anomalous thresholds}",
    eprint = "2212.06157",
    archivePrefix = "arXiv",
    primaryClass = "hep-th",
    doi = "10.1103/PhysRevD.110.025012",
    journal = "Phys. Rev. D",
    volume = "110",
    number = "2",
    pages = "025012",
    year = "2024"
}

@article{Tourkine:2023xtu,
    author = "Tourkine, Piotr and Zhiboedov, Alexander",
    title = "{Scattering amplitudes from dispersive iterations of unitarity}",
    eprint = "2303.08839",
    archivePrefix = "arXiv",
    primaryClass = "hep-th",
    reportNumber = "CERN-TH-2023-025",
    doi = "10.1007/JHEP11(2023)005",
    journal = "JHEP",
    volume = "11",
    pages = "005",
    year = "2023"
}

@article{Homrich:2019cbt,
    author = "Homrich, Alexandre and Penedones, Jo\~ao and Toledo, Jonathan and van Rees, Balt C. and Vieira, Pedro",
    title = "{The S-matrix Bootstrap IV: Multiple Amplitudes}",
    eprint = "1905.06905",
    archivePrefix = "arXiv",
    primaryClass = "hep-th",
    doi = "10.1007/JHEP11(2019)076",
    journal = "JHEP",
    volume = "11",
    pages = "076",
    year = "2019"
}

@article{Coleman:1978kk,
    author = "Coleman, Sidney R. and Thun, H. J.",
    title = "{On the Prosaic Origin of the Double Poles in the {Sine-Gordon} S-matrix}",
    reportNumber = "HUTP-78/A002",
    doi = "10.1007/BF01609466",
    journal = "Commun. Math. Phys.",
    volume = "61",
    pages = "31",
    year = "1978"
}

@article{Braden:1990wx,
    author = "Braden, H. W. and Corrigan, Edward and Dorey, P. E. and Sasaki, R.",
    title = "{Multiple poles and other features of affine Toda field theory}",
    reportNumber = "YITP-U-90-25, NSF-ITP-90-174, DTP-90-57",
    doi = "10.1016/0550-3213(91)90317-Q",
    journal = "Nucl. Phys. B",
    volume = "356",
    pages = "469--498",
    year = "1991"
}

@article{Dorey:2022fvs,
    author = "Dorey, Patrick and Polvara, Davide",
    title = "{From tree- to loop-simplicity in affine Toda theories I: Landau singularities and their subleading coefficients}",
    eprint = "2206.09368",
    archivePrefix = "arXiv",
    primaryClass = "hep-th",
    reportNumber = "SAGEX-22-25-E",
    doi = "10.1007/JHEP09(2022)220",
    journal = "JHEP",
    volume = "09",
    pages = "220",
    year = "2022"
}

@article{Fabri:2024xmi,
    author = "Fabri, Matheus and Polvara, Davide",
    title = "{One-loop integrability with shifting masses}",
    eprint = "2411.15080",
    archivePrefix = "arXiv",
    primaryClass = "hep-th",
    reportNumber = "ZMP-HH/24-27",
    month = "11",
    year = "2024"
}

@article{KORS,
  title={Location of the Nearest Singularities of the $\pi$$\pi$-scattering Amplitude},
  author={Kolkunov, VA and Okun, LB and Rudik, AP and Sudakov, VV},
  journal={JETP},
  volume={12},
  number={2},
  pages={242--244},
  year={1961}
}

@article{kolkunov1960singular,
  title={The Singular Points of Some Feynman Diagrams},
  author={Kolkunov, VA and Okun, LB and Rudik, AP},
  journal={JETP},
  volume={11},
  number={3},
  pages={634--636},
  year={1960},
 publisher={Inst. of Physics and Technology, Academy of Sciences, USSR}
}

@article{Kruczenski:2022lot,
    author = "Kruczenski, Martin and Penedones, Joao and van Rees, Balt C.",
    title = "{Snowmass White Paper: S-matrix Bootstrap}",
    eprint = "2203.02421",
    archivePrefix = "arXiv",
    primaryClass = "hep-th",
    month = "3",
    year = "2022"
}

@article{Correia:2020xtr,
    author = "Correia, Miguel and Sever, Amit and Zhiboedov, Alexander",
    title = "{An analytical toolkit for the S-matrix bootstrap}",
    eprint = "2006.08221",
    archivePrefix = "arXiv",
    primaryClass = "hep-th",
    reportNumber = "CERN-TH-2020-095",
    doi = "10.1007/JHEP03(2021)013",
    journal = "JHEP",
    volume = "03",
    pages = "013",
    year = "2021"
}

@book{Horn_Johnson_1985, place={Cambridge}, title={Matrix Analysis}, publisher={Cambridge University Press}, author={Horn, Roger A. and Johnson, Charles R.}, year={1985}}

@article{Bogner:2017xhp,
    author = "Bogner, C. and Borowka, S. and Hahn, T. and Heinrich, G. and Jones, S. P. and Kerner, M. and von Manteuffel, A. and Michel, M. and Panzer, E. and Papara, V.",
    title = "{Loopedia, a Database for Loop Integrals}",
    eprint = "1709.01266",
    archivePrefix = "arXiv",
    primaryClass = "hep-ph",
    reportNumber = "CERN-TH-2017-175, CP3-17-26, MAPHY-AVH-2017-07, MSUHEP-17-013, MPP-2017-173",
    doi = "10.1016/j.cpc.2017.12.017",
    journal = "Comput. Phys. Commun.",
    volume = "225",
    pages = "1--9",
    year = "2018"
}

@article{Matsubara-Heo:2023ylc,
    author = "Matsubara-Heo, Saiei-Jaeyeong and Mizera, Sebastian and Telen, Simon",
    title = "{Four lectures on Euler integrals}",
    eprint = "2306.13578",
    archivePrefix = "arXiv",
    primaryClass = "math-ph",
    doi = "10.21468/SciPostPhysLectNotes.75",
    journal = "SciPost Phys. Lect. Notes",
    volume = "75",
    pages = "1",
    year = "2023"
}

@article{Bargiela:2024hgt,
    author = "Bargiela, Piotr",
    title = "{Integrated unitarity for scattering amplitudes and the four-loop ladder Feynman integral}",
    eprint = "2403.18047",
    archivePrefix = "arXiv",
    primaryClass = "hep-th",
    reportNumber = "ZU-TH 18/24",
    doi = "10.1007/JHEP05(2025)004",
    journal = "JHEP",
    volume = "05",
    pages = "004",
    year = "2025"
}

@article{Vergu:2023rqz,
    author = "Vergu, C.",
    title = "{Cutkosky representation and direct integration}",
    eprint = "2311.16069",
    archivePrefix = "arXiv",
    primaryClass = "hep-th",
    doi = "10.1007/JHEP05(2024)302",
    journal = "JHEP",
    volume = "05",
    pages = "302",
    year = "2024"
}

\end{document}